\documentclass[prd,%
reprint,
superscriptaddress,
frontmatterverbose, 
nofootinbib,
amsmath,amssymb,
aps,
prstab,
prstper,
floatfix,
]{revtex4-2}

\usepackage{cancel}
\usepackage{graphicx}
\usepackage{dcolumn}
\usepackage{bm}
\usepackage{hyperref}
\usepackage[mathlines]{lineno}
\usepackage[caption=false]{subfig}
\usepackage{ulem} 
\usepackage{comment}
\usepackage{orcidlink}
\usepackage{multirow}

\usepackage{tikz}
\usetikzlibrary{positioning, shapes.geometric, calc}
\usetikzlibrary{arrows.meta}

\usepackage[
scale=0.7, marginratio={1:1, 2:3}, ignoreall,
text={7in,10in},centering,
margin=0.24in, 
total={6.5in,8.75in}, top=1.2in, left=0.6in,right = 0.6in, includefoot, 
]{geometry}

\usepackage{setspace}
\hypersetup{
	colorlinks=true, linkcolor=blue, 
    urlcolor=purple, 
	citecolor= cyan, 
    linktocpage = true
}
\usepackage{hyperref}
\usepackage{url}
\usepackage{natbib}
\usepackage{standalone}

\usepackage{enumitem}
\usepackage[caption=false]{subfig}  
\usepackage{mathrsfs}
\usepackage{bbm}
\usepackage{array} 
\usepackage{multirow}
\usepackage{booktabs} 
\usepackage{amssymb}

\usepackage{xcolor}
\usepackage{pifont}
\newcommand{\cmark}{\textcolor{green}{\checkmark}}
\newcommand{\xmark}{\textcolor{red}{\ding{55}}}

\begin{document}

\preprint{APS/123-QED}



\title{Assessing Universal Relations for Rapidly Rotating Neutron Stars: \\ Insights from an Interpretable Deep Learning Perspective}

\author{\large Grigorios Papigkiotis \orcidlink{0009-0008-2205-7426}}
\email{gpapigki@auth.gr}
\affiliation{\large Department of Physics, Aristotle University of Thessaloniki,\\ Thessaloniki 54124, Greece}
\author{\large Georgios Vardakas \orcidlink{0000-0003-1352-2062}}
\email{g.vardakas@uoi.gr}
\affiliation{\large Department of Computer Science $\&$ Engineering, University of Ioannina,\\ Ioannina 45110, Greece}
\author{\large Nikolaos Stergioulas \orcidlink{0000-0002-5490-5302}}
\email{niksterg@auth.gr}
\affiliation{\large Department of Physics, Aristotle University of Thessaloniki,\\ Thessaloniki 54124, Greece}

\date{\today}

\begin{abstract}

Relations between stellar properties independent of the nuclear equation of state offer profound insights into neutron star physics and have practical applications in data analysis by breaking the degeneracy between the parameters of interest. Commonly, these relations are derived from utilizing various realistic nuclear cold hadronic, hyperonic, and hybrid EoS models, each of which should obey the current multimessenger constraints and cover a wide range of stiffnesses. Concurrently, the field of multimessenger astronomy has been significantly enhanced by the advent of gravitational wave astronomy, which increasingly incorporates deep learning techniques and algorithms. At the same time, X-ray spectral data from NICER based on known pulsars are available, and additional observations are expected from upcoming missions. Therefore, in the coming years, a wealth of information on neutron stars is expected to be gathered from multiple observational channels. In this study, we revisit established universal relations, introduce new ones, and reassess them using a feed-forward neural network as a regression model. More specifically, we mainly propose ``deep'' EoS-insensitive hypersurface relations for rapidly rotating compact objects between several of the star's global parameters, such as the moment of inertia, the reciprocal of the moment of inertia, the rotational frequency, mass and current multipole moments and the equatorial radius, which achieve an accuracy of within $1\%$ in most cases, with only a small fraction of investigated models exceeding this threshold. While analytical expressions can be used to represent some of these relations, the neural network approach demonstrates superior performance, particularly in complex regions of the parameter space. Furthermore, we use the SHapley Additive exPlanations (SHAP) method to interpret the predictions of the suggested feed-forward network, as SHAP is grounded in a robust theoretical framework inspired by the field of cooperative Game Theory. Most importantly, these highly accurate universal relations, empowered with the interpretability description, could be used in efforts to constrain the high-density equation of state in neutron stars, with the potential to enhance our understanding as new observables emerge.

\end{abstract}

\maketitle

\section{Introduction}

Over the last few years, a wealth of observational data on neutron stars (NSs) has emerged from diverse sources, including gravitational wave ground-based detectors such as Advanced LIGO \cite{aasi2015advanced} and Advanced Virgo \cite{acernese2014advanced}, as well as electromagnetic radiation, such as the NICER mission that focuses on observing X-ray pulse profile emissions emanating from hotspots on the surfaces of NSs \cite{gendreau2012neutron, arzoumanian2014neutron, gendreau2017searching, bogdanov2019constraining, riley2021nicer, choudhury2024nicer, kurpas2024detection, chatziioannou2024neutron}. A significant milestone in astronomy occurred in 2017 with the detection of gravitational and electromagnetic signals from the merger of two NSs, known as the GW170817 event \cite{abbott2017gw170817, coulter2017swope, Abbot2017multi,abbott2017gravitational, cowperthwaite2017electromagnetic, soares2017electromagnetic,nicholl2017electromagnetic,margutti2017electromagnetic,margutti2017electromagnetic,chornock2017electromagnetic, abbott2019properties}. Since then, the LIGO-Virgo collaboration (with the later addition of Kagra \cite{akutsu2021overview}) has announced several binary neutron stars (BNS) and neutron star-black hole  (NSBH) systems \cite{abbott2020gw190425, abbott2021observation, abbott2021gwtc, abbott2023gwtc,williams2024beyond}. However, unlike the first BNS detection, none of these events were accompanied by electromagnetic counterparts.  Recently, the LVK global network announced the detection of an NS and a BH with a mass within the mass gap \cite{ligo2024observation}. This observation suggests an increase in the expected rate of NSBH  mergers with electromagnetic counterparts, providing additional evidence supporting the existence of compact objects within the conjectured lower mass gap.

These advances not only opened a new window for understanding and determining the global properties of these unique cosmic laboratories \cite{bauswein2017neutron,radice2018gw170817,margalit2017constraining,rezzolla2018using, ruiz2018gw170817}, but also provided a new perspective for unraveling the poorly understood fundamental characteristics of the high-dense nuclear matter that exists in their cores. This uncertainty is reﬂected by the fact that it is essential to establish the relation between the NS's global properties (such as mass and radius) and the underlying stellar-core microphysics governed by the equation of state (EoS). Assuming the nuclear matter within the star is well described by a perfect fluid, the intrinsic fundamental properties are encapsulated by a barotropic EoS, identifying the relation between the pressure and the energy density of matter \cite{ozel_masses_2016, burgio2021modern,compose2022compose,oertel2017equations,lattimer2021neutron,chatziioannou2024neutron}.

The precise determination of NS global properties such as masses and equatorial radii, is vital for unveiling insights into the EoS \cite{baym2018hadrons, lattimer2021neutron, chatziioannou2024neutron, miller2019psr, miller2021radius, drischler2021limiting, imam2024implications}. While accurate mass measurements have been achieved through double pulsar systems using radio astronomy, equatorial radii remain challenging to determine due to substantial systematic errors \cite{miller2019psr,miller2021radius,chatziioannou2024neutron,coleman2016observational,riley2019nicer,riley2021nicer,choudhury2024nicer}. At the same time, in the realm of multimessenger astronomy, a wealth of astrophysical observations has driven extensive efforts to constrain the EoS, leading to a deeper understanding of the fundamental particle interactions within NSs' interiors. These efforts encompass utilizing NICER measurements of mass and radius \cite{riley2019nicer, miller2019psr, miller2021radius, chatziioannou2024neutron, pang2021nuclear, choudhury2024nicer, kurpas2024detection}, assessing tidal deformability through gravitational waves (GWs) \cite{van2017upper, hinderer2008tidal, binnington2009relativistic, damour2009relativistic, chatziioannou2020neutron,dietrich2021interpreting,agathos2015constraining,chatziioannou2015probing,paschalidis2018implications,abbott2019ligo_prl,kumar2019inferring,kumar2019inferring,carson2019equation,tan2020neutron,narikawa2021gravitational,bauswein2019equation,gamba2023resonant,ripley2024constraint,williams2024phenomenological,carson2019future, huxford2024accuracy,francesco2023nuclear}, and applying joint constraints \cite{raaijmakers2020constraining, biswas2022constraining, biswas2021impact, biswas2022bayesian, traversi2020bayesian, xie2019bayesian, biswas2021towards, dietrich2020multimessenger, landry2020nonparametric, raaijmakers2021constraints, bauswein2019equation, ripley2024constraint, williams2024phenomenological, carson2019future, huxford2024accuracy, huxford2024accuracy, ho2023new, al2021combining, al2021combining}. Notably, the first BNS detection has stimulated further studies in this field \cite{jiang2019equation, fattoyev2018neutron, most2018new, abbott2018gw170817, landry2018constraints, annala2018gravitational, lim2018neutron, kumar2019inferring, ho2023new, ripley2024constraint, williams2024phenomenological, huxford2024accuracy}.


 Meanwhile, a wide variety of nuclear-matter EoS models have been proposed, each offering unique predictions regarding the properties of NS sequences. These models rely on distinct assumptions concerning NS crust and core composition, nucleon interaction properties, and theoretical methodologies for addressing the many-body problem. Consequently, the observed global features of NSs serve as invaluable constraints for refining and validating these theoretical EoS candidates \cite{lattimer2021neutron, compose2022compose, haensel_neutron_2007}. Based on the observational data related to the macroscopic properties of NSs, common Bayesian methods have been implemented to infer the EoS \cite{miller2021radius,raaijmakers2020constraining, biswas2022constraining, biswas2021impact, biswas2022bayesian, traversi2020bayesian, xie2019bayesian, biswas2021towards, dietrich2020multimessenger, al2021combining, kruger2023rapidly, beznogov2024bayesian}. Furthermore, various studies have also utilized Machine Learning (ML) and Deep Learning (DL) techniques to investigate or reconstruct the nuclear matter EoS based on the NS observable quantities (see e.g., the Refs. \cite{fujimoto2020mapping, fujimoto2018methodology, ferreira2021unveiling, Morawski2020,soma2022neural, soma2023reconstructing, lobato2022cluster, lobato2022unsupervised, ferreira2022extracting, krastev2023deep, di2025explainable, farrell2023deducing, carvalho2023decoding} for a review).

Although the internal structure and global properties (both static and dynamic) of an NS depend sensitively on the specific choice of the EoS, it is quite fascinating that various EoS-independent relations have been discovered. These relations apply to both stationary isolated NSs and fully dynamical and merging BNS systems, see e.g., the Refs \cite{riahi2019universal,rezzolla_physics_2018,yagi2017approximate, manoharan2021universal} for a review. Typically, they connect various physical quantities of stellar configurations and are considered universal in the sense that they are insensitive to EoS models to an order of $\mathcal{O}(1\%)$ accuracy \cite{yagi2017approximate}. Also, the universal behavior between observable quantities is highly sensitive to parameter selection \cite{rezzolla_physics_2018}. However, such relations are typically derived empirically by fitting a diverse array of EoS models, obtained from either phenomenological (agnostic) or theoretical nuclear frameworks. As a result, their applicability is limited to the nuclear physics represented within the selected set of EoSs. Deviations from these relations might suggest the presence of new physics \cite{wijngaarden2022probing, bauswein2019identifying, most2019signatures}. In addition, the breakdown of universal behavior in a catalog of observations can also be used to identify outliers, which may be attributed to quark matter \cite{chatziioannou2020studying} or NSBH binaries \cite{chen2020distinguishing}.

The impact of EoS-independent relations transcends the realms of NS physics, offering insights into fundamental properties while enabling practical data analysis applications. Initially, these relations identified a crucial link between the normalized moment of inertia and stellar compactness \cite{ravenhall1994neutron}. Subsequent advancements refined this understanding, particularly in estimating NS radii by analyzing mass and moment of inertia data obtained from pulsars in binary systems \cite{lattimer2001neutron, bejger2002moments, lattimer2005constraining}. Furthermore, additional relations concerning the star's reduced moment of inertia $\bar{I}$ were proposed, addressing slowly and rapidly rotating configurations \cite{breu2016maximum}.

Additionally, a significant portion of these relations pertains to the multipole moments of NSs. More specifically,  a quadratic correlation between the star's quadrupole moment $Q$ and its angular momentum $J$ has been unveiled \cite{laarakkers1999quadrupole}. Also, further investigations have identified a cubic dependence between the spin octupole moment $S_3$ and the star's angular momentum $J$, suggesting a Kerr-like behavior \cite{pappas2012revising,pappas2012multipole}. By employing the Hartle-Thorne slow-rotation approximation, a universal connection between the reduced quadrupole moment $\bar{Q}$ and the inverse compactness has been established for slowly rotating NSs \cite{urbanec2013quadrupole}. Building upon this, a new association between the quadrupole deformation $\bar{Q}$ and the star's stellar compactness has been revealed for slowly rotating NSs as well as quark stars \cite{yagi2013love}.

On top of that, Yagi and Yunes discovered the I-Love-Q relations, showing high accuracy for slowly rotating NSs \cite{yagi2013love, yagi2017approximate}. Recent studies have updated the original I-Love-Q relations and introduced new ones incorporating parameters such as $I$, $\Lambda$, and $C$ (see e.g., Refs. \cite{jiang2020analytic, lowrey2025improved, aranguren2023revisiting, chatterjee2025insights, suleiman2024quasiuniversal,  saes2022equation, godzieba2021updated, legred2024assessing, nath2023love,kumar2024universal, roy2025icq}). These relations serve as a valuable tool for eliminating degeneracies when modeling gravitational wave signals from inspiraling binaries \cite{yagi2013love_b, maselli2013equation, manoharan2021universal, xie2023breaking}. In general, the signal emitted during the coalescence of a BNS is influenced by a set of tidal deformability parameters and the rotational quadrupole deformation of each star \cite{yagi2013love,yagi2013love_b,yagi2016binary}. Therefore, relations involving NS tidal parameters streamline the analysis of GW data by reducing the number of free parameters, thereby improving our capacity to infer astrophysical properties from observed signals \cite{yagi2013love, yagi2013love_b, yagi2017approximate, chatziioannou2018measuring, yagi2014multipole,yagi2016binary,abbott2018gw170817,raithel2018tidal,benitez2021investigating, xie2023breaking}. For rapidly rotating NSs, the universality of these relations weakens unless parameters are carefully chosen 
\cite{doneva2013breakdown,pappas2014effectively,chakrabarti2014q, papigkiotis2023universal}. Thus, with the appropriate feature selection, such relations maintain their robustness even at high rotation rates. On that topic, the Pearson correlation analysis applied in Ref. \cite{papigkiotis2023universal} introduced a new way to identifying suitable parameters for characterizing how these features depend on the star’s key observable quantities. Additionally, a cross-validation method was employed to validate these EoS-insensitive relations within the scope of data analysis, using a supervised ML approach with polynomial functions as the underlying models.

Further relations include those for the binding energy of rotating and non-rotating configurations \cite{cipolletta2017last}, the last stable circular orbit around rapidly rotating NSs \cite{luk2018universal}, approximately EoS-independent parametrizations regarding the average NS stiffness I-Love-$<c_s^2>$ \cite{saes2024approximately}, and three-parameters relations involving higher-order multipole moments \cite{pappas2014effectively,stein2014three,chatziioannou2014toward}. Especially, relations between the NS multipole moments \cite{yagi2014effective, pappas2014effectively, stein2014three, chatziioannou2014toward} have led to a generalization of the no-hair theorem for BHs to the three-hair relations for Newtonian configurations \cite{stein2014three}. In addition, recent studies proposed EoS-insensitive formulas related to the rotating star's surface estimation associated with the equatorial radius \cite{morsink2007oblate,silva2021surface} or the effective acceleration due to gravity \cite{algendy2014universality} using fitting coefficients dependent on stellar compactness and dimensionless spin. Building on this contribution, Ref. \cite{papi_vard_surface} proposed a DL methodology featuring accurate universal relations for the star’s surface, its logarithmic derivative, and its effective acceleration due to gravity. In addition, supplementary relations were also proposed, providing valuable fitting functions for key global properties associated with the star's surface.

Although low-order analytical expressions have been proposed for most universal relations, the neural network approach offers a robust and flexible alternative, often achieving higher accuracy. In this study, we revisit some established EoS-insensitive relations and investigate some new ones using a deep learning architecture as a regression model. More specifically, we deployed a feed-forward neural network to perform regression. Furthermore, we employed the SHapley Additive exPlanations (SHAP) method to interpret the predictions of the proposed feed-forward network \cite{lundberg2017unified}. SHAP is grounded in a robust theoretical framework derived from cooperative Game Theory, making it an excellent tool for understanding the contributions of individual parameters to the model's output \cite{shapley1953value, maschler2020game}. By leveraging this method, we can gain insights into how each input parameter influences the ML model's decisions, thereby enhancing the transparency and interpretability of the neural network's predictions. The primary investigation covers an extended ensemble of astrophysically relevant rotating models, including both quite slowly and rapidly rotating configurations, with frequencies ranging from a few hundred $\mathrm{Hz}$ up to  $\sim \ 1.87 \ \mathrm{kHz}$ (exceeding the values observed in known pulsars).

The plan of the paper is as follows. In Sec. \ref{sec:numerical_setup}, we provide a concise overview of the numerical setup employed to generate our ensemble of equilibrium NS models. In Sec.  \ref{sec:deep_learning_and_SHAP}, we detail the proposed DL architecture used to obtain our relations and provide an overview of the SHAP method applied to interpret the DL model's predictions. In Sec. \ref{sec:univ_rel_with_ANNs}, we present our ``deep'' EoS-insensitive parametrizations, whereas in Sec. \ref{sec:conclusions} we summarize our findings and present some concluding remarks. It is important to highlight that the suggested universal relations are mainly formulated as hypersurfaces, with each quantity of interest expressed as a function of other physical parameters. Finally, Appendix \ref{app:EoS_list} lists the EoS models used in this study, while Appendices \ref{app:train_prop_characteristics} and \ref{app:model_characteristics} provide the training characteristics of each ANN model and the corresponding distributions of relative deviations across EoS categories in the test set. Unless specified otherwise, we use geometric units with $G=c=1$.

\section{\label{sec:numerical_setup} Rotating Stars and EoS Ensemble}

One of the primary challenges in nuclear astrophysics is to unravel the properties of ultradense, cold matter at limits that exceed the nuclear saturation density $\rho_0 \approx 2.8 \times 10^{14}\ \mathrm{g}/\mathrm{cm}^3$. The main uncertainties in the bulk properties of NSs originate from unknown particle interactions in high-density regions within the star's core. Beyond the nuclear saturation density, typical of standard symmetric nuclear matter, the structure and composition of stellar configurations become increasingly ambiguous \cite{lattimer2021neutron, compose2022compose}. Different EoSs have substantially distinct bulk properties that affect the modeling of both static and rotating NS sequences. The EoS is essential for characterizing the macroscopic parameters of NS physics in $\beta$ equilibrium and is a critical input for solving Einstein’s field equations \cite{friedman_rotating_2013}.

To address the uncertainty of the EoS at ultrahigh densities in equilibrium, various models are proposed in the literature based on diverse many-body nonrelativistic and relativistic theories \cite{haensel_neutron_2007, dutra2012skyrme, dutra2014relativistic, lattimer2021neutron, compose2022compose}. Incorporating these theoretical assumptions, we utilized a realistic catalog of EoS models from the CompOSE database \cite{compose, compose2022compose}. The utilized EoS ensemble includes hadronic, hyperonic, and hybrid models, provided in tabulated form, which comprehensively describe the NS interior, encompassing the crust and the stellar core. In total, we have used 70 realistic EoS models of cold nuclear matter in a tabulated form. The detailed list of EoS samples for each NS category, along with their specific theoretical description and the associated microphysics composition, can be found in Appendix B of Ref. \cite{papi_vard_surface}. At the static limit, the selected EoS models satisfy the stipulated constraints determined by the lower bounds on the maximum nonrotating mass for PSR $\mathrm{J}0348+0432$ ($M = 2.01^{+0.04}_{-0.04} \ M_{\odot}$) \cite{demorest2010two, antoniadis_massive_2013}, and PSR $\mathrm{J}0740+6620$ ($M = 2.08^{+0.07}_{-0.07} \ M_{\odot}$)  \cite{fonseca2021refined} within $2\sigma$ and $1\sigma$ credibility, respectively. Furthermore, they also yield nonrotating maximum mass NS with a radius $R_{M_{max}} \geq 9.60^{+0.14}_{-0.03} \ \mathrm{km}$, as indicated by the GW170817 BNS merger analysis \cite{bauswein_neutron-star_2017, friedman_astrophysical_2020-2}. Additionally, regarding the static limit, none of the selected EoS models surpass a maximum mass of $2.32 \ M_{\odot}$ within the $2\sigma$ bound, assuming that the final remnant of GW170817 was a BH \cite{dietrich2020multimessenger, rezzolla_using_2018}. It is crucial to highlight that the whole EoS ensemble verified the physical acceptability conditions (see e.g., Refs. \cite{papigkiotis2023universal, haensel_neutron_2007,papi_vard_surface}, for a review), ensuring $\beta$-equilibrium.

To integrate the nonlinear Einstein field equations along with the hydrostationary equilibrium equation \cite{friedman_rotating_2013, paschalidis2017rotating}, various numerical methods have been developed \cite{wilson1972models, bonazzola1974exact-2, friedman1989implications, komatsu_rapidly_1989-2, komatsu_rapidly_1989_II-2, stergioulas1994comparing}. With the EoS ensemble in hand, we now proceed to calculate rapidly rotating stellar configurations using the RNS code developed by Stergioulas and Friedman \cite{rns, stergioulas1994comparing}. The code provides equilibrium NS solutions by solving Einstein’s equations in the presence of a perfect fluid, employing the Komatsu-Eriguchi-Hachisu (KEH) scheme \cite{komatsu_rapidly_1989-2, komatsu_rapidly_1989_II-2} with the enhancements from Cook, Shapiro, and Teukolsky's (CST) modifications \cite{friedman_rotating_2013}. All these methods utilize the line element of a stationary and axisymmetric spacetime, expressed in quasi-isotropic coordinates as:
\begin{align}
\label{eq:ds_rot_2}
\nonumber \mathrm{ds}^2=&-e^{(\gamma+\rho)}\mathrm{dt}^2+e^{(\gamma-\rho)}r^2\sin^2\theta(\mathrm{d}\phi-\omega \mathrm{dt})^2\\
&+e^{2a}(\mathrm{dr}^2+r^2\mathrm{d}\theta^2),
\end{align}
where the metric potentials $\gamma, \rho, \omega, \ \mathrm{and} \ \alpha$ are functions of the quasi-isotropic coordinates $(r, \theta)$. Assuming uniform rotation, the star's angular velocity $\Omega$, as defined by an observer at infinity, remains constant. The equation governing hydrostatic equilibrium for a stationary, axisymmetric, and uniformly rotating NS is given by \cite{friedman_rotating_2013},
\begin{equation}
\label{hydro_uniform}
\frac{\nabla_a P}{\epsilon + P}=\nabla_{a} \ln u^t,
\end{equation}
where $\epsilon$ and $P$ are scalar quantities denoting the fluid's total energy density and pressure,  while
\begin{equation}
\label{u_t}
u^t=\frac{e^{-(\rho+\gamma)/2}}{\sqrt{1-(\Omega-\omega)^2r^2\sin^2(\theta) \ e^{-2\rho}}},
\end{equation}
is the four-velocity of a fluid element.

For a given EoS under a specific rotation law (we assume uniform rotation), the code produces unique equilibrium solutions for the interior and the exterior region by specifying the central energy density $\epsilon_c$ and the axial ratio $r_{\mathrm{pole}}/r_{\mathrm{eq}}$ between the polar and equatorial coordinate radii. The numerical integration is performed on a discrete grid with the quasi-isotropic radial coordinate $r$ compactified and equally spaced in the interval $s\in [0,1]$, where $s\equiv r/(r+r_{\mathrm{eq}})$. The angular coordinate $\mu = \cos(\theta)$ is equally spaced in the interval $\mu \in[0,1]$. The computational grid is fixed so the star's center is at $s = 0$, the surface at $s=1/2$, and infinity at $s=1$. The equatorial plane is at $\mu = 0$, and the pole is located at $\mu=1$ \cite{friedman_rotating_2013, butterworth_structure_1976, komatsu_rapidly_1989-2}. Half of the grid corresponds to the star's interior, while the remaining is allocated to the vacuum exterior. 

NS models are then computed along sequences varying the central energy density and the axial ratio \cite{friedman_rotating_2013, stergioulas1994comparing}. In addition to numerically calculating the star's metric functions in the interior and exterior regions, the RNS code also determines the fluid configuration and various equilibrium quantities. These comprise the star's gravitational mass $M$, the proper mass $M_p$, the baryonic mass $M_b$, the angular momentum $J$, the equatorial radius $R_{\mathrm{eq}}$, the moment of inertia $I$, the ratio of kinetic to gravitational energy $T/|W|$, and the spacetime's Geroch-Hansen multipole moments ($M_0\equiv M, S_1\equiv J, M_2, S_3$, and $M_4$.) \cite{geroch_multipole_1970_I,geroch_multipole_1970_II,hansen_multipole_1974,fodor_multipole_1989,rezzolla_physics_2018, Pappas:2012ns, Pappas:2012qg}. For each EoS included in our ensemble, we computed an extensive sample of astrophysically relevant rotating stellar models in equilibrium with central densities $\epsilon_c \sim [3.93 \times10^{14}-2.74\times10^{15}] \ \mathrm{g}/\mathrm{cm}^3$ and masses ranging from $\sim 0.9 \ M_{\odot}$ up to each rotating sequence star's maximum mass $M_{\mathrm{max}}$. Our entire sample includes $1109271$ models of rotating configurations, with frequencies spanning from a few hundred  $\mathrm{Hz}$ ($f\sim 172.5 \ \mathrm{Hz}$) up to the $f\sim 1.87 \ \mathrm{kHz}$ limit. This ensemble provides a comprehensive dataset for a thorough investigation. Finally, it should be highlighted that for NS models with axial ratio $r_{\mathrm{pole}}/r_{\mathrm{eq}}\in [0.90,0.98]$ a high-resolution grid size of $\mathrm{MDIV}\times \mathrm{SDIV}=751\times 1501$ is employed, where $\mathrm{MDIV}$ is the number of points in the angular $\mu$ direction, whereas $\mathrm{SDIV}$ is the number of points in the compactified radial $s$ direction. This choice of high resolution is preferred for more accurate calculations of the star's higher-order multipole moments. On the contrary, for an axial ratio less than $0.90$, a sufficient grid size of $\mathrm{MDIV}\times \mathrm{SDIV}=261\times 521$ is preferred to accelerate each stellar model's convergence.

In this work, we utilize a comprehensive set of dimensionless quantities to characterize rotating stellar configurations. The normalized moment of inertia is defined as $\bar{I} = I c^4 / (G^2 M^3)$. A key parameter in describing stellar rotation is the dimensionless angular momentum $\chi = (c/G)(J/M^2)$, which allows for the definition of the reduced quadrupole deformation $\bar{Q} = -(c^4/G^2)(M_2/M^3)(1/\chi^2)$, where $M_2$ denotes the Geroch–Hansen mass quadrupole moment. Additional spin-related quantities include the reduced spin parameter $\sigma = \Omega^2 R_{\mathrm{eq}} / (G M)$ and the kinetic-to-gravitational energy ratio $\mathcal{E} = T / |W|$. Alternative spin normalizations are also employed, such as $M \times \hat{f} \equiv M \times \Omega G / (2\pi c^3)$ and $R_{\mathrm{eq}} \times f / c$, which offer dimensionless parameters of the spin frequency. The stellar compactness is defined as $C = G M / (R_{\mathrm{eq}} c^2)$. To further encapsulate rotational effects, we introduce the reciprocal normalized moment of inertia\footnote{The notation $\mathcal{D}$ follows the convention introduced in Ref. \cite{papigkiotis2023universal}.} $\mathcal{D} \equiv (G / 2\pi c^3)(M \Omega / \chi) = (G^2 / 2\pi) (M^3 / I c^4)$. The analysis also includes higher-order multipole moments, such as the Geroch–Hansen current octupole $S_3$ and its dimensionless version $\bar{S}_3 = - (c^5/G^3)(S_3/M^4)(1/\chi^4)$, as well as the mass hexadecapole moment $M_4$ and its reduced form $\bar{M}_4 = (c^6/G^4)(M_4/M^5)(1/\chi^4)$. Lastly, we adopt normalized forms for the equatorial radius and gravitational mass, defined as $\bar{\mathcal{R}} = R_{\mathrm{eq}}[\mathrm{km}]/(1.0 \ \mathrm{km})$ and $\bar{M} = M / M_{\odot}$, respectively.


\section{\label{sec:deep_learning_and_SHAP} Deep Learning \& Explainable Decisions}


\subsection{\label{sec:shapley_values} Mapping explainability using the SHAP method}

With the growth of multimessenger astronomy, ML has emerged as a powerful tool that is revolutionizing the way of analyzing and interpreting massive amounts of data. By facilitating the development of algorithms capable of recognizing intricate patterns in complex datasets, ML enables us to extract meaningful insights, make accurate predictions, and discover data patterns that might otherwise remain hidden \cite{bishop2006pattern,prince2023understanding}.

Unfortunately, recent ML problems are becoming increasingly challenging and require more advanced and complex models to tackle them \cite{lecun2015deep}. However, model complexity typically presents a trade-off between explainability and performance in decision-making. For instance, simpler models like those employing linear regression are highly interpretable but typically fall short in predictive performance compared to DL ones. In addition, DL models often sacrifice interpretability due to their inherent complexity, and nonlinear transformations, among other factors \cite{guidotti2018survey}. On top of that, contrary to the common perception that DL models operate as ``black-boxes'', they are, in fact, well-defined systems, and we can provide clear explanations for how the input parameters influence their behavior. It should be noted that in several cases, understanding why a model made a certain prediction can be crucial.
Over the years, various techniques have been developed to interpret the predictions of complex models, including LIME, DeepLIFT, and SHAP \cite{lundberg2017unified}. Among these, SHapley Additive exPlanations (SHAP) is a widely recognized method inspired by the concept of Shapley values from the field of cooperative game theory \cite{shapley1953value, maschler2020game}. 

Building on this foundation, it is essential to demonstrate how Shapley values, denoted as $\phi_i(x)$, provide a systematic quantification of the {\it contribution} of each feature $f_i$ to the model’s prediction for a given input $x$. At first, let us consider a set of features $D = \{f_1,\dots,f_d\}$ with dimension $|D| = d$. For each subset of features $S \subseteq D$, we can define a model $\mathcal{F}_{S}: \mathbb{R}^{|S|} \rightarrow \mathbb{R}$, which only takes the values of the features in $S$. Building on this framework, we define
\begin{equation}
    \mathcal{F} : \bigcup_{S \subseteq [d]} (S \times \mathbb{R}^{|S|}) \to \mathbb{R}, 
\end{equation}
where $\mathcal{F}(\emptyset) = 0$ and $[d] = \{1, \dots, d\}$. The Shapley value for feature $f_i$ is then defined as:
\begin{align}
\nonumber
\label{eq:Shapley-Value}
& \nonumber \phi_i(x) = \\&
\sum_{S \subseteq [d] \setminus \{f_i\}}
\frac{|S|!(d-|S|-1)!}{d!}
\Big(\mathcal{F}_{S \cup \{f_i\}}(x_{S \cup \{f_i\}}) - \mathcal{F}_S(x_S)\Big)
\end{align}
where the summation iterates over all subsets $S$ that do not contain $f_i$. Note that $x_S$ denotes the values of the features in $S$, and $\mathcal{F}_{S}(x_S)$ is the corresponding model output. Therefore, this computes the average marginal contribution of feature $f_i$ across all possible subsets. This approach provides a rigorous framework for assessing the {\it contribution} of each feature to the ML model's predictions.

It is important to note that SHAP\footnote{For the experiments, we utilized the Python \href{https://shap.readthedocs.io/en/latest/}{SHAP} library.} builds on the solid theoretical foundations of Shapley values and integrates them into the machine learning framework to explain individual model predictions. More specifically, the SHAP value for each feature $f_i$ in a given input feature vector $x$ quantifies its {\it contribution} to the model’s prediction. This contribution is defined relative to the model’s expected value, as illustrated in the equation \cite{lundberg2017unified}:

\begin{equation}
\label{eq:SHAP}
\hat{\mathcal{F}}(x) = \mathbb{E}_{X}(\hat{\mathcal{F}}(X)) + \sum_{j=1}^{d} \phi_j(x), 
\end{equation}
where $\hat{\mathcal{F}}(x)$ represents the model's prediction, $\mathbb{E}_{X}(\hat{\mathcal{F}}(X))$ is the model's expected value, and $\phi_j$\footnote{In general, SHAP value $\phi_i(x)$ is a function of the model's feature vector $x$. Nevertheless, for simplicity, we adopt the more straightforward notation $\phi$ as a definition.} are the individual SHAP values. In Fig. \ref{fig:SHAP}, as an indicative clarification example, we present the contribution of each SHAP value $\phi_i$ regarding the specific input vector $x_i = (f_{i,1}, f_{i,2}, f_{i,3}, f_{i,4})$, containing four features. Each SHAP value contributes positively (represented by a green-colored vector) or negatively (represented by a red-colored vector) to the model's inference, $\hat{\mathcal{F}}(x_i)$, relatively to $\mathbb{E}_{X}(\hat{\mathcal{F}}(X))$. These values serve as useful tools to explain the model's prediction ability, offering interpretability to the deep learning model's predictions.

\begin{figure}[ht]
\centering
\begin{tikzpicture}[scale=0.8]
\draw[thick, ->] (-1,0) -- (8,0);
\draw[thick, ->] (-1,0) -- (-1,4);

\node[anchor=north] at (3.5, -1.0) {Model's Output};
\node[anchor=south, rotate=90] at (-2.0, 2.0) {Features};

\draw[dashdotted, thick] (1, 0) node[anchor=north] {$\mathbb{E}_{X}(\hat{\mathcal{F}}(X))$} -- (1, 4);
\draw[dashdotted, thick] (6, 0) node[anchor=north] {$\mathcal{\hat{F}}(x_i)$} -- (6, 4);

\draw[-{Stealth[length=3mm]}, ultra thick, green] (1, 0.5) -- ++(1, 0) node[midway, above, black] {$\phi_1$};
\draw[-{Stealth[length=3mm]}, ultra thick, red] (2, 1.5) -- ++(-2, 0) node[midway, above, black] {$\phi_2$};
\draw[-{Stealth[length=3mm]}, ultra thick, green] (0, 2.5) -- ++(3, 0) node[midway, above, black] {$\phi_3$};
\draw[-{Stealth[length=3mm]}, ultra thick, green] (3, 3.5) -- ++(3, 0) node[midway, above, black] {$\phi_4$};

\node[anchor=east] at (-1.2, 0.5) {$f_1$};
\node[anchor=east] at (-1.2, 1.5) {$f_2$};
\node[anchor=east] at (-1.2, 2.5) {$f_3$};
\node[anchor=east] at (-1.2, 3.5) {$f_4$};

\draw[dashed] (2, 0.5) -- (2, 1.5);
\draw[dashed] (0, 1.5) -- (0, 2.5);
\draw[dashed] (3, 2.5) -- (3, 3.5);
\end{tikzpicture}

\caption{Illustration of the SHAP values {\it contribution} concerning four features $f_1,\cdots, f_4$. Each SHAP value $\phi_i$ associated with the corresponding feature $f_i$ contributes positively (green-colored vector) or negatively (red-colored vector) to the estimator's result $\hat{\mathcal{F}}(x_i)$ relatively to $\mathbb{E}_{X}(\hat{\mathcal{F}}(X))$. The associated inference is consistent with Eq. (\ref{eq:SHAP}), considering as input vector $x$ the specific vector $x_i$.}
\label{fig:SHAP}
\end{figure}

In this work, SHAP is applied in the framework of NS universal relations, where physical quantities associated with the star's stellar global parameters are used as feature inputs. The resulting SHAP value attributions are defined for a specific trained DL model and its corresponding feature set. Therefore, SHAP values are model-dependent interpretability scores, defined with respect to a specific trained model. For instance, retraining can lead to different learned optimal $\theta^{\star}$ parameters (e.g., due to stochastic optimization or convergence to a different local optimum), and consequently to some changes in the SHAP attributions across features. Having a well-trained model, it is important to note that SHAP values are not measures of the model's accuracy; predictive performance should instead be assessed separately using standard statistical evaluation measures. Depending on the parametrization, one or more features may dominate the average attribution, or several may contribute comparably. Comparable SHAP values across features do not imply higher or lower model's inference accuracy; rather, they reflect how the associated optimal weights distribute sensitivity across inputs. In any case, inference accuracy ultimately depends on the quality of the training data, the suitability of the parametrization, and the characteristics of the model chosen to verify the data. Therefore, SHAP is not intended to assess model performance, but rather to interpret how trained DL models distribute importance across input features.

In summary, interpreting the results of a DL model using SHAP values provides valuable insights into the factors influencing its predictions. SHAP values offer a systematic approach to quantifying the contribution of each input feature to the model’s output, thereby aiding in understanding its decision-making process. By analyzing SHAP values, we can identify which features have the most significant impact (in terms of magnitude) and how they contribute (positively or negatively) relative to the model’s expected value. This interpretability framework not only enhances our understanding of the model’s behavior but also supports its validation and helps identify potential areas for improvement or further investigation. For instance, if a feature has a very small mean SHAP value, one might consider excluding that input and retraining the model with a simpler parametrization. Such steps, however, should be approached with caution. Features are fundamental to model inference, and an EoS-insensitive parametrization allows the DL model to capture patterns in the data and achieve reliable performance. Removing a feature can alter the optimization landscape, potentially making training less stable and reducing the model's predictive accuracy or generalization. Therefore, any feature exclusion should be carefully validated by benchmarking against standard evaluation metrics, ideally using cross-validation.






\subsection{\label{sec:deep_learning} Deep Learning Architecture, Training and Testing}

Systematic investigation of the NS global properties for rotating stellar models irrespective of the EoS revealed that conventional methods are insufficient for representing the entire spectrum of rapidly rotating configurations with extremely high accuracy. The complexity of the problem necessitates exploring the use of artificial neural networks (ANNs) within a supervised learning framework. While traditional approaches typically rely on nonlinear polynomial expressions truncated at a fixed order, this limits their ability to capture higher-order effects. To overcome this limitation, we employ a neural network capable of modeling such higher-order interactions, which are otherwise neglected in conventional analytical formulations. ANNs are well-established as powerful learners that have significantly advanced the state-of-the-art in addressing various learning tasks in data science \cite{hornik1989multilayer,cybenko1989approximation,hornik1991approximation,lecun2015deep}.

Building on the preceding discussion, our approach primarily relies on a fully connected DL architecture to uncover patterns for more accurate data inference. According to the Universal Approximation Theorem \cite{hornik1989multilayer}, a feed-forward neural network with only a single hidden layer containing a finite number of neurons can approximate any continuous fitting function on a compact input space. Notably, this approximation can achieve any desired degree of accuracy, provided that the network has a sufficiently large number of neurons and appropriate activation functions.

Determining the optimal mathematical model in advance is not feasible. However, we can formulate plausible hypotheses regarding its structure based on relevant prior information derived from the available data. For instance, consider a dataset $X$ consisting of pairs of $(x_i, z_i)$ data, where $i=1,..., n$. Here, $x_i$ represents the input feature vector, while $z_i$ denotes the corresponding dependent variables, denoted as labels. We intend to identify the most concise model $\hat{z}(\theta) = \hat{\mathcal{F}}_{\theta}(x)$ that adequately describes the data. In this pursuit, we first introduce the Huber Loss function \cite{prince2023understanding}, defined as:
\begin{equation}
\mathcal{L}_{\delta}\left(z, \hat{z}(\theta)\right) = \sum_{i=1}^{n} \mathcal{L}_i, 
\end{equation}
where each term $\mathcal{L}_i$ is given by:
\begin{equation}
\mathcal{L}_i =
\begin{cases} 
\frac{1}{2} |z_i - \hat{z}_i(\theta)|^2 & \text{, } |z_i - \hat{z}_i| \leq \delta, \\
\\
\delta \left(|z_i - \hat{z}_i(\theta)| - \frac{1}{2}\delta \right) & \text{, } |z_i - \hat{z}_i| > \delta,
\end{cases}
\end{equation}
combining the strengths of typical MSE, and $L_1$ objective functions. In the preceding definition, $z$ represents the data's actual target values, whereas $\hat{z}$ signifies the predictions made by the model. In addition, we employ a hyperparameter $\delta = 1$ within the Loss function $\mathcal{L}_{\delta}(\theta)$. This selection of objective function utilizes the advantages of $L_1$ Loss and MSE Loss; the delta-scaled $L_1$ region makes it less sensitive to outliers compared to MSE Loss, while the MSE region ensures smoothness, especially around zero, surpassing $L_1$ Loss in this aspect.

Then, a gradient descent optimization procedure is necessary to be employed to minimize the model’s objective function and derive the optimal parameters $\theta ^ \star$ represented by,
\begin{equation}
    \label{eq:optimization}
    \theta^{\star} = \arg\min\limits_{\theta} \mathcal{L}_{\delta}(\theta).
\end{equation}
Minimizing the Loss function and ensuring the selected model captures the intricacies of the training data are essential for effective learning. The ultimate goal lies in the model's ability to generalize beyond the training data \cite{bishop2006pattern,prince2023understanding}. A  generalized model can accurately predict outcomes on new data not seen during training. This capability distinguishes a robust model from one constrained in its ability to describe newly encountered data. However, it should be noted that the model's generalization ability is not about unbounded prediction, but about robust pattern recognition within the statistically validated domain of the original data collection.

For designing a DL architecture that satisfies our requirements, we utilized the  \href{https://pytorch.org/}{PyTorch} library \cite{paszke2019pytorch}. For each EoS of cold, dense nuclear matter included in our catalog, we have used a random selection of 25\% rotating stellar models as a test set\footnote{{\bf random.seed(42)} method implemented in Python was used to reproduce the same sample of random NS configurations for testing. This particular choice ensures that the results are consistent across different runs.}. In our systematic investigation, we employed an ANN architecture described by an input layer encompassing the features\footnote{Depending the case, the specific characteristics of these features will be clarified in the Sec. \ref{sec:univ_rel_with_ANNs}.} $x=(f_1,..., f_d)$, followed by five hidden layers denoted as $H_1, \ldots, H_5$, while concluding in a singular output layer represented by $\hat{z}$. The number of the associated neurons regarding each hidden layer is shown in Table (\ref{tab:hidden_layers_struct}). 
\begin{table}[!th]
  \caption{\label{tab:hidden_layers_struct} ANN hidden layers structure. For each neuron, we have used the non-linear $\mathrm{GeLU}$ function as an activation function.}
  \begin{ruledtabular}
      \begin{tabular}{ccc} 
        \textbf{Hidden Layer} & \textbf{No. Neurons} & \textbf{ Activation Function} \\
        \hline
        $H_1$ & 120 & $\mathrm{g(x)} = \mathrm{GeLU(x)}$ \\
        $H_2$ & 75 & $\mathrm{g(x)} = \mathrm{GeLU(x})$ \\
        $H_3$ & 50 & $\mathrm{g(x)} = \mathrm{GeLU(x)}$ \\
        $H_4$ & 25 & $\mathrm{g(x)} = \mathrm{GeLU(x)}$ \\
        $H_5$ & 10 & $\mathrm{g(x)} = \mathrm{GeLU(x)}$ \\
      \end{tabular}
  \end{ruledtabular}
\end{table}
In each hidden layer, we have utilized the GeLU activation function \cite{hendrycks2023gaussian}, 
\begin{align}
    \mathrm{g(x)} = \mathrm{x} \Phi(\mathrm{x}),
\end{align}
to introduce the non-linearity. Here, $\Phi(\mathrm{x})$ corresponds to the cumulative distribution function for the Gaussian distribution. The ANN design, defined by the configuration of hidden layers, neuron counts, and activation functions, establishes an efficient framework specifically optimized for the accurate inference of the key parameter underlying our application. Through experimentation, this specific model choice resulted in more stable and consistent learning tailored to the specific characteristics of our datasets. In addition, it is worth mentioning that at the end of the final hidden layer, in most cases, we incorporate a modified sigmoid function as the activation function, defined as,
\begin{align}
    \tilde{\sigma}(\mathrm{x}) = z_{\mathrm{min}} + (z_{\mathrm{max}}-z_{\mathrm{min}}) \ \sigma({\mathrm{x})},
\end{align}
given that $\sigma(\mathrm{x}) = 1/(1+e^{-\mathrm{x}})$ is the known sigmoid function. The values $z_{\mathrm{min}}$ and $z_{\mathrm{max}}$ correspond to the minimum and maximum values regarding the dataset associated with the quantity we seek to infer. This modified activation function selection ensures that the model's inference is well-suited to the specific characteristics of each dataset. Specifically, it maps the model's output during training to the $[z_{\mathrm{min}},z_{\mathrm{max}}]$ range, making it the optimal choice for this particular case study. However, while the modified sigmoid activation function is generally a beneficial choice, there are instances where models performed equally well without it in the final layer (see e.g., Appendix \ref{app:train_prop_characteristics} and Table \ref{tab:ann_architectures} for a review).

Due to the non-linear nature of the optimization process, we employed the Adamax optimizer \cite{kingma2015adam} during back-propagation to extract the optimal $\theta^\star$ parameters. It should be mentioned that we used the Xavier uniform initialization
algorithm \cite{pmlr-v9-glorot10a} for setting the model’s initial $\theta$ parameters.

Before training, as an initial prepossessing step, we applied min-max scaling defined as,  
\begin{equation}
f^{\prime}_i= \frac{f_i - \mathrm{min}(f_i)}{\mathrm{max}(f_i) - \mathrm{min}(f_i)}.
\end{equation}
to map each input feature $f_i$ employed to the interval $[0,1]$. This process is carried out on the entire dataset, which includes the associated quantities. Feature scaling is essential to ensure that all features contribute equally, preventing the dominance of those with larger values. This practice enhances the convergence and overall performance of the algorithms used. Either way, this process is an important step to render the learning process faster. Depending on the universal relation, we employ an adaptive learning rate, a tailored learning rate strategy, a case-favored mini-batch size, and a specific number of epochs for each training phase. In each case, the training time is linearly proportional to the number of epochs.


Across all of the regression models examined, we employed comprehensive evaluation measures to assess the performance of each ANN across the associated test sets (see e.g., \cite{papigkiotis2023universal,papi_vard_surface} for a detailed review). 
The statistical measures utilized include Mean Absolute Error (MAE), Max Residual Error (MRE), Mean Squared Error (MSE), Max Relative Deviation ($\mathrm{d}_{\text{max}}$), Mean Absolute Percentage Error (MAPE), Explained Variance, and Determination Coefficient ($R^2$)\footnote{For more details about the statistical evaluation functions the interested reader can be directed to this link: \href{https://scikit-learn.org/stable/modules/model_evaluation.html}{https://scikit-learn.org/stable/modules/model\_evaluation.html}}. Notably, the evaluation measures in the validation set show a slight degradation in performance compared to the training set, as expected. In each case, as is standard practice, we report the evaluation measures based on the performance observed on the test datasets.


\section{\label{sec:univ_rel_with_ANNs} ``Deep'' Universal relations \& INFERENCE EXPLAINABILITY}

In this section, we present our findings, which encompass improvements to established universal relations as well as the investigation of entirely new ones. More specifically, the correlation analysis described in Ref. \cite{papigkiotis2023universal} identified several well-known candidates from the literature, such as the $\bar{I}-\bar{Q}$ and the $\bar{S_3}-\bar{Q}$ pairs, as well as some new ones. It also suggested that the fit could be enhanced in some cases by incorporating additional parameters. Focusing on the applicability of EoS-independent relations, we reexamine the issue of EoS-independence throughout the parameter space.

Building upon prior works \cite{papigkiotis2023universal,papi_vard_surface}, we utilized a DL architecture, as depicted in Table (\ref{tab:hidden_layers_struct}), and followed the fitting procedure, the formulation, and the optimization process outlined in SubSec. \ref{sec:deep_learning}. Once more, the decision to employ a neural network stemmed from the objective of deriving more accurate universal relations for rapidly rotating NSs, incorporating a larger set of parameters derived from numerical data. Notably, neural networks significantly accelerate the inference process, providing rapid and efficient solutions that would otherwise be computationally intensive and time-consuming. In addition, we utilized the SHAP method \cite{lundberg2017unified} to obtain valuable insights in interpreting the DL model's inference results. In that direction, we investigate how the employed astrophysical set of parameters, serving as the model's input features, aids the network in discerning underlying patterns, thereby offering substantive insights into its predictive ability.

\begin{figure*}[!thb]
    \centering
    \includegraphics[width=0.46\textwidth]{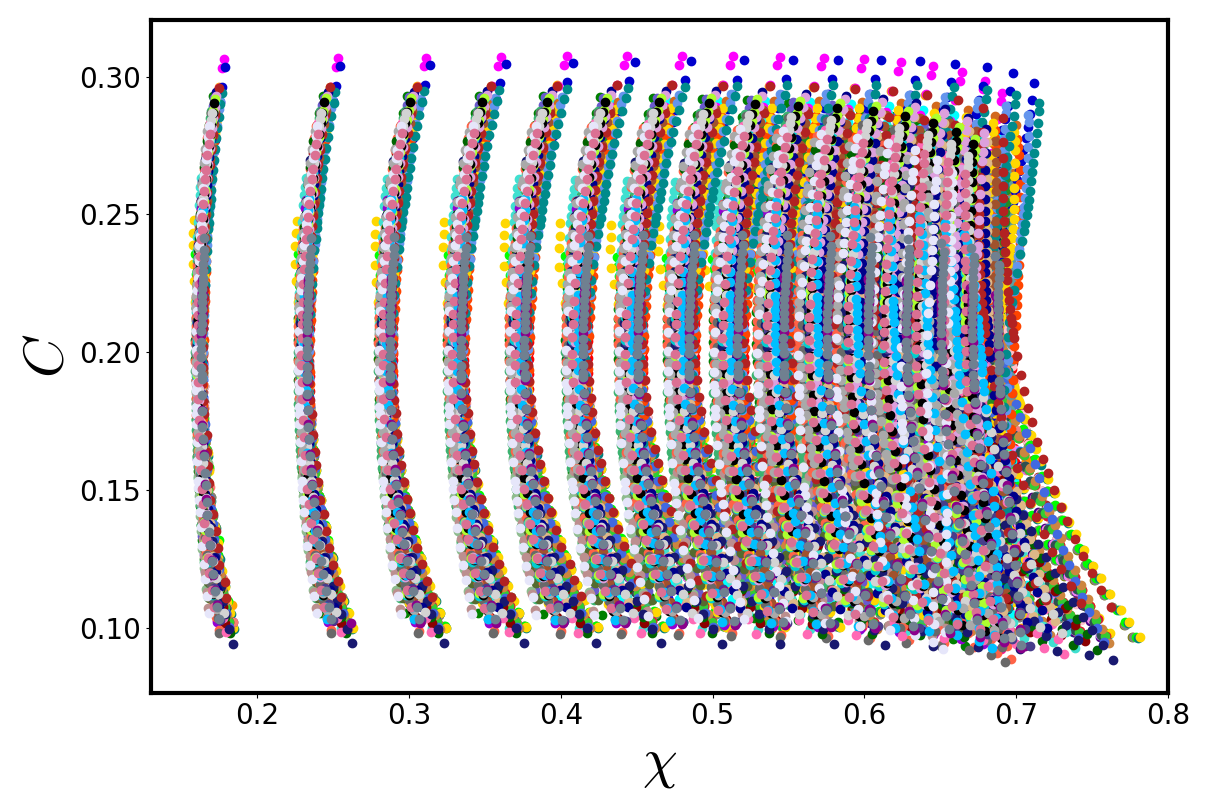}\hfill
    \includegraphics[width=0.46\textwidth]{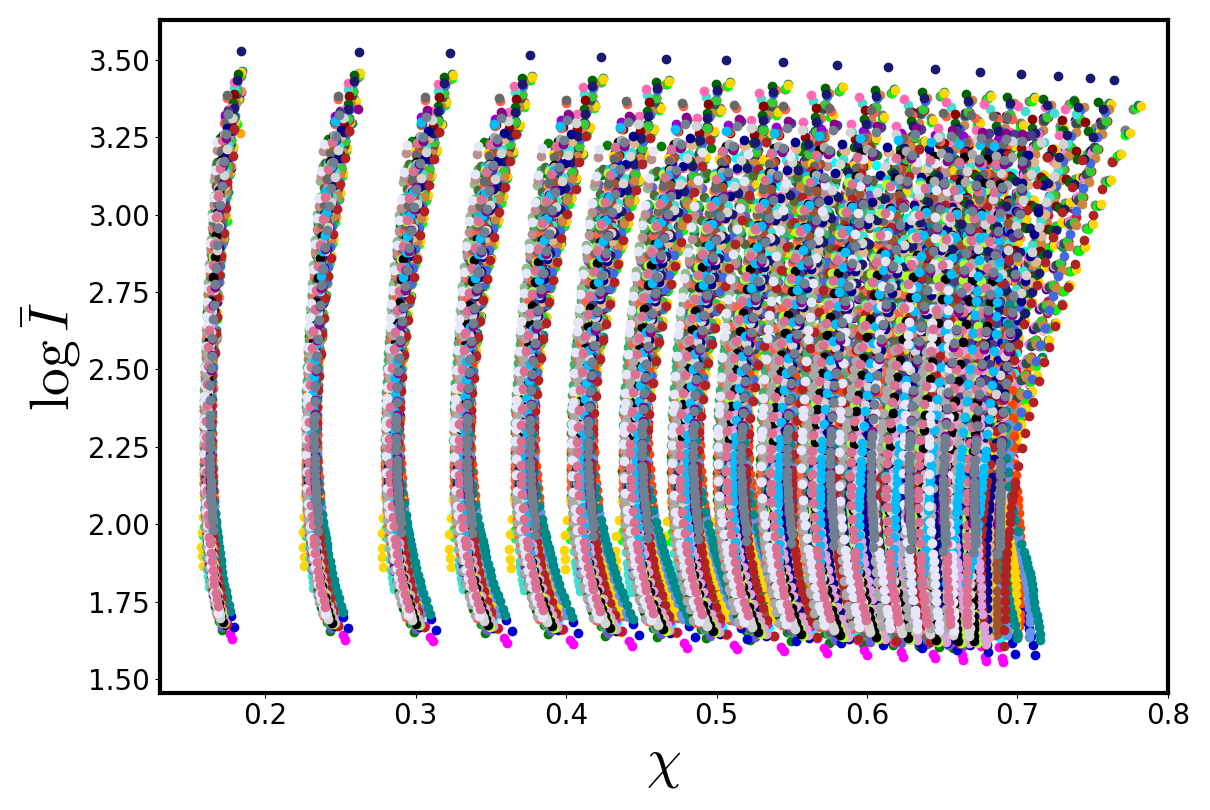}\hfill
    \includegraphics[width=0.46\textwidth]{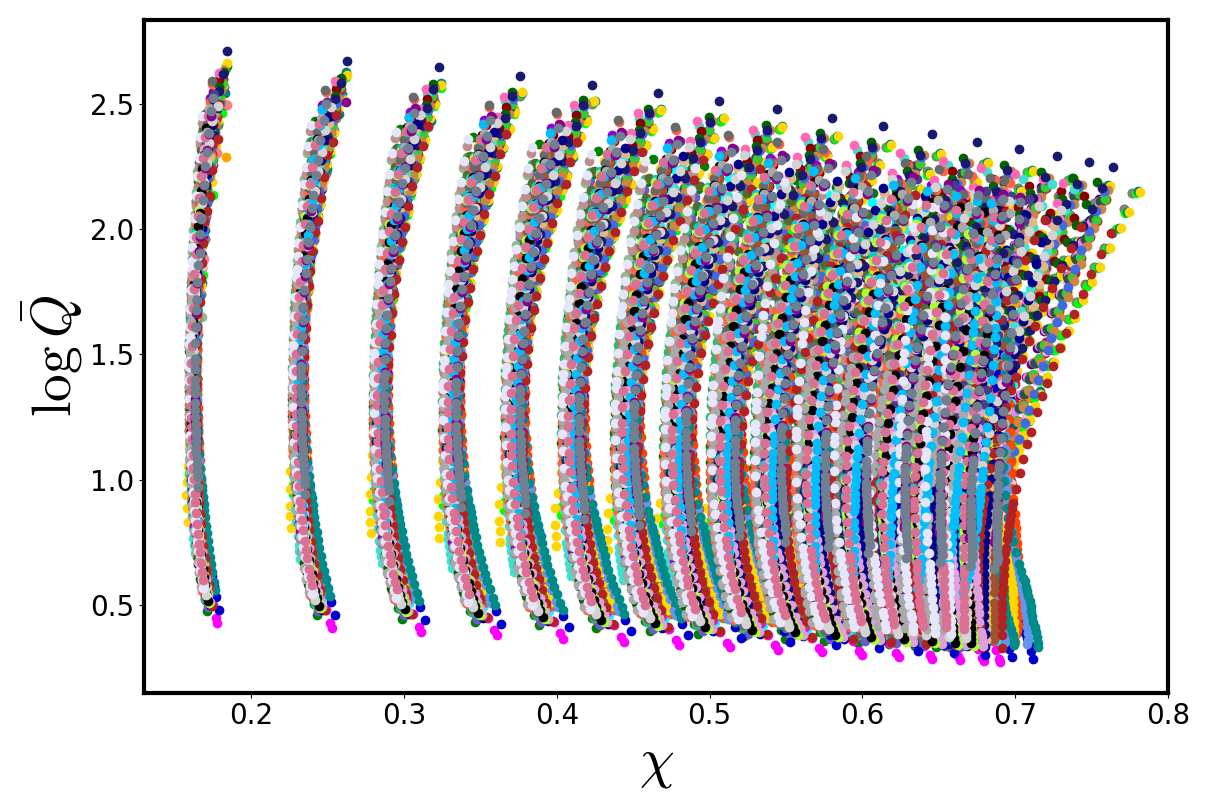}\hfill
    \includegraphics[width=0.46\textwidth]{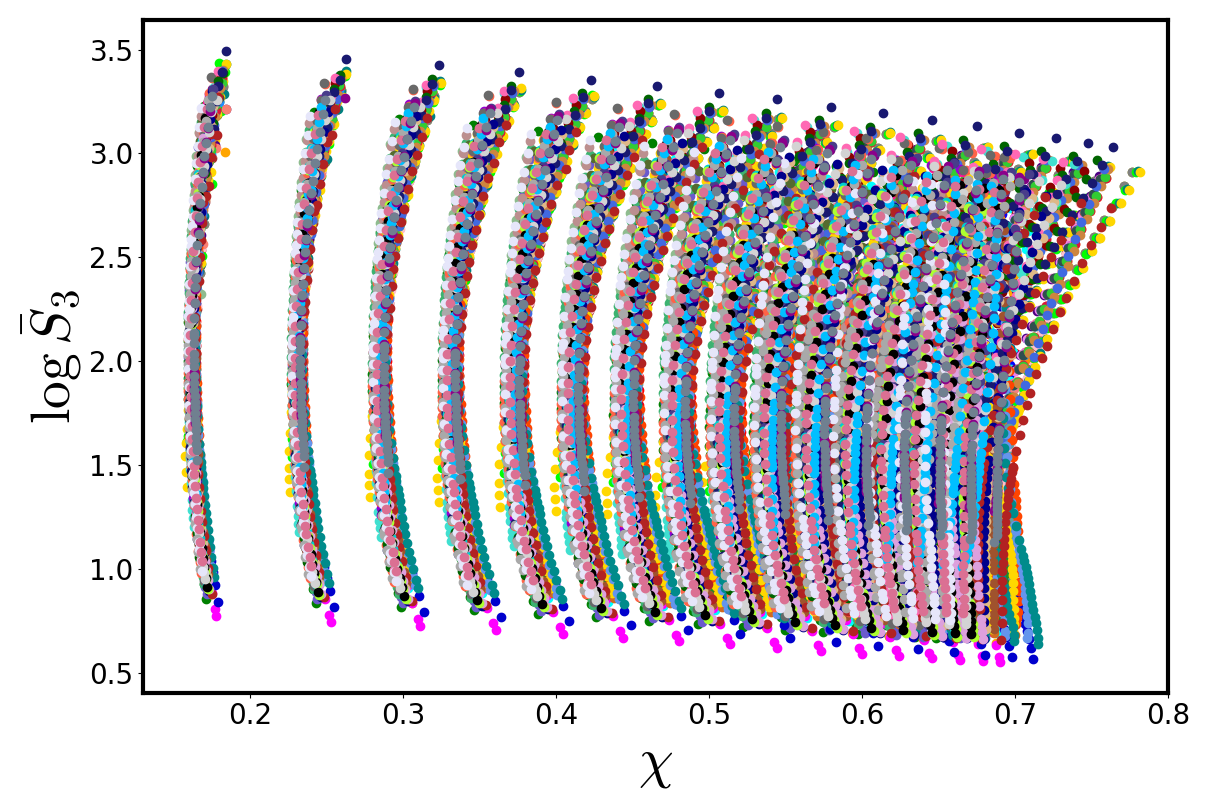}\hfill
    \caption{\label{fig:params_representations} Distributions of the $\chi-C$, $\chi-\log\bar{I}$, $\chi-\log\bar{Q}$, and $\chi-\log\bar{S}_3$ parameter spaces, encompassing a wide range of rotation rates and degree of stiffness. Each color represents the EOS mapping as highlighted in Fig. \ref{fig:color_band} of Appendix \ref{app:EoS_list}. }
\end{figure*}

It is important to mention that the complete dataset consists of $1109271$ rotating NS configurations. For each EoS, we allocate 75\% of the data for training and reserve the remaining for testing, adhering to the way outlined in Sec. \ref{sec:deep_learning}. Overall, regarding the various parameters characterizing the rotating NS ensemble, we provide the respective ranges for each relevant observable quantity in both the training and the test set for reference (see e.g, Table (\ref{tab:parameters_range}) for a review).
\begin{table}[!th]
  \caption{\label{tab:parameters_range} Respective ranges for each feature included in rotating NS ensemble relating to the associated training and test sets. }
  \begin{ruledtabular}
      \begin{tabular}{c|rr|rr} 
        \multirow{2}{*}{\textbf{Feature}} & \multicolumn{2}{c|}{\textbf{Training Set}} & \multicolumn{2}{c}{\textbf{Test Set}}\\ 
        \cline{2-5}
        & \textbf{min} & \textbf{max} & \textbf{min} & \textbf{max} \\
        \hline 
        $\bar{M}$ & 0.909 & 2.809 &0.913 & 2.809 \\ 
        $\bar{\mathcal{R}}$ & 9.939 & 19.405 &  9.971 & 19.414 \\ 
        $C\times 10^{-1}$ & 0.876 & 3.075 & 0.884 & 3.075 \\
        $\chi \times 10^{-1}$ & 1.406 & 7.818 & 1.406 & 7.802 \\ 
        $\bar{I}$ & 4.737 & 33.836 &   4.743& 33.869 \\
        $\mathcal{D}\times 10^{-2}$ & 0.470 & 3.360 &  0.470 & 3.356\\
        $\bar{Q}$ & 1.313 & 14.740 & 1.315 & 14.645 \\
        $\bar{S_3}$ & 1.736 & 32.106 & 1.740 & 31.900 \\
      \end{tabular}
  \end{ruledtabular}
\end{table}
Furthermore, Fig. \ref{fig:params_representations} shows the $\chi-C$, $\chi-\log\bar{I}$, $\chi-\log\bar{Q}$, and $\chi-\log\bar{S}_3$ representations for each EoS, based on an indicative subsample of $40000$ NS configurations from the full dataset. For visualization purposes, this subsample corresponds to a more distinct set of polar-to-equatorial radius ratios $r_{\mathrm{pole}}/r_\mathrm{eq}$ than that employed in the full dataset.

The investigated ``deep'' EoS-insensitive relations are primarily expressed as hypersurfaces, where each quantity of interest is modeled as a function of multiple underlying physical parameters. The presentation of the results will be organized in the following way. In SubSec. \ref{subsubsec:ANN_for_Ibar}, an EoS-independent relation regarding the star's normalized moment of inertia $\bar{I}$ is presented, while in SubSec. \ref{subsubsec:ANN_for_Irecip}, two universal parametrizations concerning the star's reciprocal of the normalized moment of inertia, and the star's reduced spin are suggested. Then, in SubSec. \ref{subsubsec:ANN_for_Qbar}, an EoS-insensitive relation regarding the star's quadrupole deformation $\bar{Q}$ is provided, whereas in SubSec(s). \ref{subsubsec:ANN_for_S3bar}, and \ref{subsubsec:ANN_for_M4bar} universal relations for higher multipole moments are proposed. Finally, in SubSec. \ref{subsubsec:ANN_for_Req}, we introduce a universal parametrization that links the star's equatorial radius $R_{\mathrm{eq}}$ to the star's mass and the spacetime's higher multipole moments.

Last but not least, we define a relation between the NS parameters as EoS-insensitive when the selected features at the DL model's inference yield relative errors in the training and test sets of $\lesssim \mathcal{O}(10\%)$. Crucially, our regression ``deep'' models accurately reproduce the data on the test set, showcasing a state-of-the-art generalization capability beyond the training data. For each universal relation examined, the ANN model's input layer features and targets, the associated number of training epochs, and the corresponding learning rate strategy are detailed in Appendix \ref{app:train_prop_characteristics}. In each instance, both the indicative numerical code and the optimal parameters $\theta^\star$ associated with the trained model are publicly available in Ref. \cite{deep_univ_gitrepo}.

\subsection{Relations for the star's Moment of Inertia}

\subsubsection{\label{subsubsec:ANN_for_Ibar}\bf{Relation for the star's normalized moment of Inertia}}
We first focus on the rotating star's moment of inertia $I$, a key global parameter featured in one of the established EoS-insensitive relations. For a rotating stellar configuration, the moment of inertia is defined as $I = J/\Omega$, where the parameters $J$, and $\Omega$ represent the star's angular momentum and the angular frequency, respectively. The mass of the NS greatly affects its moment of inertia, especially in the star's outer layers. Therefore, when the star's mass and radius increase, its moment of inertia usually becomes larger \cite{breu2016maximum}.

Above all, the star's moment of inertia strongly depends on the EoS. 
In the case of slow rotation, there are well-established EoS-insensitive relations for the reduced $\bar{I}$ with accuracy $\mathcal{O} (1\%)$ \cite{yagi2014effective, yagi2013love,yagi2013love_b,yagi2014multipole,yagi2017approximate}. In addition, significant efforts have also been devoted to estimating this global parameter for rapidly rotating NSs (see, for example, the Refs. \cite{papigkiotis2023universal, pappas2014effectively, chakrabarti2014q,papi_vard_surface} for a review). 
However, in the case of rapid rotation, there is still a significant demand for a highly accurate universal formula to infer the stellar object's moment of inertia. This formula must accommodate a broad range of EoS candidates, ensuring compatibility with multimessenger astronomy data, and encompasses various microphysical structures. Therefore, for our extended ensemble of realistic EoSs, it is essential to establish an accurate and effective universal relation for $\bar{I}$, encompassing a wide range of rotation rates, incorporating the rapid rotation case.

Most importantly, the universal behavior among observable quantities is highly sensitive to the star's global parameters selection and their dependence \cite{rezzolla_physics_2018}. This way, utilizing the most representative ones is crucial for better estimating the star's global characteristics (quantities we seek to infer). 
Expanding on the prior insights \cite{papigkiotis2023universal, pappas2014effectively, chakrabarti2014q}, it is well-established that the normalized moment of inertia $\bar{I}$ is related to the star’s spin parametrization and the relativistic quadrupole deformation. 

While many NSs have slow rotation rates, the most rapidly spinning pulsar observed to date \cite{hessels2006radio} can achieve a rotational frequency of $f \approx 716 \ \mathrm{Hz}$. Thus, examining how the spin parameter affects the properties of the $\bar{I}-\bar{Q}$ relation is crucial. For the EoS-insensitive relation, we have chosen to express the normalized moment of inertia $\bar{I}$ as a function of the parameters $\chi$ and $\bar{Q}$. These parameters correspond to the input feature vector fed into the DL architecture outlined in Table (\ref{tab:hidden_layers_struct}). This choice is driven by the intention to observe how the neural network responds to an already well-established correlation, to achieve superior performance. Regarding the rotation parametrization, it is also notable that instead of using $\chi$ one may also employ alternative rotation features like the reduced spin parameter $\sigma$, the fraction of kinetic to gravitational energy $\mathcal{E}$, or even alternative spin normalizations like $M\times \hat{f}$, and $R_{\mathrm{eq}}\times f/c$, as suggested in \cite{papigkiotis2023universal, chakrabarti2014q}.

Using the dataset of NS models, we then train the proposed ML model, aiming to predict the normalized moment of inertia $\bar{I}$ as the associated inference target $\hat{z}$. The model's optimal parameters (weights) $\theta^\star$ can be determined through the stochastic optimization procedure described in the previous Sec. \ref{sec:deep_learning}. The resulting $2$--D surface that best fits the data is formulated as\footnote{Throughout this investigation, the trained ANN model evaluated at its optimal parameters, $\theta^\star$, is denoted using calligraphic or Latin letter symbols (e.g., $\hat{\mathcal{F}}_{\theta^\star}({\bf x})$). This notation is consistently adopted across all regression models presented in this work.},
\begin{equation}
    \label{eq:Ibar_fit}
    \bar{I}_{\mathrm{model}} = \hat{\mathcal{I}}_{\theta^\star}(\chi, \bar{Q}).
\end{equation}
The evaluation measures for the optimal model's weights $\theta^\star$ associated with Eq. (\ref{eq:Ibar_fit}) on the validation set are presented in Table (\ref{tab:Ibar_eval_meas}).
\begin{table}[!h]
    \footnotesize
    \caption{\label{tab:Ibar_eval_meas} Evaluation measures for the universal representation given by Eq. (\ref{eq:Ibar_fit}) regarding the test dataset.}
    \begin{ruledtabular}
        \begin{tabular}{ccccccc}
            MAE & MRE & MSE & $\mathrm{d}_{\text{max}}$ & MAPE & ExpVar & $R^2$ \\
              $\times 10^{-2}$&  $\times 10^{-1}$&  $\times 10^{-4}$ &  ($\%$) &  $\times 10^{-3}$ ($\%$) &  $\times 10^{-2}$&$\times 10^{-2}$   \\    
            \hline
            2.399  & 6.553  & 9.709  &  2.76    & 2.624  &  99.997 &  99.997 \\

        \end{tabular}
    \end{ruledtabular}
\end{table}
In its current form, the regression formula (\ref{eq:Ibar_fit}) effectively reproduces the data across the test dataset, demonstrating high accuracy with relative deviation better than $2.76\%$.
This model consistently delivers satisfactory results regardless of the specific EoS chosen, establishing itself as a robust universal relation.

\begin{figure}[!thb]
    \includegraphics[width=0.46\textwidth]{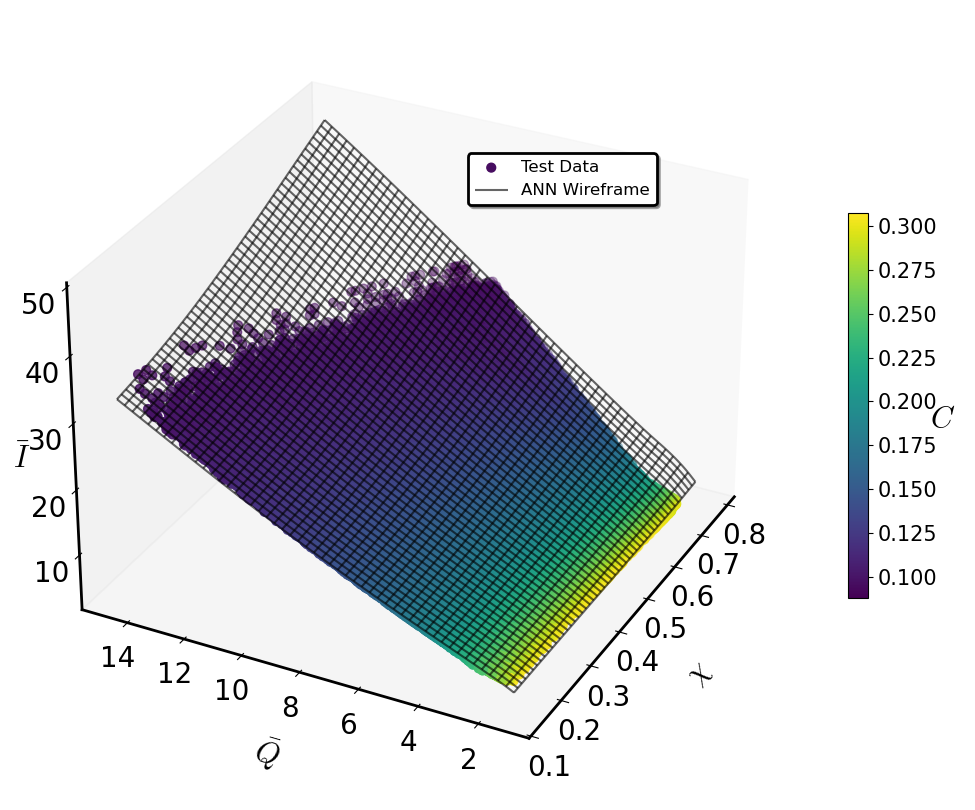}\hfill
    \caption{\label{fig:Ibar_representation} Reduced moment of inertia $\bar{I}$ as a function of the dimensionless parameters $\chi$, and $\bar{Q}$. In this illustration, the colored black grid is associated with the ANN wireframe produced by Eq.  (\ref{eq:Ibar_fit}). In addition, the colored variation of the presented data corresponds to the star's stellar compactness $C$, as highlighted in the accompanying vertical colored bar.}
\end{figure}

For visualization, the predicted surface (\ref{eq:Ibar_fit}) that optimally represents the data in the test dataset is depicted in Fig. \ref{fig:Ibar_representation}.
\begin{figure}[!htb]
   \includegraphics[width=0.46\textwidth]{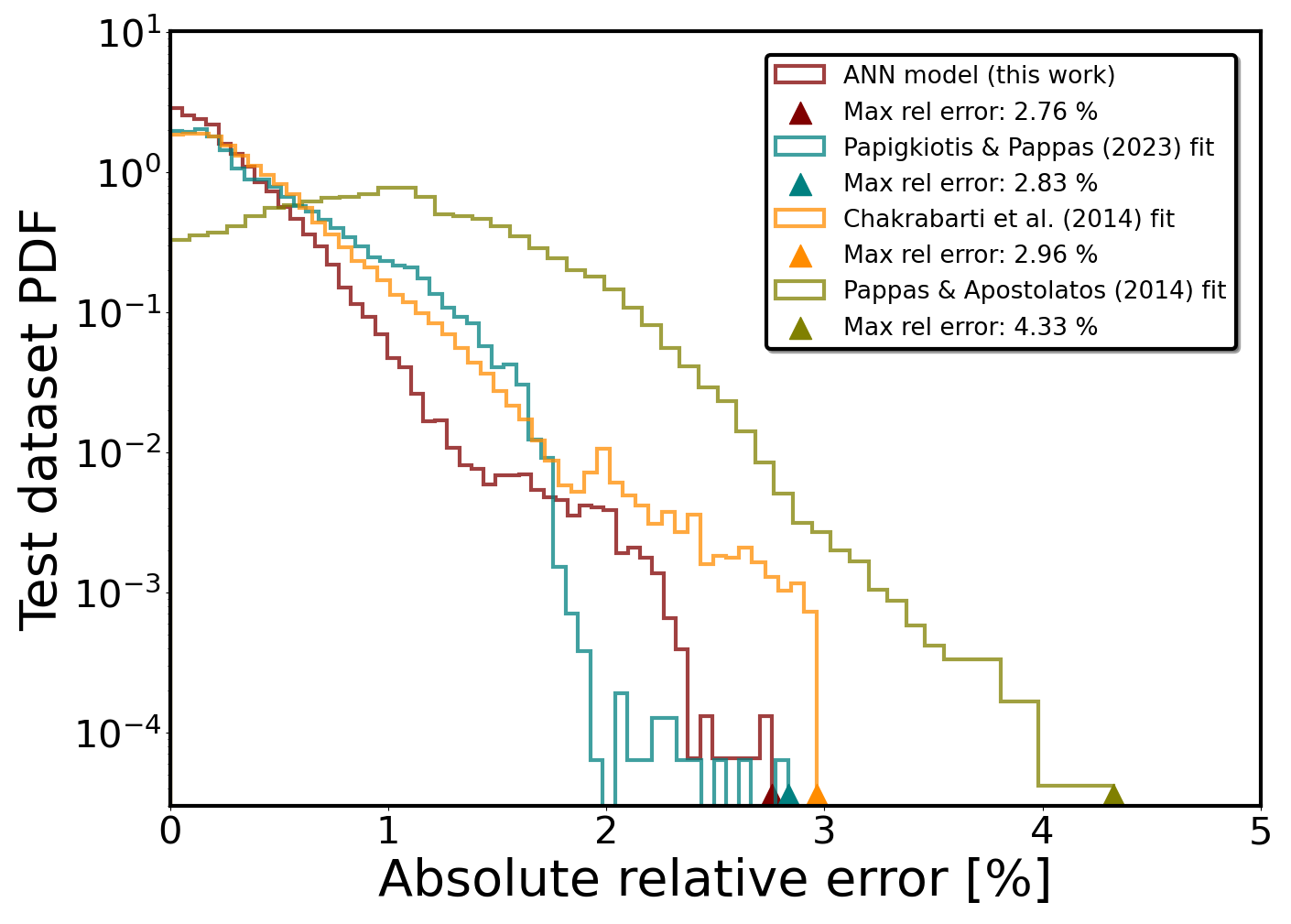}
    \caption{\label{fig:Ibar_fit_relative_errors} 
    Absolute relative error distribution ($100\% \times ({\bar{I}}_{\mathrm{model}} - \bar{I})/ \bar{I})$ derived evaluating the suggested regression model (\ref{eq:Ibar_fit}) on the test set. 
     The vertical axis corresponds to the probability density function (PDF) of the test dataset, shown on a logarithmic scale. In addition, relative deviations for fitting functions proposed in the literature are shown in different colors, providing a basis for comparative analysis.}
\end{figure}
In the constructed model's wireframe representation\footnote{To visualize the ANN-predicted relation between the input parameters and the target quantity, we constructed a wireframe surface as follows: a uniform mesh grid was generated over the 2D input space ($\chi,\bar{Q}$), using the {\it mesh\_grid\_funct} in python, which samples values within the domain of the test dataset. For each grid point, the trained neural network predicted the corresponding target value (e.g., $\bar{I}$), producing a 2D array of model estimates via the ANN model function. The resulting surface was plotted as a semi-transparent wireframe over the 3D scatter plot of the actual test data, allowing for direct visual comparison between the model predictions and the ground truth.}, the test data lies approximately on a single universal plane, while the varying colors of the depicted data correspond to different values of the star's compactness $C$, as indicated in the accompanying vertical colored bar. As anticipated, stellar modes that are less compact, corresponding to higher values of the reduced quadrupole deformation, demonstrate larger normalized moments of inertia. Additionally, the predicted wireframe includes regions that lie outside the physically meaningful parameter space. When the model is evaluated on inputs beyond the domain of the test dataset, it produces a smooth extrapolation of the surface—despite the fact that no physically viable stellar models exist in that regime. This behavior does not indicate a flaw in the model but simply reflects the smooth continuation of the learned mapping beyond the physical domain, where no viable stellar models exist.

In Fig. \ref{fig:Ibar_fit_relative_errors}, we present the distribution of relative errors of the fit for the test set. It should be noted that most of the model's predictions exhibit a relative error better than $ 1\%$. Moreover, it is noteworthy that among the entire test set, only $3562$ rapidly rotating stellar configurations have a relative deviation $> 1\%$. Most of these are associated with massive NS models exhibiting quadrupole deformation $\bar{Q}\in [1.465,3.511]$ and stellar compactness $C\in [0.193,0.294]$, regardless of the dimensionless spin parameter $\chi$. In the same figure, we also present the absolute fractional differences corresponding to the fitting functions previously established in the literature and are associated with rapidly rotating NS configurations \cite{papigkiotis2023universal, chakrabarti2014q, pappas2014effectively}. The suggested ANN regression approach reliably reproduces the test data, leading to improved inference performance. While the overall accuracy of our model is only marginally better, yet comparable to the fitting functions proposed in Refs. \cite{chakrabarti2014q, papigkiotis2023universal}, it significantly outperforms the method of Ref. \cite{pappas2014effectively}. This performance advantage becomes particularly evident when examining the proportion of test cases with larger relative deviations. Specifically, only $1.26\%$ of the NS configurations in the test set exhibit relative deviations exceeding $1\%$ with our model, compared to $4.83\%$ for the model in Ref. \cite{chakrabarti2014q}, $7.31\%$ for Ref. \cite{papigkiotis2023universal}, and a substantially higher $47.13\%$ for Ref. \cite{pappas2014effectively}.

A key aspect of the analysis involves assessing the maximum relative deviation of the proposed regression model (\ref{eq:Ibar_fit}) across different EoS categories within the test set. To this end, Appendix. \ref{app:model_characteristics} presents violin plots that illustrate the distribution of absolute fractional differences for each category, offering a detailed perspective on the model’s accuracy and variability (see e.g., Fig. \ref{fig:violin_plots} for a review). Among the EoS categories, the hybrid class displays the highest level of deviations, with a maximum relative error of $2.76\%$. In comparison, the hadronic and hyperonic classes exhibit lower peak deviations, around $2.14\%$ and $1.93\%$, respectively. These values of largest deviations correspond to the EoS models {\it DS(CMF)-2 Hybr, SkMp}, and {\it OPGR(DDH$\delta$Y4)}.

In any case, the regression formula (\ref{eq:Ibar_fit}) achieves improved accuracy in estimating the normalized moment of inertia of the rotating stellar model while effectively reducing systematic biases, demonstrating the strong generalization capability of our model. In addition, it should be noted that similar performance results and universal dependence can also anticipated when training the DL network using as a spin feature one of the parameters $\sigma$, $\mathcal{E}$, $M\times \hat{f}$, and $R_{\mathrm{eq}}\times f/c$ as an alternative of $\chi$.

\begin{figure}[!htb]
   \includegraphics[width=0.4\textwidth]{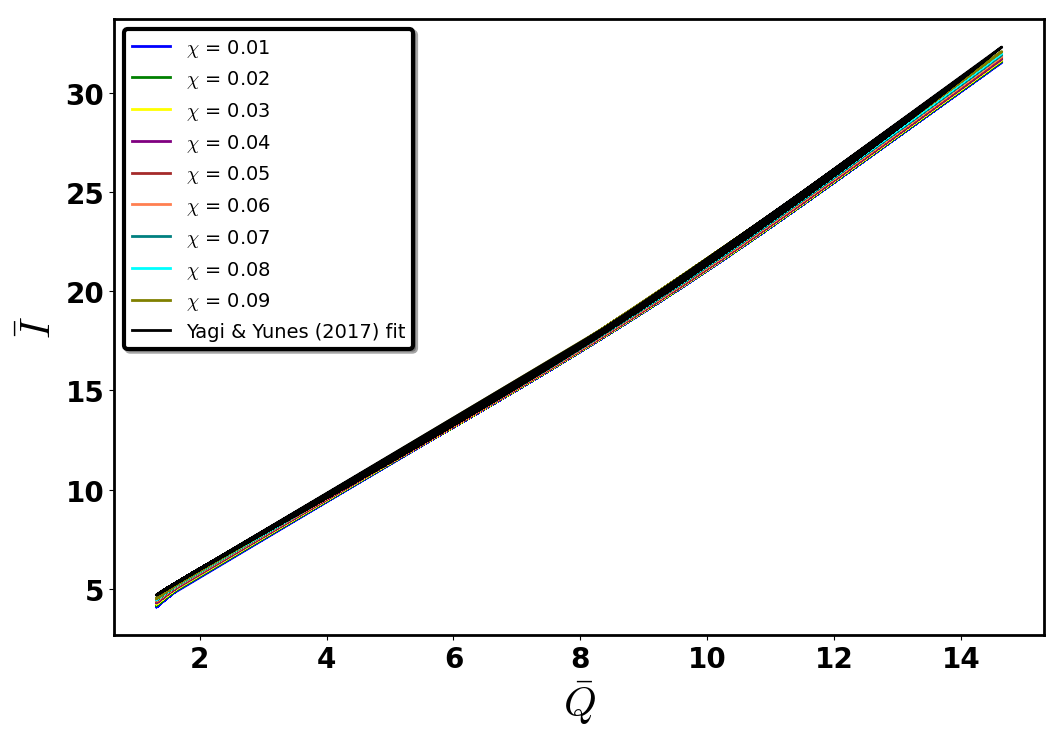}\hfill
    \caption{\label{fig:Ibar_slow_fits} $\bar{I} = \bar{I}(\bar{Q})$ theoretical curves for a discrete sample of spin parameter values $\chi \in [0.01,0.09]$ associated with slowly rotating NS configurations. The regression model (\ref{eq:Ibar_fit}) compared against the fitting function proposed in \cite{yagi2017approximate} satisfactorily reproduces the $\bar{I}-\bar{Q}$ universal behavior in the slowly rotating limit.}
\end{figure}

Additionally, investigating the model's ability to accurately reproduce the behavior of slowly rotating NSs is quite compelling. Despite not being trained on spin values associated with slowly rotating stellar configurations, the theoretical formula (\ref{eq:Ibar_fit}) accurately reflects the slowly rotating case described by the Hartle-Thorne approximation as detailed in \cite{yagi2017approximate}, 
demonstrating the model's extrapolation robustness. This behavior is illustrated in Fig. \ref{fig:Ibar_slow_fits}, which presents the model's theoretical predictions across a range of spin values $\chi \in [0.01, 0.09]$. In all instances, the corresponding curves from the $\bar{I}-\bar{Q}$ representation tend to coincide, leading to the well-known universal behavior for slowly rotating NS configurations. Furthermore, it should be emphasized that for the highest quadrupole deformation values ($\bar{Q} \gtrsim 13.5$) and the lowest ($\bar{Q} \lesssim 2.5$), the DL model predictions align well with the Yagi and Yunes formula \cite{yagi2017approximate} as the spin increases. 

Having trained the ANN with the ``key'' features of $\chi$ and $\bar{Q}$ that characterize a rotating NS, our focus now shifts to interpreting its predictions. While complex ML architectures may sacrifice simplicity for enhanced performance, we utilize the SHAP method to provide interpretability. As discussed in Sec. \ref{sec:shapley_values}, this framework equips us with tools to interpret and explain the DL model's predictions, providing a means to verify the reliability and robustness of the suggested model's predictions.

\begin{figure}[!htb]
    \includegraphics[width=0.46\textwidth]{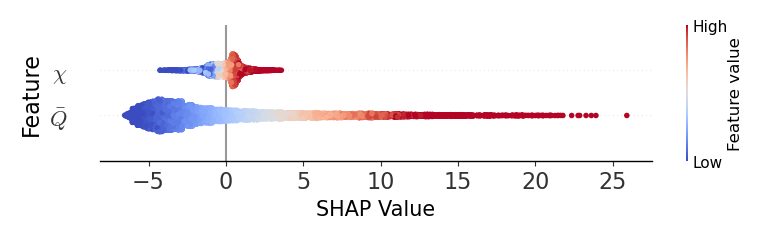}\hfill

    \includegraphics[width=0.46\textwidth]{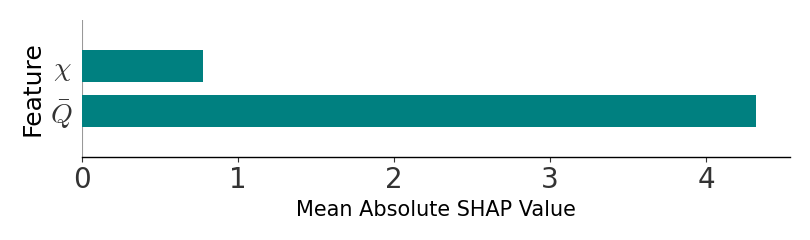}\hfill
    \caption{\label{fig:Ibar_shap_figs}
   SHAP analysis, based on Eq. (\ref{eq:Shapley-Value}), illustrating the contribution of each feature to the predictions of the trained model (\ref{eq:Ibar_fit}) for the selected subset of NS configurations. The color scale indicates the magnitude of each feature's value. Top panel: Contribution of each feature to the model's prediction results. Bottom panel:  Feature importance based on average SHAP values indicating the parameters' average impact on model predictions.}
\end{figure}

Building upon this, we employed the trained DL model (\ref{eq:Ibar_fit}) to interpret its predictions on a test subset $X_1$ consisting of $10000$ NS configurations. Given the computational demands of the SHAP procedure \cite{lundberg2017unified, shapley1953value}, this data sample was randomly selected\footnote{As before, the {\bf random.seed(42)} function in Python was applied to ensure reproducibility across experimental runs.} from the initial test dataset to expedite the analysis. We have to note that the selected dataset encompasses values of $\chi \in [0.141,0.774]$ and $\bar{Q} \in [1.343,14.370]$, ensuring adequate representation of the parameter space. 
Subsequently, we compute the SHAP values for this subset to interpret the contribution of each input feature to the model’s predictions. Fig. \ref{fig:Ibar_shap_figs} presents the SHAP values, as defined in Eq. (\ref{eq:Shapley-Value}), illustrating the influence of each parameter on the predictions for every stellar model in the examined subset.

At first glance, in Fig. \ref{fig:Ibar_shap_figs} (top panel), it is clear that each feature can have both a positive and negative influence. Notably, very high values of $\chi$ and $\bar{Q}$ positively {\it impact the estimator's prediction compared to its expected value} $\mathbb{E}_X(\hat{\mathcal{I}}_{\theta^\star})$\footnote{For the rest of this work, when we refer to the feature's positive or negative impact on the model's inference, the relative influence is associated to the comparison with respect to the estimator's expected value (see e.g., Eq. (\ref{eq:SHAP}) and Fig. \ref{fig:SHAP} for a review).}. These NS configurations correspond to less compact stellar objects with larger equatorial radii, resulting in higher moments of inertia. Conversely, the extremely lower values concerning the features of parameter space contribute negatively to the estimator's prediction. These stars correspond to highly compact (most massive) NS configurations, characterized at most by smaller radii and, consequently, smaller moments of inertia (see also Fig. \ref{fig:Ibar_representation} for a representative review).

Furthermore, as shown in the bottom panel, we present the mean absolute SHAP values for each feature, averaged across the NS models in the test subset. The rotation parameter $\chi$ corresponds to a mean value of $0.78$, whereas the quadrupole deformation parameter $\bar{Q}$ attains a substantially higher value of $4.32$. These findings indicate that both parameters play a role in exerting a measurable influence on the model’s predictive performance. Specifically, $\chi$ characterizes the rotational state of the star, enabling the network to capture rotational patterns within the data while $\bar{Q}$ encapsulates the deformation induced by rotation, thereby providing complementary structural information. This dual dependence renders $\bar{Q}$ as a particularly informative parameter for pattern recognition—especially in the slow-rotation limit and along NS sequences at fixed $\chi$, where the established EoS-independence between $\bar{Q}$ and $\bar{I}$ becomes prominent. The increased mean SHAP value attributed to $\bar{Q}$ indicates that the trained model relies more strongly on this parameter when forming its predictions, particularly in regimes of higher rotation where deformation effects are more significant. In any case, the interplay between $\chi$ and $\bar{Q}$ strengthens the model’s predictive capability, ensuring robust universal inference across a diverse range of stellar configurations.

\subsubsection{\label{subsubsec:ANN_for_Irecip} \bf{Relation for the star's reciprocal of the normalized moment of Inertia}}

The stellar object's rotational frequency $f$ stands out as a pivotal observational parameter for compact objects, often being the most straightforward quantity to measure. Consequently, establishing a relation that connects the star's rotational frequency with other fundamental parameters becomes highly advantageous. In this endeavor, we employ the reciprocal of the normalized moment of inertia $\mathcal{D}$ as a parameter of interest. As first proposed in Ref. \cite{pappas2015unified}, it is worth emphasizing that the normalized parameter $\mathcal{D}$ plays a meaningful role, as it effectively serves as a reduced frequency in place of the spin frequency $f$, offering a direct connection between rotational dynamics and the star’s moment of inertia.

Motivated by previous works \cite{papigkiotis2023universal, pappas2015unified}, we use the suggested neural network to infer the normalized parameter $\mathcal{D}$ as a function dependent on both the dimensionless spin parameter $\chi$ and the spacetime's reduced quadrupole deformation $\bar{Q}$. However, as highlighted in \cite{papigkiotis2023universal}, employing $\log\bar{Q}$ as a feature significantly improves the inference of $\mathcal{D}$ compared to $\bar{Q}$, particularly in cases involving rapid rotation. This prior knowledge of optimal and informative features can be harnessed by feeding the network at the input layer with a feature vector containing the star's global parameters $\chi$ and $\log \bar{Q}$. Using the suggested DL architecture (\ref{tab:hidden_layers_struct}) to perform regression, the predicted $2$--D surface that verifies the data has functional form,
\begin{equation}
    \label{eq:D_freq_fit}
    \mathcal{D}_{\mathrm{model}} = \hat{\mathcal{D}}_{\theta^\star}(\chi, \log \bar{Q} ).
\end{equation}
The evaluation measures for the best network's weights $\theta^\star$ associated with the parametrization (\ref{eq:D_freq_fit}) on the test set are presented in Table (\ref{tab:D_freq_eval_meas}).

\begin{table}[!h]
    \footnotesize
    \caption{\label{tab:D_freq_eval_meas} Evaluation measures for the parametrization given by Eq. (\ref{eq:D_freq_fit}) regarding the test set.}
    \begin{ruledtabular}
        \begin{tabular}{ccccccc}
            MAE & MRE & MSE & $\mathrm{d}_{\text{max}}$ & MAPE & ExpVar & $R^2$ \\
            $\times 10^{-5}$&  $\times 10^{-4}$&  $\times 10^{-9}$ &  ($\%$) &  $\times 10^{-3}$ ($\%$) &  $\times 10^{-2}$&$\times 10^{-2}$   \\    
            \hline
            5.238 & 6.368   & 6.194  &   2.70   & 2.619  & 99.987  &  99.987 \\

        \end{tabular}
    \end{ruledtabular}
\end{table}
In its current state, the trained regression model (\ref{eq:D_freq_fit}) accurately verifies the data in the test dataset, achieving high accuracy with relative deviation better than $2.70\%$. Therefore, this model yields reliable outcomes irrespective of the chosen EoS, establishing its robustness as an EoS-insensitive formulation.

\begin{figure}[!thb]
    \includegraphics[width=0.46\textwidth]{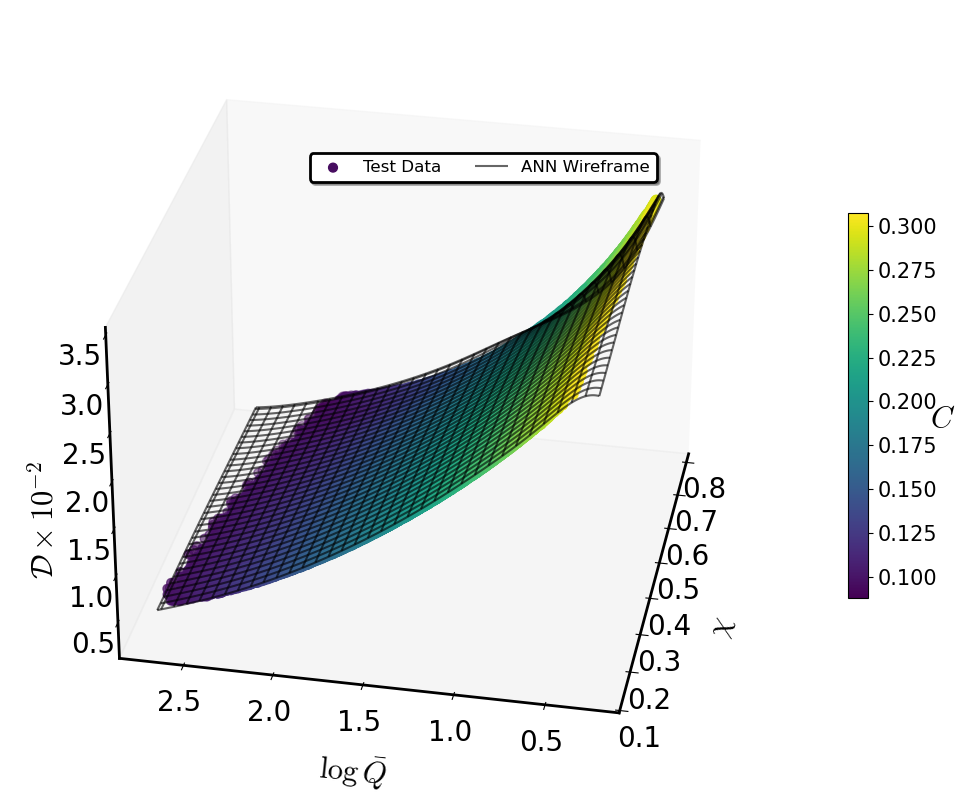}\hfill
    \caption{\label{fig:D_representation} Normalized frequency $\mathcal{D}$ as a function of the dimensionless parameters $\chi$, and $\log\bar{Q}$. The black grid is associated with the predicted ANN wireframe produced by the model (\ref{eq:D_freq_fit}). In addition, the color variation in the presented test data corresponds to the star's compactness $C$, as indicated by the accompanying vertical colored bar.}
\end{figure}

In Fig. \ref{fig:D_representation}, we present the ANN predicted surface (\ref{eq:D_freq_fit}) as a wireframe that best reproduces the data in the test set.
The colors in the illustrated test data representation denote different values of the star's stellar compactness, as indicated by the vertical colored bar. Regardless of the specific value of the dimensionless angular momentum $\chi$, the most compact stellar configurations—with smaller $\bar{Q}$ values—tend to exhibit higher normalized $\mathcal{D}$ parameters, which in turn lead to lower normalized moments of inertia $\bar{I}$. This complements the description provided in the previous SubSec. \ref{subsubsec:ANN_for_Ibar}. Overall, this illustration offers a detailed view of how the ANN model accurately represents the test dataset and the variability across different stellar compactness values. Furthermore, the predicted wireframe extends into regions beyond the physically meaningful parameter space, where no realistic stellar models are available. Despite this, the network yields smooth and coherent extrapolations, underscoring the robustness and consistency of the learned representation.

\begin{figure}[!htb]
   \includegraphics[width=0.46\textwidth]{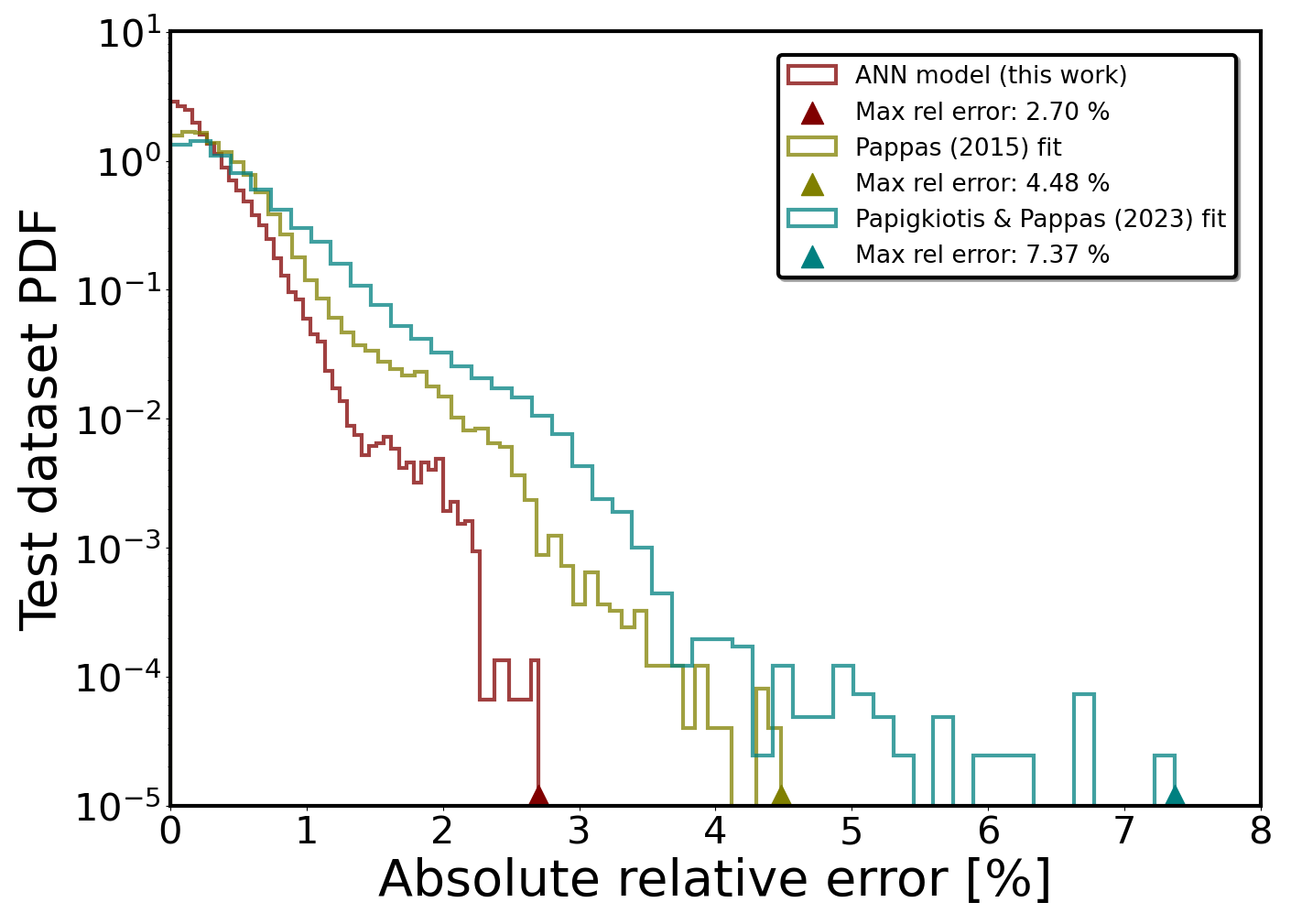}
    \caption{\label{fig:D_fit_relative_errors} 
    Absolute fractional difference $100\% \times ({\mathcal{D}}_{\mathrm{model}} - \mathcal{D})/ \mathcal{D}$.
    The maroon histogram is the relative error distribution associated with our regression model (\ref{eq:D_freq_fit}), while the other two correspond to the fits proposed in the literature.}
\end{figure}

Notably, Fig. \ref{fig:D_fit_relative_errors} presents the histogram of relative errors, further highlighting the accuracy of the suggested model in capturing the data pattern.
While our model achieves the reported maximum relative deviation, it is noteworthy that only $1.34\%$ of the NS configurations in the test ensemble exhibit deviations exceeding $1\%$. The vast majority of these compact objects correspond to high-mass stellar objects, characterized by $C \in [0.193, 0.297]$ and $\bar{Q} \in [1.432, 3.492]$, regardless of their rotational state. Furthermore, in the same panel, we illustrate the absolute fractional differences between the predictions of the suggested DL regression model and those derived from previously established fitting functions \cite{papigkiotis2023universal, pappas2015unified}. As evidenced by the histogram distributions, the suggested model achieves enhanced accuracy and generalization across the entire ensemble of NS configurations, surpassing the performance of existing methods throughout the examined test dataset. Regarding the variance of errors, it is also worth noting that, among the different EOS categories, the hybrid class exhibits the highest maximum deviation, while the hadronic and hyperonic classes show lower maximum fractional differences, around $2.24\%$ and $2.11\%$, respectively (see e.g., Appendix \ref{app:model_characteristics} for a review) It is worth emphasizing that the largest deviations are associated with the {\it DS(CMF)-2 Hybr, SkMp}, and {\it OPGR(DDH$\delta$Y4)} EoS models, as previously identified.

In all cases, the regression model (\ref{eq:D_freq_fit}) provides a comprehensive universal framework for deriving the reciprocal of the star's normalized moment of inertia $\mathcal{D}$ or the star's rotation frequency $f$, given the parameters $M$, $\chi$, and $\bar{Q}$ of the compact object and its surrounding spacetime. 
Moreover, it is worth highlighting that the EoS-independent parametrization (\ref{eq:D_freq_fit}) offers better inference performance compared to the previously introduced relation (\ref{eq:Ibar_fit}) in SubSec. \ref{subsubsec:ANN_for_Ibar}, particularly in predicting the reduced moment of inertia.

\begin{figure}[!htb]
   \includegraphics[width=0.46\textwidth]{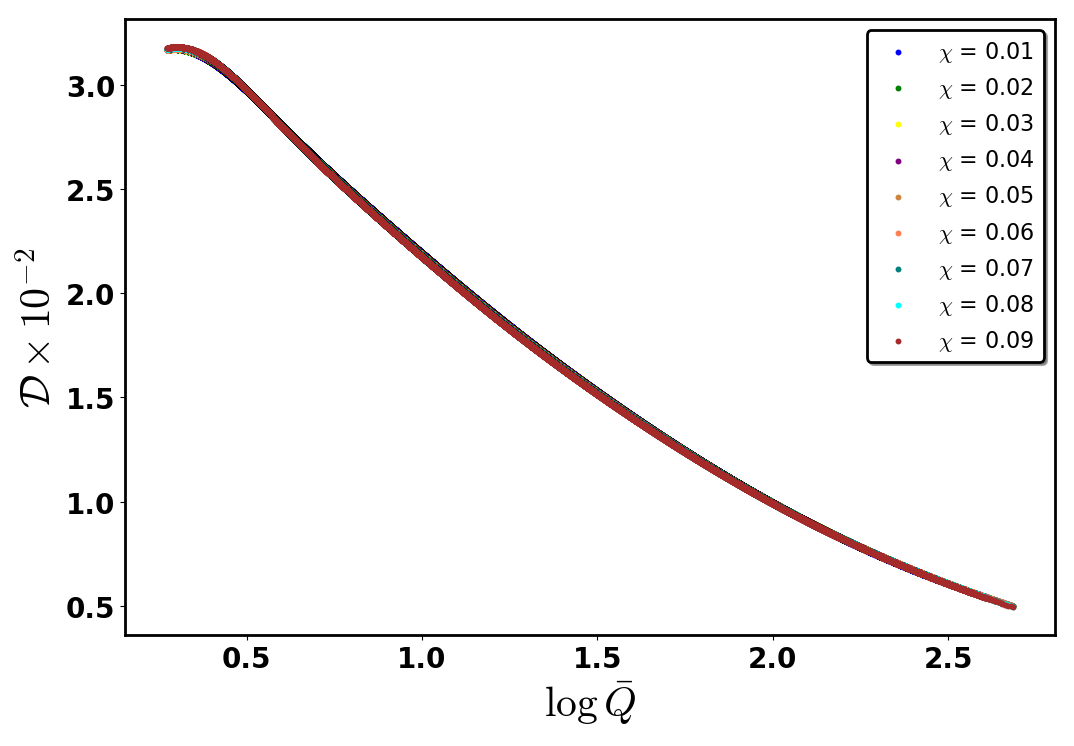}\hfill
    \caption{\label{fig:D_slow_fits} Indicative $\mathcal{D} = \mathcal{D}(\log \bar{Q})$ theoretical curves for a discrete sample of dimensionless angular momentum values $\chi \in [0.01,0.09]$ correspond to slowly rotating NS configurations. The regression model (\ref{eq:D_freq_fit}) fully reproduces the $\mathcal{D}-\bar{Q}$ EoS-insensitive behavior in the slowly rotating limit.}
\end{figure}

Moreover, delving into the model's capacity to faithfully replicate the behavior of slowly rotating NSs presents an intriguing investigation. Despite not being explicitly trained on spin values associated with slowly rotating stellar configurations, the theoretical formula (\ref{eq:D_freq_fit}) effectively mirrors the characteristics described by the Hartle--Thorne approximation for an array of discrete spins $\chi \in [0.01, 0.09]$, as shown in Fig. \ref{fig:D_slow_fits}. This underscores the model's robustness and its ability to generalize across varying rotation rates. Importantly, across all slow spin rates examined, the $\mathcal{D}$ parameter, has a dependence of $\bar{Q}$, illustrating a consistent universal behavior for slowly rotating NS configurations similar to the better-known $\bar{I}-\bar{Q}$ universal relation.

\begin{figure}[!htb]
    \includegraphics[width=0.46\textwidth]{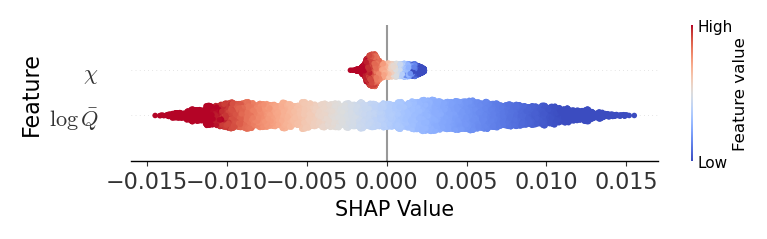}\hfill
    \includegraphics[width=0.46\textwidth]{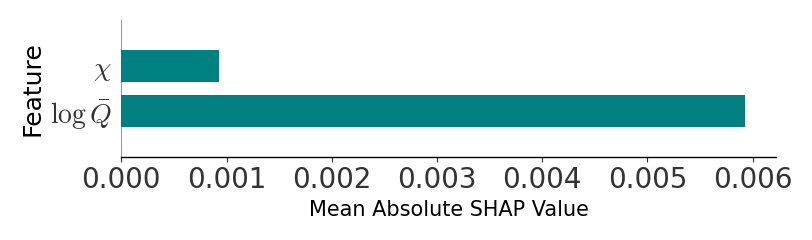}\hfill
    \caption{\label{fig:D_fit_shap_figs} 
    Same as Fig. \ref{fig:Ibar_shap_figs} for the trained DL model (\ref{eq:D_freq_fit}).}
\end{figure}

Following the methodology presented in SubSec. \ref{subsubsec:ANN_for_Ibar}, we utilize the SHAP method to provide interpretability for the predictions of the trained DL model. The SHAP values illustration as depicted in Fig. \ref{fig:D_fit_shap_figs} (top and bottom panel) is similar to the corresponding representation provided in Fig. \ref{fig:Ibar_shap_figs}. It should be noted that the test subset used for computing the SHAP values of the parameters $\chi$ and $\log \bar{Q}$ for each NS configuration corresponds to the same $X_1$ dataset defined in the previous subsection.

For each particular NS configuration on the test subset $X_1$, we investigate the influence of the parameter space values to provide an interpretation of the model’s output. From Fig. \ref{fig:D_fit_shap_figs} (top panel), it is evident that extremely high parameter values of $\chi$, and $\log \bar{Q}$ contribute negatively to the model’s expected value $\mathbb{E}_X(\hat{\mathcal{D}}_{\theta^\star})$. These NS configurations generally correspond to less compact stars, characterized by smaller values of parameter $\mathcal{D}$ and consequently larger moments of normalized inertia. Conversely, significantly small values of spin feature $\chi$ primarily contribute positively, while the lower quadrupole $\bar{Q}$ values, associated with more compact NS models, can impact either positively or negatively. The combination of features with low parameter values corresponds to the most massive NS models, characterized by smaller normalized moments of inertia $\bar{I}$.

In Fig. \ref{fig:D_fit_shap_figs} (bottom panel), the mean absolute SHAP value for each feature across the NS models in the test subset is displayed. The corresponding mean value for the rotation parameter $\chi$ is $9.28\times 10^{-4}$, whereas the quadrupole deformation parameter $\bar{Q}$ exhibits a significantly higher value of $5.93\times 10^{-3}$. The spin parameter $\chi$ enables the model to capture rotational patterns across the NSs ensemble, while the logarithmic quadrupole deformation $\log \bar{Q}$ offers deeper insights into the star’s structure by encoding information related to both rotation and the EoS. Taken together, these features are informative for the DL model (\ref{eq:D_freq_fit}), enabling it to identify hidden patterns in the data and, therefore, produce accurate predictions of the normalized parameter $\mathcal{D}$ across a diverse ensemble of EoSs and NS configurations.

\begin{figure}[!htb]
   \includegraphics[width=0.46\textwidth]{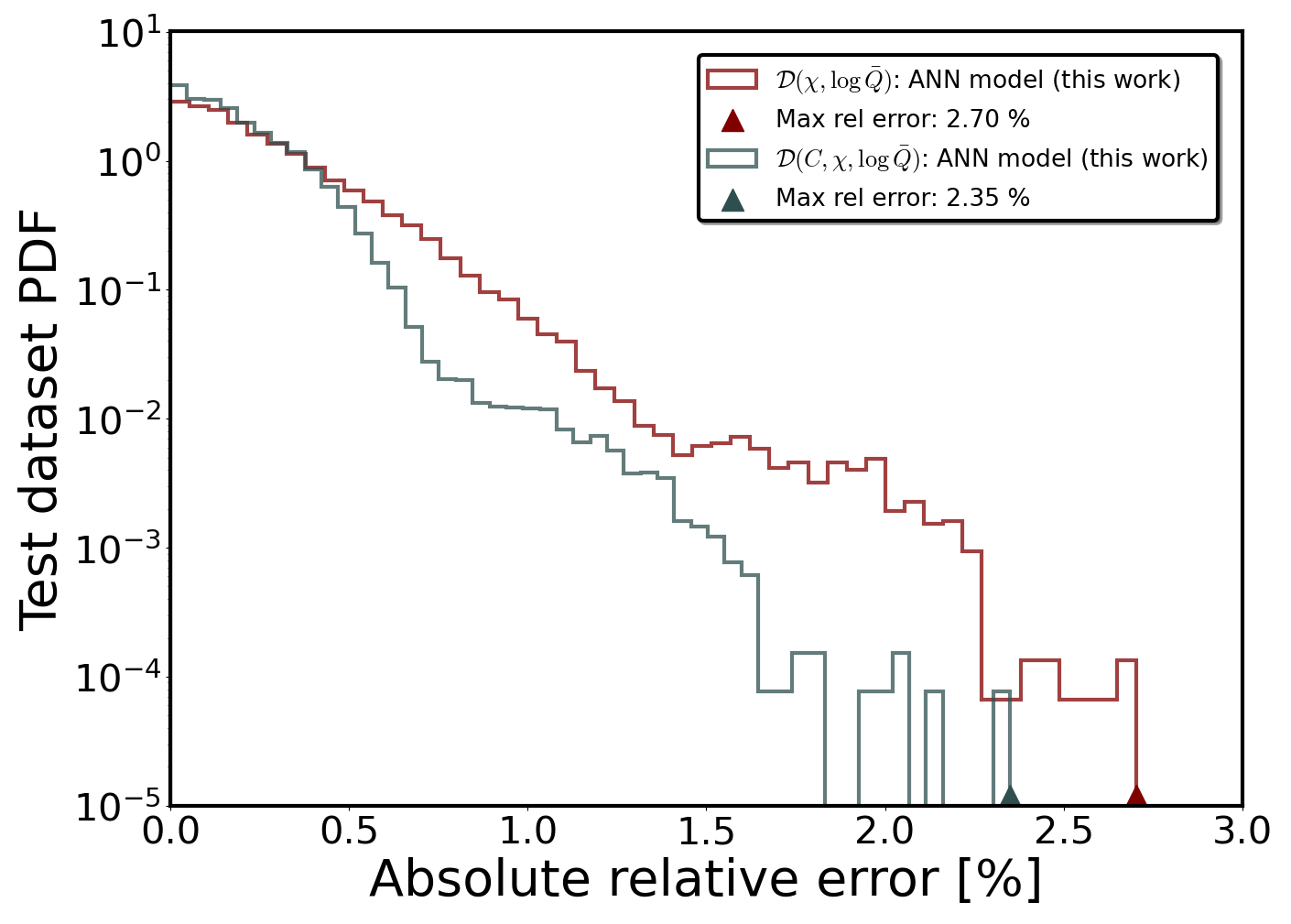}
    \caption{\label{fig:D_fit_relative_errors_2} 
     Relative error distributions on the test dataset for the proposed ANN models described by Eqs. (\ref{eq:D_freq_fit}) and (\ref{eq:D_freq_fit_2}).}
\end{figure}

Complementary to the above, we focus on developing a parametrization for $\mathcal{D}$ that includes stellar compactness alongside the existing input parameters. This expanded set is anticipated to yield a more universal and informative—albeit more intricate— hyperstructure. The new suggested regression model that achieves the closest agreement with the data takes the following functional form:
\begin{equation}
    \label{eq:D_freq_fit_2}
    \mathcal{D}_{\mathrm{model}} = \hat{\mathbb{D}}_{\theta^\star}(C, \chi, \log \bar{Q} ).
\end{equation}
Table \ref{tab:D_freq_eval_meas_2} summarizes the performance measures of the regression model defined in Eq. (\ref{eq:D_freq_fit_2}) as evaluated on the test set. In this formulation, the DL model accurately reproduces the data across the test set, demonstrating improved performance over the previous model with relative deviations $\leq 2.35 \%$ and consistently better evaluation measures.
\begin{table}[!h]
    \footnotesize
    \caption{\label{tab:D_freq_eval_meas_2} Evaluation measures for the parametrization given by Eq. (\ref{eq:D_freq_fit_2}) regarding the test set.}
    \begin{ruledtabular}
        \begin{tabular}{ccccccc}
            MAE & MRE & MSE & $\mathrm{d}_{\text{max}}$ & MAPE & ExpVar & $R^2$ \\
            $\times 10^{-5}$&  $\times 10^{-4}$&  $\times 10^{-9}$ &  ($\%$) &  $\times 10^{-3}$ ($\%$) &  $\times 10^{-2}$&$\times 10^{-2}$   \\    
            \hline
            3.564  &  4.970  & 2.447  & 2.35 & 1.923  & 99.995  & 99.995 \\

        \end{tabular}
    \end{ruledtabular}
\end{table}
For completeness, the histogram of the relative errors is presented in Fig. \ref{fig:D_fit_relative_errors_2}, illustrating the overall distribution of the model's deviations.
In addition, to enable a direct comparison, this figure also includes the relative deviations corresponding to the ANN model given in Eq. (\ref{eq:D_freq_fit}). Unlike the previous model that excluded stellar compactness, the new parametrization significantly improves accuracy, reducing the fraction of data points with deviations over $1\%$ to just $0.31\%$. As with the previous model, the small fraction of cases with larger deviations predominantly corresponds to high-mass stars, independent of their rotation. Most of these NS configurations are associated with $C$ and $\bar{Q}$ values lying within the ranges $C\in[0.184,0.247]$ and $\bar{Q}\in[1.869, 4.729]$, respectively. In addition, a closer analysis of the error distribution across different EoS categories reveals that the hybrid class is associated with the highest maximum relative deviations. Among all tested EoS models, the {\it DS(CMF)-4 Hybr} EoS displays the most pronounced deviation. In contrast, the hadronic and hyperonic categories show notably smaller maximum errors, approximately $2.03\%$ and $1.79\%$, respectively (see e.g.,  Appendix \ref{app:model_characteristics} for the corresponding violin plots). These values are associated with the {\it Rs} and {\it OPGR(DDH$\delta$Y4)} EoS models.

\begin{figure*}[!thb]
    \centering
    \includegraphics[width=1\textwidth]{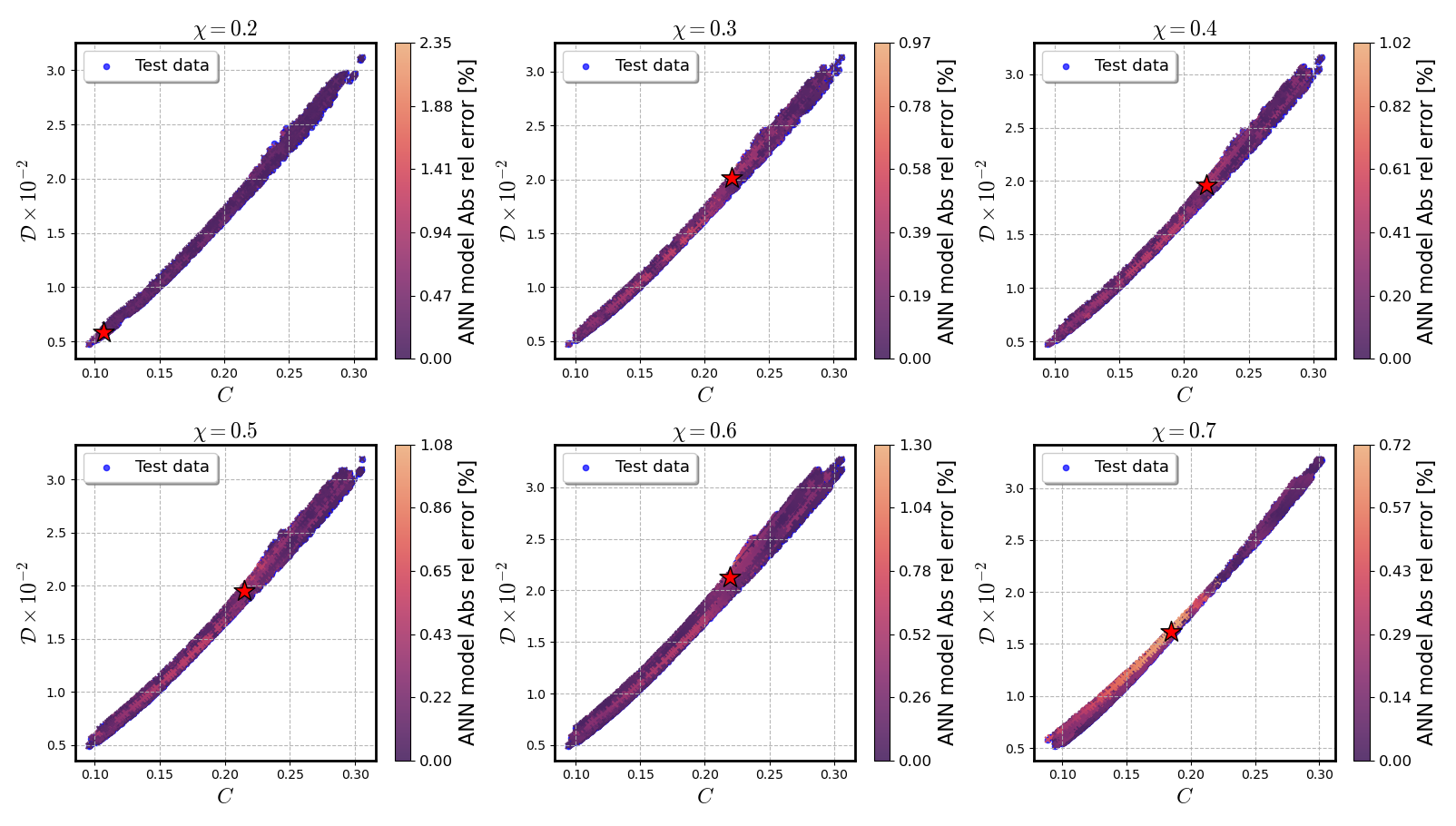}\hfill

    \caption{\label{fig:D_fit_2_curves} $\mathcal{D}-C$ plane: Predicted $\mathcal{D}(C, \log\bar{Q})$ values from the regression model (\ref{eq:D_freq_fit_2}) evaluated on the test dataset for discrete spin values $\chi \in [0.2, 0.7]$. Data points are color-coded by the ANN model’s absolute relative error [\%], while the red star marks the configuration with the maximum relative error. The predictions align closely with the test data, exhibiting sub-percent deviations across most of the investigated parameter space. These results highlight the model’s robustness and reliability under varying spin conditions.}
\end{figure*}

To further evaluate the performance of the regression model in Eq. (\ref{eq:D_freq_fit_2}), Fig. \ref{fig:D_fit_2_curves} illustrates the predicted $\mathcal{D}(C,\log \bar{Q})$ data across the test dataset, evaluated for an array of fixed $\chi$ values. Furthermore, the observed variation in the predictions—consistent with the underlying data distribution—stems from the model’s inference within a lower-dimensional projection of the examined quasi-universal space. Across all cases, each theoretical representation shows strong agreement with the underlying data, reinforcing the quasi-universal nature of the newly presented $\mathcal{D}$–$C$ correlation across the employed ensemble of EoSs.

\begin{figure}[!htb]
    \includegraphics[width=0.46\textwidth]{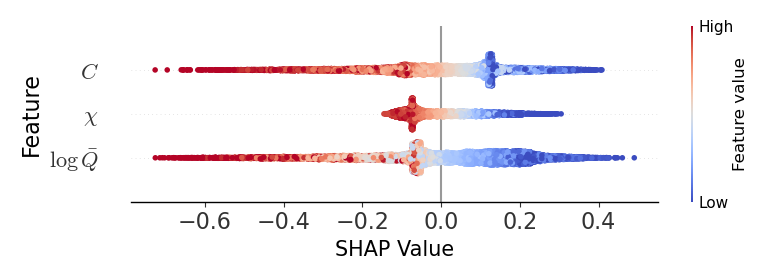}\hfill
    \includegraphics[width=0.46\textwidth]{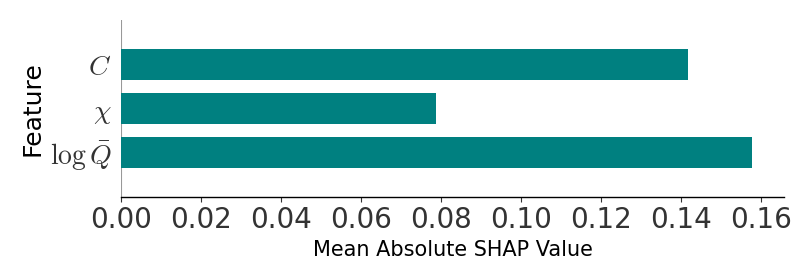}\hfill
    \caption{\label{fig:D_fit_2_shap_figs} 
    Same as Fig. \ref{fig:Ibar_shap_figs} for the trained DL model (\ref{eq:D_freq_fit_2}).}
\end{figure}

In addition, using SHAP values, we provide an interpretation of how the employed parameters contribute to the model’s predicted output. As shown in the top panel of Fig. \ref{fig:D_fit_2_shap_figs}, extremely high values of $\chi$, $C$, and $\log \bar{Q}$ tend to reduce the model's inference, while lower values of these parameters contribute positively. This fact aligns with the earlier findings, where high values of $\chi$ and $\log \bar{Q}$ were associated with negative contributions. The key difference here lies in the behavior of the $C$ parameter: the higher values of $C$ lead to a negative contribution in the model's expected value, just as we observed for $\chi$ and $\log \bar{Q}$. In contrast, lower values of $C$ lead to a positive contribution.
In the bottom panel of Fig. \ref{fig:D_fit_2_shap_figs}, the mean absolute SHAP values for each feature across the NS models in the test subset are shown. The corresponding values for the parameters $\chi$, $C$, and $\log \bar{Q}$ are $7.86 \times 10^{-2}$, $1.42 \times 10^{-1}$, and $1.58 \times 10^{-1}$, respectively. Collectively, these parameters provide an informative feature space that the DL model (\ref{eq:D_freq_fit_2}) uses to identify underlying patterns in the data. Above all, the suggested model provides accurate predictions of the normalized parameter $\mathcal{D}$ across the utilized ensemble of EoSs and NS configurations.

\begin{figure}[!htb]
    \includegraphics[width=0.46\textwidth]{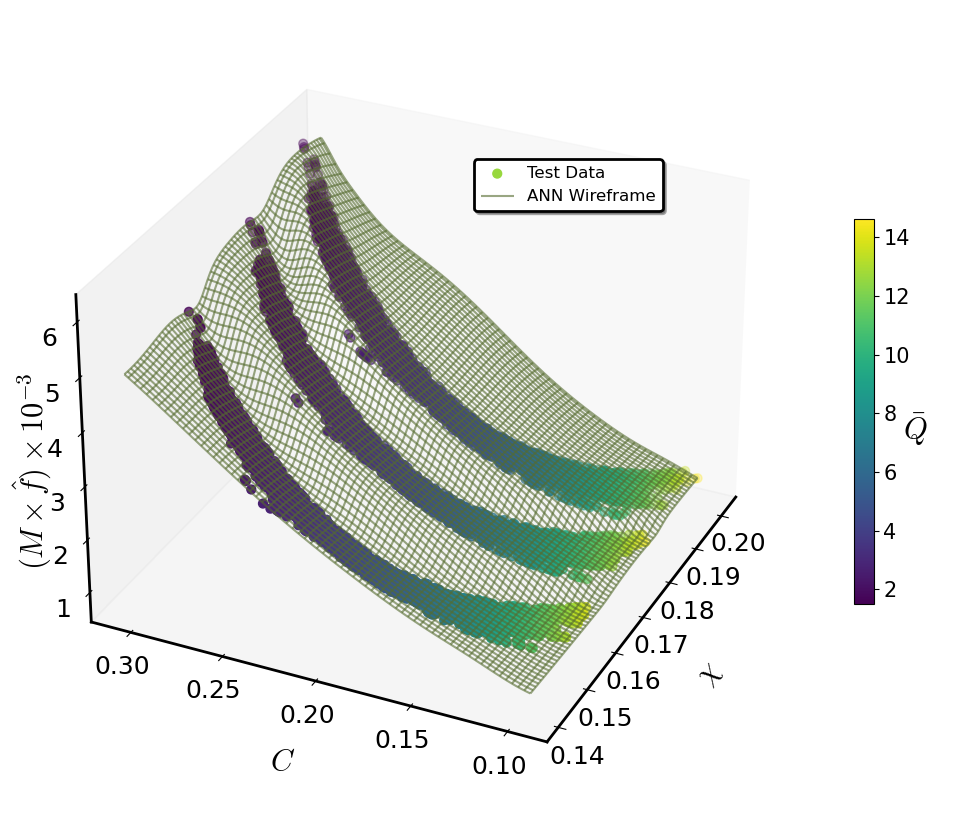}\hfill
    \includegraphics[width=0.46\textwidth]{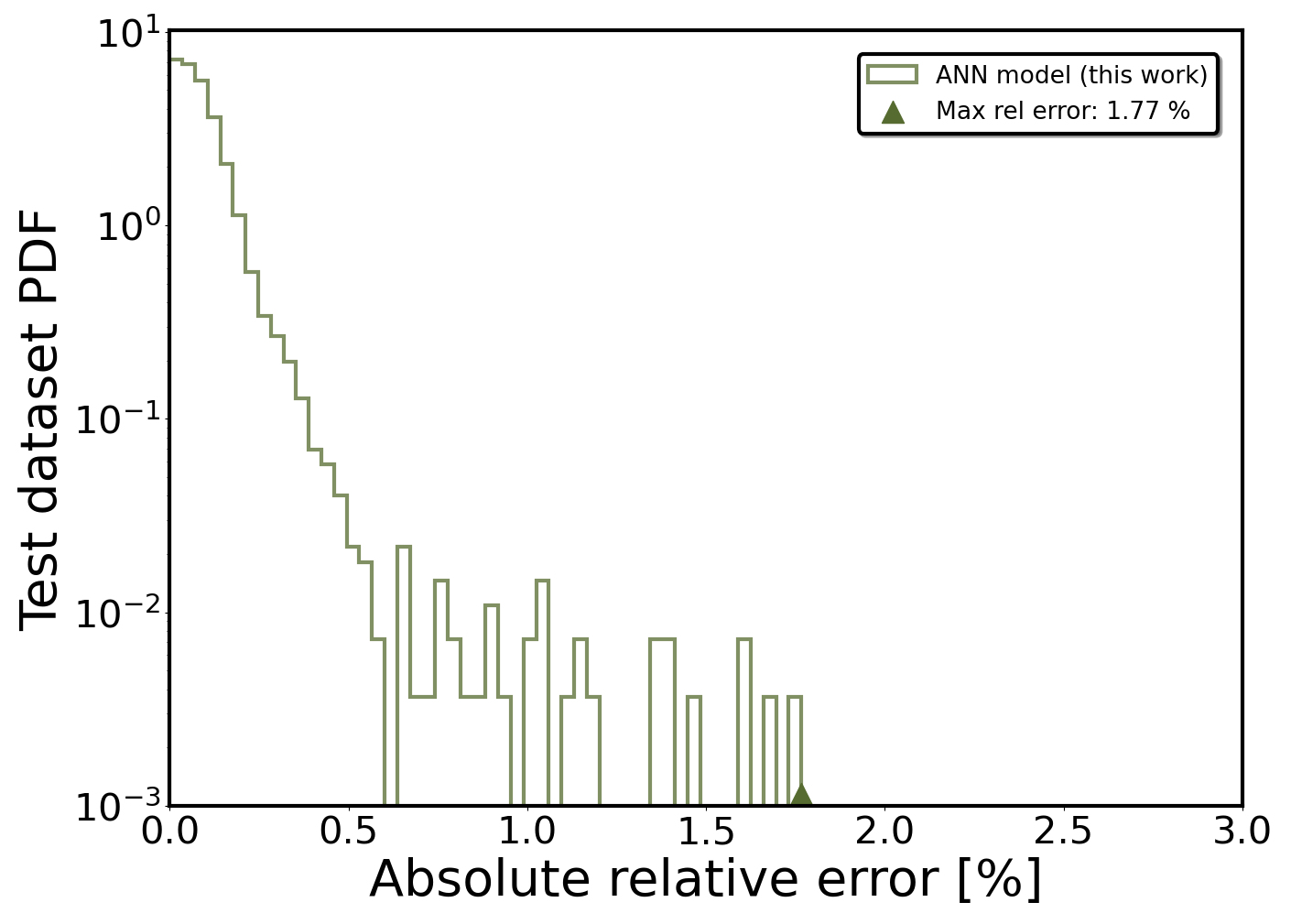}\hfill
    \caption{\label{fig:Mxf_fit_figs} 
    Top panel: Normalized frequency $M \times \hat{f}$ plotted as a function of the dimensionless parameters $\chi$ and $C$. The grid represents the predicted wireframe obtained from the regression model given by Eq. (\ref{eq:Mxf_fit}). The color scale encodes the quadrupole deformation $\bar{Q}$ of each stellar configuration in the test set, as shown by the adjacent vertical color bar. Bottom panel: Distribution of the absolute relative errors between the test data and the predictions from the suggested regression model, evaluated over the test sample of relatively slowly rotating NS configurations discussed in the text.}
\end{figure}

Given the crucial role of $\mathcal{D} -\bar{Q}$ connection in the model's learning ability, it prompts the question of whether there exists a well-behaved correlation that includes other stellar parameters besides $\bar{Q}$ that could be considered at least partially universal. Focusing on relatively slowly rotating NSs with $f\in [172.5, 581.4] \ \mathrm{Hz}$, $\chi \in [0.1406, 0.2024]$ and $C \in [0.095,0.3065]$, we now direct our attention to expressing the star's dimensionless frequency $M \times \hat{f}$ as a function of the parameters $\chi$ and $C$ using the suggested DL model. The trained ANN model that accurately describes the data should be,

\begin{equation}
    \label{eq:Mxf_fit}
    \left( M\times \hat{f}\right)_{\mathrm{model}}  = \hat{\mathcal{F}}_{\theta^\star}(\chi, C).
\end{equation}
We utilize the initial training and validation sets for training and testing, excluding parameters that lie outside the ranges typical of relatively slowly rotating stellar configurations. The regression model (\ref{eq:Mxf_fit}) that best reproduces the particular test data and the corresponding relative deviations histogram are illustrated in Fig. \ref{fig:Mxf_fit_figs} top and bottom panels, respectively. It should be noted that the apparent wireframe deformation in Fig.  \ref{fig:Mxf_fit_figs} (top panel) reflects the visualization of the highly nonlinear mapping properties of the trained ANN model across the investigated $(\chi, C)$ parameter space. More specifically, even when the model is evaluated strictly within the trained input domain (i.e., without extrapolation) on a uniform mesh grid, local regions of rapid variation appear visually warped.

In any case, the model (\ref{eq:Mxf_fit}) quite well reproduces the data in the test set, showcasing satisfactory precision with relative deviation better than $1.77\%$. It should be highlighted that only $0.24\%$ of the relatively slowly rotating NS configurations exhibit relative deviations exceeding $1\%$. These data are primarily associated with less compact stars, characterized by $C \lesssim 0.14$ and $\bar{Q} \gtrsim 8.12$. Concerning the variability of errors across the examined EoS categories, it is noteworthy that the hybrid EoS category exhibits the highest deviation, whereas the hadronic and hyperonic classes display lower maximum relative deviations of about $1.69\%$ and $1.37\%$, respectively (see Appendix \ref{app:model_characteristics} for a review). Among the employed EoS categories, the highest deviations are found to be linked to the {\it DS(CMF)-2 Hybr, GPPVA(TW)}, and {\it DS(CMF)-1} EoS models.

Considering the investigated range of relatively slow rotation rates, the suggested parametrization reliably delivers satisfactory results regardless of the EoS selected, thereby establishing itself as a credible universal relation. This relation is particularly significant, as it enables a reliable estimation of the star’s moment of inertia through the expression $J = I \Omega$, given that the parameters $\chi$, $M$, and $C$ are known. However, it is important to note that this $M\times\hat{f}-\chi-C$ universality breaks down for higher rotation rates, where higher relative errors emerge.

\begin{figure}[!htb]
    \includegraphics[width=0.46\textwidth]{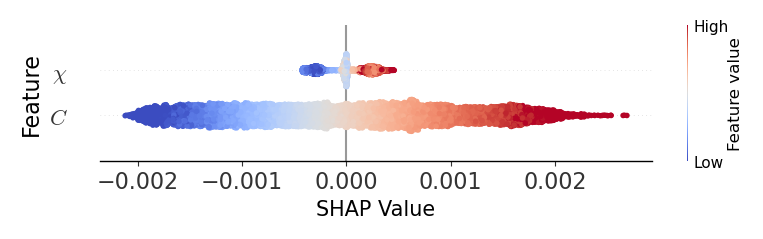}\hfill
    \includegraphics[width=0.46\textwidth]{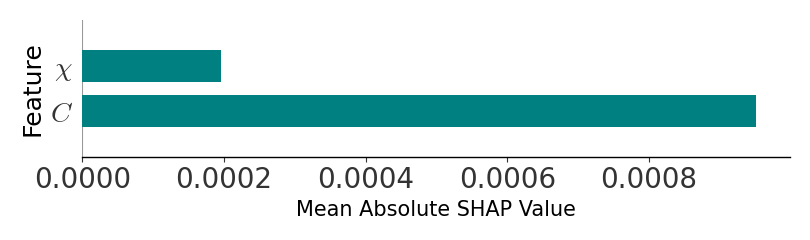}\hfill
    \caption{\label{fig:Mxf_shap_figs} 
    Same as Fig. \ref{fig:Ibar_shap_figs} for the trained DL model (\ref{eq:Mxf_fit}).}
\end{figure}

Using the test set comprising relatively ``slowly'' rotating NS configurations, we then proceed to analyze how different values within the parameter space influence the model’s prediction. As illustrated in Fig. \ref{fig:Mxf_shap_figs} (top panel), both the rotation parameter $\chi$ and the compactness $C$ exhibit influence that can be positive or negative, indicating contributions above or below the model’s expected value $\mathbb{E}_X(\hat{\mathcal{F}}_{\theta^\star})$. In general, NSs that are more compact or exhibit higher rotation rates tend to have a positive impact on the estimator’s expected value $\mathbb{E}_X(\hat{\mathcal{F}}_{\theta^\star})$, while less compact stars or those with very low spin contribute negatively.
Furthermore, Fig. \ref{fig:Mxf_shap_figs} (bottom panel) presents the average absolute SHAP values for each feature across the ensemble of relatively slowly rotating stellar models. The mean values for the dimensionless spin parameter $\chi$ and the compactness $C$ are $1.95 \times 10^{-4}$ and $9.52 \times 10^{-4}$, respectively. While $\chi$ supports the model in identifying rotational patterns within the data, the compactness $C$ contributes more substantially to the model’s inference, as reflected by its higher average absolute SHAP value. At fixed $\chi$, this result supports a quasi-universal correlation between the $M\times\hat{f}$ and $C$ parameters in the slow-rotation regime, consistent with the established quasi-universal $\bar{I}-C$ and $\mathcal{D}-C$ relations. In any case, both parameters play integral roles in the model's ability to describe the data, offering complementary physical information critical to accurately estimating the normalized spin $ M\times\hat{f}$.

\subsection{Relations for the star's multipole moments}
The relativistic multipole moments characterize the spacetime’s structure and physical properties, including the exterior geometry of the star and the geodesics. For a rapidly rotating NS, nonspherical deformation is significant, and contributions from higher-order multipole moments to the exterior vacuum solution surrounding the stellar object become more important. The odd mass multipole moments and the even current moments vanish due to symmetry considerations. However, the non-vanished ones are influenced by the star’s internal microphysics structure, which in turn is determined by the unknown EoS. Therefore, it is essential to investigate  EoS-insensitive relations concerning the star's multipole moments \cite{rezzolla_physics_2018, Pappas:2012ns, pappas2014effectively, laarakkers1999quadrupole, morsink1999relativistic, pappas2015unified, yagi2017approximate,papigkiotis2023universal}.

\subsubsection{\label{subsubsec:ANN_for_Qbar}\bf{Relations for the star's reduced Quadrupole deformation}}

In the relativistic regime, we delve into an EoS-independent approach for parametrizing the Geroch-Hansen reduced quadrupole deformation, $\bar{Q}$, providing a more EoS-insensitive framework for analysis. The universal behavior among observable quantities is highly sensitive to parameter selection, making it crucial to employ the most representative features for accurately estimating a star's global properties \cite{rezzolla_physics_2018}.

We first look for an alternative parametrization to the $\bar{Q}(C,\sigma)$ presented in Ref. \cite{papigkiotis2023universal}, investigating potential enhancements. 
In this pursuit, the parameters $C$, $\chi$, and $\mathcal{D}$ are employed as input features to the DL model. These parameters are carefully selected based on their physical relevance. For example, it is known that quadrupole deformation exhibits a distinct spin dependence when described in terms of stellar compactness \cite{yagi2014multipole}. In addition, the inclusion of $\mathcal{D}$ as a parameter effectively embeds the $\bar{I}-\bar{Q}$ relation—an established universal correlation for slowly rotating NSs \cite{yagi2017approximate}, and a valid approximation at fixed spin in the rapid rotation regime \cite{papigkiotis2023universal, chakrabarti2014q, pappas2014effectively}—thereby enhancing the model’s capacity to estimate the reduced quadrupole deformation under varying rotational conditions.

\begin{figure}[!htb]
   \includegraphics[width=0.46\textwidth]{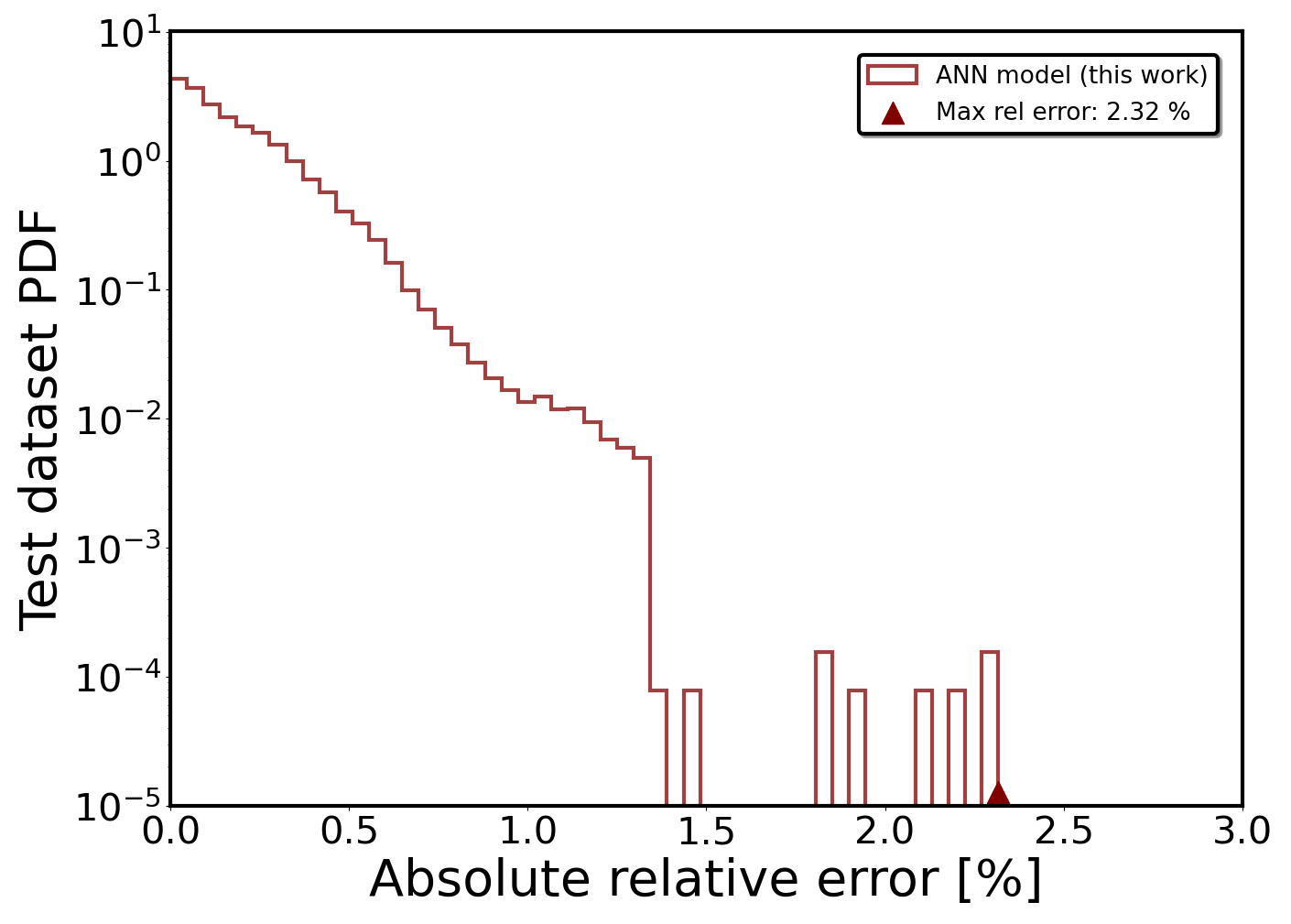}
    \caption{\label{fig:Q_fit_relative_errors} 
    Absolute fractional difference $100\% \times ({\bar{Q}}_{\mathrm{model}} - \bar{Q})/ \bar{Q})$. The colored histogram corresponds to the relative error distribution associated with the suggested model (\ref{eq:Qbar_fit}).}
\end{figure}

Building on the preceding framework, the resulting ANN model that best reproduces the data is expressed as,
\begin{equation}
    \label{eq:Qbar_fit}
    \bar{Q}_{\mathrm{model}} = \hat{\mathcal{\bar{Q}}}_{\theta^\star}(C,\chi, \mathcal{D}).
\end{equation}
The evaluation measures associated with the trained model (\ref{eq:Qbar_fit}) on the validation set are presented in Table (\ref{tab:Qbar_eval_meas}).
\begin{table}[!h]
    \footnotesize
    \caption{\label{tab:Qbar_eval_meas} Evaluation measures for the parametrization given by Eq. (\ref{eq:Qbar_fit}) regarding the test set.}
    \begin{ruledtabular}
        \begin{tabular}{ccccccc}
           MAE & MRE & MSE & $\mathrm{d}_{\text{max}}$ & MAPE & ExpVar & $R^2$ \\
            $\times 10^{-3}$&  $\times 10^{-1}$&  $\times 10^{-5}$ &  ($\%$) &  $\times 10^{-3}$ ($\%$) &  $\times 10^{-2}$&$\times 10^{-2}$   \\    
            \hline
            6.120& 2.774 & 7.522 &  2.32    & 1.871 &  99.998 & 99.998  \\

        \end{tabular}
    \end{ruledtabular}
\end{table}
It is evident that the regression formula (\ref{eq:Qbar_fit}) accurately reproduces the data on the test set, exhibiting good accuracy with a relative error better than $2.32\%$. The histogram displaying the distribution of relative errors is shown in Fig. \ref{fig:Q_fit_relative_errors}.
The resulting optimal model demonstrates robust performance regardless of the EoS selection, establishing it as an effective EoS-insensitive relation. It is also important to highlight that only $0.33\%$ of the total set of stellar models exhibit relative errors larger than $1\%$. These cases are predominantly associated with NSs possessing intermediate compactness values $C \in [0.184, 0.237]$, reduced parameter $\mathcal{D}\in[1.446,2.317]\times 10^{-2}$ and dimensionless moment of inertia within the range $\bar{I} \in [4.743, 11.01]$, regardless of spin. With regard to the distribution of errors across the different EoS categories, the hybrid class shows the highest maximum deviation. This value is associated with the {\it DS(CMF)-7 Hybr} EoS model. In contrast, the hadronic and hyperonic classes yield lower maximum relative deviations, approximately $2.20\%$ (EoS {\it Rs}) and $1.84\%$ (EoS {\it OPGR(DDH$\delta$Y4)}), respectively (see e.g., Appendix \ref{app:model_characteristics} for a review).

Furthermore, it should be noted that instead of the dimensionless spin parameter $\chi$, one may utilize an alternative rotational parametrization such as the reduced spin parameter $\sigma$. This substitution yields equally reliable results and preserves the universal representation of the relation. In this context, the expression $\bar{Q}(C, \sigma, \mathcal{D})$ can be regarded as an extended refinement of the parametrization introduced in \cite{papigkiotis2023universal}. Comparable results are also anticipated for alternative rotation parametrizations, such as $\mathcal{E}$, $M \times \hat{f}$, and $R_{\mathrm{eq}} \times f/c$.

\begin{figure*}[!thb]
    \centering
    \includegraphics[width=1\textwidth]{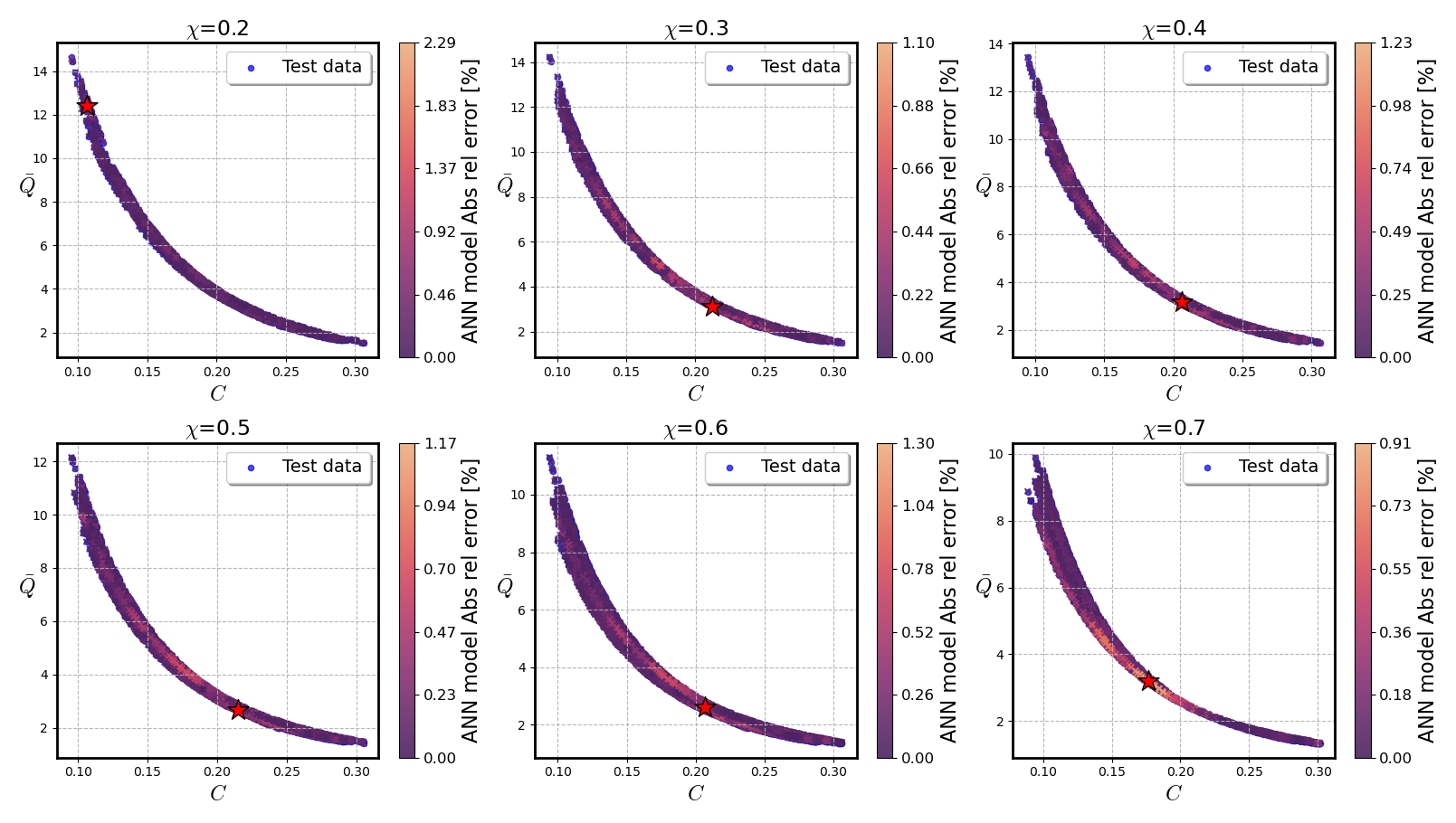}\hfill

    \caption{\label{fig:Q_fit_contours} $\bar{Q}-C$ plane: Predicted $\bar{Q}(C,\mathcal{D})$ data from the regression model (\ref{eq:Qbar_fit}) for a discrete set of spin values, $\chi \in [0.2, 0.7]$. Data points are color-coded by the ANN model’s absolute relative error [\%], while the red star marks the configuration with the maximum relative error. The predictions align closely with the test data, exhibiting sub-percent deviations over most of the investigated parameter space. This demonstrates the model’s robustness and reliability across different spin conditions.}  
\end{figure*}

\begin{figure*}[!thb]
    \centering
    \includegraphics[width=1\textwidth]{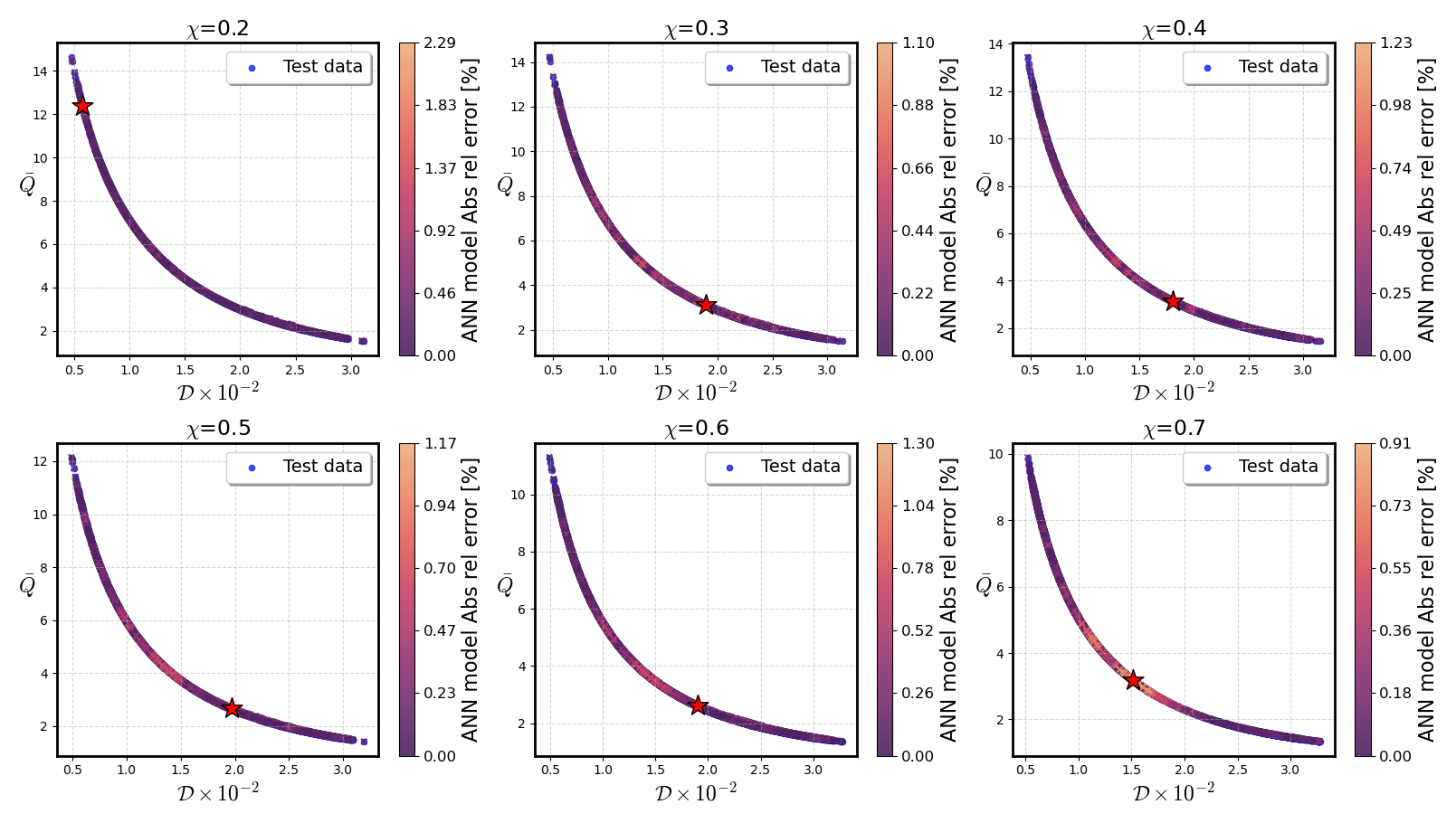}\hfill

    \caption{\label{fig:Q_D_fit_curves} $\bar{Q}-\mathcal{D}$ plane: Predicted $\bar{Q}(C, \mathcal{D})$ data from the regression model (\ref{eq:Qbar_fit}) for a discrete set of spin values, $\chi \in [0.2, 0.7]$. Data points are color-coded by the ANN model’s absolute relative error [\%], while the red star marks the configuration with the maximum relative error. The predictions align closely with the test data, exhibiting deviations below one percent for most cases. This consistent performance demonstrates the model’s robustness under varying spin conditions.}
\end{figure*}

As further validation of the effectiveness of the regression model given in Eq. (\ref{eq:Qbar_fit}), Figs. \ref{fig:Q_fit_contours} and \ref{fig:Q_D_fit_curves} display the corresponding predicted data over the test dataset, evaluated for a discrete array of fixed $\chi$ values. Each  $\bar{Q}(C)$ and $\bar{Q}(\mathcal{D})$ data representation that arises from the DL model demonstrates excellent agreement with the underlying test data, reinforcing the quasi-universal nature of the $\bar{Q}-C$ relation as well as the universal connection of the $\bar{Q}-\mathcal{D}$ parameters across a broad set of EoS categories for rapidly rotating NS star sequences as long as the spin parameter $\chi$ remains fixed.
In addition, this analysis offers a complementary perspective on the well-known $\bar{I}-C$ quasi-universal relation, the $\mathcal{D}-C$ representation serving as a complement, and the $\bar{I}-\bar{Q}$ universal relation for sequences of rapidly rotating stars with fixed spin values \cite{yagi2017approximate, rezzolla_physics_2018, roy2025icq}. 

\begin{figure}[!htb]
    \includegraphics[width=0.46\textwidth]{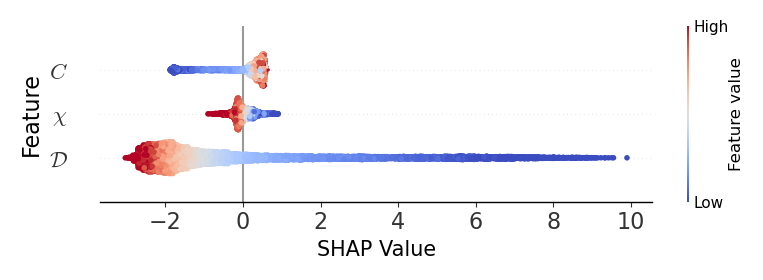}\hfill
    \includegraphics[width=0.46\textwidth]{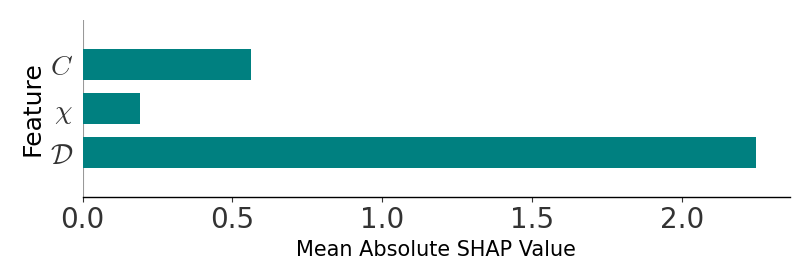}\hfill
    \caption{\label{fig:Q_shap_figs} Same as Fig. \ref{fig:Ibar_shap_figs} for the trained DL model (\ref{eq:Qbar_fit}).}
\end{figure}

Using the already defined test subsample $X_1$, we further investigate the influence of the model's (\ref{eq:Qbar_fit}) dependent parameters on the final output. As shown in Fig. \ref{fig:Q_shap_figs} (top panel), each parameter can contribute either positively or negatively relative to the model’s expected value $\mathbb{E}_X(\hat{\bar{Q}}_{\theta^\star})$. For each $\bar{Q}$ estimation, higher values of parameters $\chi$ and $\mathcal{D}$ are associated with negative contributions, while higher values of $C$ exert a positive impact. In contrast, less compact stellar models generally have a negative impact, while models with low dimensionless angular momentum and $\mathcal{D}$ tend to exert a positive effect.

In Fig. \ref{fig:Q_shap_figs} (bottom panel), the mean absolute SHAP value for each parameter of interest across the NS models in the test subset is displayed. The average values of the parameters $\chi$, $C$, and $\mathcal{D}$ are $0.19$, $0.56$, and $2.25$, respectively. While the parameters $C$ and $\chi$ help the DL model to capture the physical structure and rotational dynamics of the stellar configurations, the parameter $\mathcal{D}$ emerges as the most influential feature, as indicated by its average SHAP value contribution. This result is anticipated, given that the reciprocal of the star's moment of inertia and the quadrupole deformation are intrinsically connected through the star’s rotational and structural characteristics. In particular, this intrinsic connection is evident in the slow-rotation regime where the well-established $\bar{I}–\bar{Q}$ relation reflects its common dependence. Likewise, for sequences at fixed spin $\chi$, as shown previously, EoS-insensitive $\mathcal{D}-\bar{Q}$ and $\bar{I}–\bar{Q}$ relations persist (see e.g. Fig. \ref{fig:Q_D_fit_curves} for a review). Although the mean contribution of the $\chi$ feature appears small compared to the other parameters in the corresponding figure, it plays a crucial role in capturing the universal combination among the examined quantities. As such, its inclusion is essential for preserving the universal character of $\bar{Q}$ in the predictive model. Notably, training the suggested model using only $C$ and $\mathcal{D}$ as parameters results in a noticeable increase in the relative errors of approximately $20\%$, indicating a breakdown of universality. In all cases, the parameters $(C, \chi, \mathcal{D})$ encode essential physical properties, and their combined utilization empowers the suggested DL model in Eq. (\ref{eq:Qbar_fit}) to deliver accurate and robust predictions of $\bar{Q}$ across a diverse ensemble of NS models in a way that is independent of the underlying EoS.

\subsubsection{\label{subsubsec:ANN_for_S3bar}\bf{Relations for the star's reduced Octupole deformation}}
In addition to the mass quadrupole moment, the spin octupole moment $S_3$ associated with the next higher-order contribution to the rotation multipole moments sequence after angular momentum (known as current moments) is also significant for NSs. Studies have demonstrated that stellar configurations adhere to a universal three-parameter property, where multipole moments beyond the quadrupole depend on the first three moments: mass, dimensionless angular momentum, and reduced quadrupole \cite{pappas2014effectively, stein2014three, yagi2014effective, rezzolla_physics_2018}. Building on this foundation, we further extend the investigation to explore potential EoS-insensitive relations involving the Geroch-Hansen octupole moment $S_3$.

\begin{figure}[!htb]
    \includegraphics[width=0.46\textwidth]{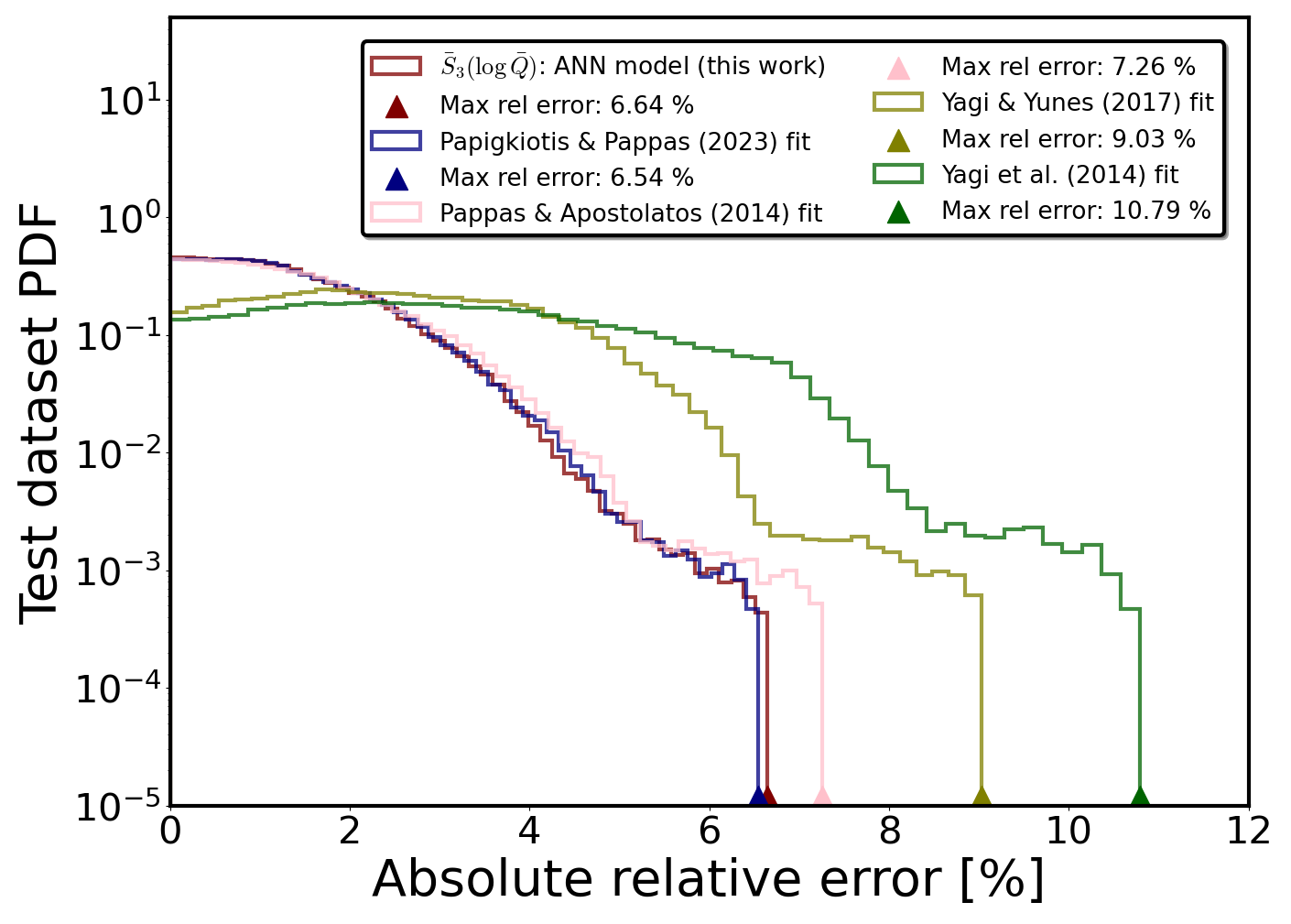}\\
    \includegraphics[width=0.46\textwidth]{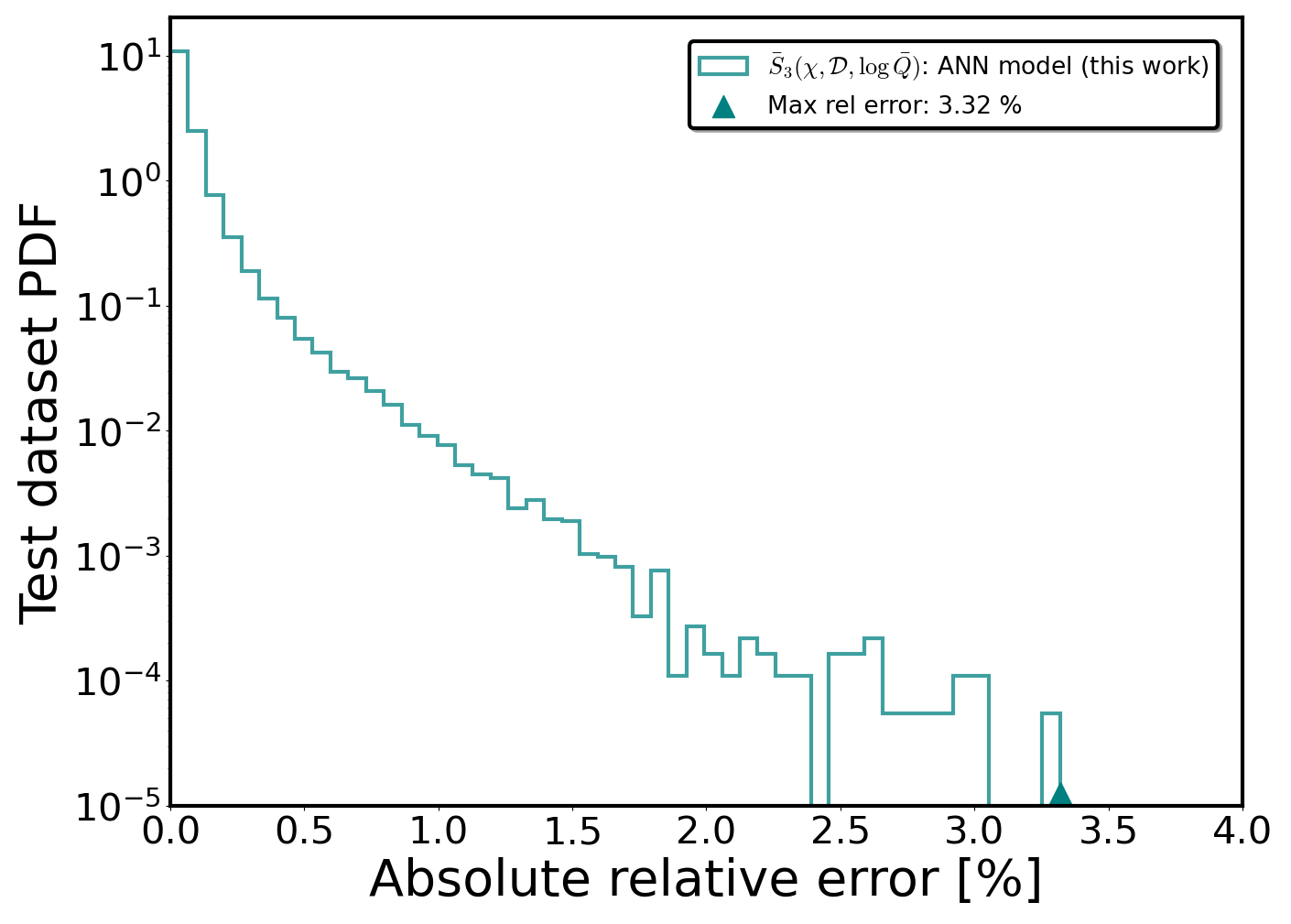}
    \caption{\label{fig:S3bar_logQbar_fit_relative_errors} 
    Top panel: Absolute relative error histogram $100\% \times ({\bar{S}}_{3,\mathrm{model}} - \bar{S}_3)/ \bar{S}_3$ derived using the suggested DL model (\ref{eq:S3bar_curve_fit}) on the test dataset. Alongside the relative errors of the investigated DL model, the other colored histograms also depict the error distributions associated with previously established analytical fits from the literature, providing a basis for comparative analysis. Bottom panel: Same as the top panel for the regression model $\bar{S}_{3}(\chi,\mathcal{D},\log\bar{Q})$ provided via Eq. (\ref{eq:S3bar_surface_fit}).}
\end{figure}

At first, it is worth highlighting the well-established correlation between the reduced spin octupole moment $\bar{S}_3$ and the reduced quadrupole deformation $\bar{Q}$, extensively explored in the literature \cite{yagi2017approximate, pappas2014effectively, yagi2014effective, rezzolla_physics_2018}. Building upon this, Ref. \cite{papigkiotis2023universal} suggested an effective new parametrization, expressing $\bar{S}_3$ as a function of $\log \bar{Q}$. This refined approach demonstrates improved performance over previous methods, substantially reducing the associated relative errors. Furthermore, employing the proposed DL architecture to fit the data yields the best model 
\begin{equation}
    \label{eq:S3bar_curve_fit}
    \bar{S}_{3,\mathrm{model}} = \hat{\bar{\mathcal{S}}}_{\theta^\star}(\log \bar{Q}),
\end{equation}
with accuracy comparable to that of the previously discussed analytical fit, further reinforcing the effectiveness of this approach in capturing the $\bar{S}_3-\bar{Q}$ relation, irrespective of the chosen EoS. The reported accuracy on the test dataset for the DL model is approximately $6.64\%$. Additionally, Fig. \ref{fig:S3bar_logQbar_fit_relative_errors} (top panel) presents the error variance in comparison with alternative analytical formulations from the literature, further indicating the effectiveness of the $\bar{S}_3-\log \bar{Q}$ representation. Between the two $\bar{S_3}-\log \bar{Q}$ fits with comparable overall performance, it is noteworthy that only $1\%$ of the stellar models in the test set exhibit relative deviations exceeding $4\%$ for the ANN-based model, compared to $1.20\%$ of the corresponding stellar configurations for the associated analytical fit. 
In terms of error distribution across the different EoS categories, relative deviations exceeding approximately $4\%$ are predominantly associated with the hybrid EoS class. In contrast, the hadronic and hyperonic categories exhibit lower maximum relative errors, both close to $4\%$ (see e.g., Appendix \ref{app:model_characteristics} for the relative violin plots). The most significant deviations are associated with {\it DS(CMF)-2 Hybr, MTVTC}, and {\it DNS}, corresponding to the hybrid, hadronic, and hyperonic EoS categories, respectively.

Moreover, building on the $\bar{S}_3-\chi$, $\sigma$, and $\mathcal{E}-\bar{Q}$ parametrizations introduced in Ref.~\cite{papigkiotis2023universal}, which incorporate higher-order spin corrections to the $\bar{S}_3-\bar{Q}$ relation, it is worthwhile to investigate whether including $\mathcal{D}$ alongside the reduced quadrupole can further enhance predictive accuracy while preserving the EoS-insensitive characteristics of the parameters combination.
At this point, our analysis shifts on the parameters $\bar{S}_3$, $\chi$, $\mathcal{D}$, and $\log \bar{Q}$. The resulting regression model that accurately captures the underlying data structure is presented by the functional form
\begin{equation}
    \label{eq:S3bar_surface_fit}
    \bar{S}_{3,\mathrm{model}} = \hat{\bar{\mathbb{S}}}_{\theta^\star}(\chi, \mathcal{D},\log \bar{Q}).
\end{equation}
Evaluation measures for this regression model, assessed on the validation set, are summarized in Table \ref{tab:S3bar_surface_fit_eval_meas}.

\begin{table}[!h]
    \footnotesize
    \caption{\label{tab:S3bar_surface_fit_eval_meas} Evaluation measures for the parametrization given by the Eq. (\ref{eq:S3bar_surface_fit}) regarding the test dataset.}
    \begin{ruledtabular}
        \begin{tabular}{ccccccc}
            MAE & MRE & MSE & $\mathrm{d}_{\text{max}}$ & MAPE & ExpVar & $R^2$ \\
            $\times 10^{-3}$&  $\times 10^{-1}$&  $\times 10^{-4}$ &  ($\%$) &  $\times 10^{-4}$ ($\%$) &  $\times 10^{-2}$&$\times 10^{-2}$   \\    
            \hline
            7.261 & 7.632  &  4.649 &  3.32    & 6.934  & 99.998  & 99.998  \\

        \end{tabular}
    \end{ruledtabular}
\end{table}

In this form, the ANN hypersurface (\ref{eq:S3bar_surface_fit}) accurately reproduces the data on the test dataset, demonstrating high accuracy with relative error $\leq 3.32\%$.  Fig. \ref{fig:S3bar_logQbar_fit_relative_errors} (bottom panel) presents a histogram illustrating the distribution of relative errors, providing an insight into the model's performance.

The suggested model performs consistently well across different EoS selections, establishing it as a robust EoS-insensitive relation. Notably, just $671$ NS models, amounting to only $0.24\%$ of the entire test dataset, exceed a relative error of $1\%$, highlighting the reliability of the suggested ANN model across a wide array of stellar configurations. Furthermore, the error variance appears to be randomly distributed, showing no systematic dependence across specific regions of the ($\chi, \mathcal{D}, \log \bar{Q}$) parameter space. A detailed examination of the error distribution across various EoS categories indicates that the hadronic class yields the highest maximum relative deviation. On the contrary, the hyperonic and hybrid classes exhibit comparatively lower maximum errors, on the order of $3.02\%$ and $2.77\%$, respectively (see e.g., Appendix \ref{app:model_characteristics} for a review). In addition, the most pronounced deviations are sequentially associated with the EoS models {\it TM1e, OPGR(DDH$\delta$Y4}), and {\it DS(CMF)-1 Hybr}.

\begin{figure*}[!thb]
    \centering
    \includegraphics[width=1\textwidth]{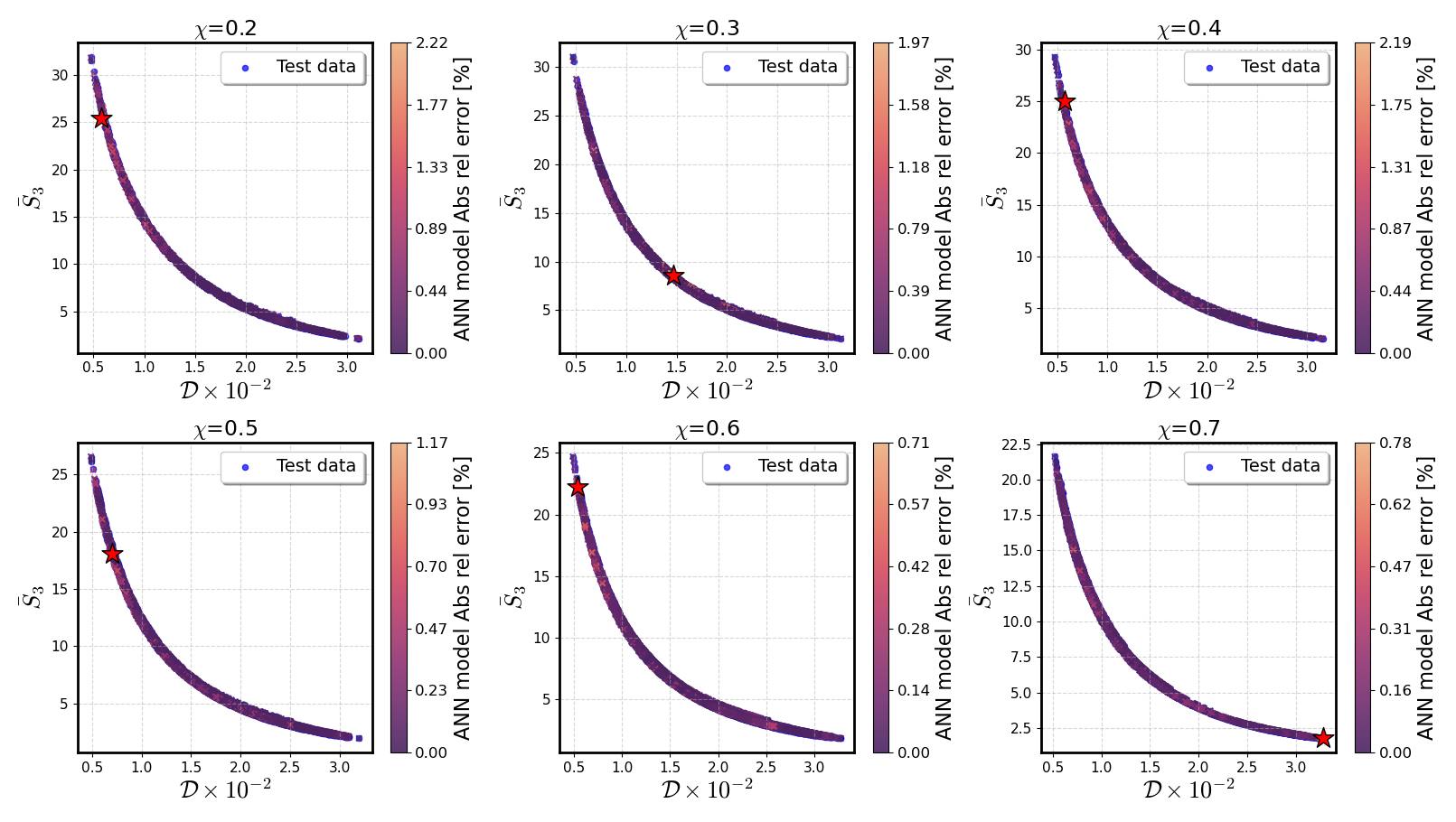}\hfill
    \caption{\label{fig:S3_D_fit_curves} $\bar{S}_3 - \mathcal{D}$ plane: Predicted $\bar{S}_3(\mathcal{D},\log\bar{Q})$ data from the regression model (\ref{eq:S3bar_surface_fit}), evaluated over a discrete set of spin values $\chi \in [0.2, 0.7]$. Data points are color-coded by the ANN model’s absolute relative error [\%], while the red star marks the configuration with the maximum relative error. In any case, the model predictions align closely with the test data, showing deviations below one percent in most of the investigated parameter space. This further indicates the model's robustness and reliability across different conditions.}
\end{figure*}

As further validation of the effectiveness of the regression model proposed in Eq. (\ref{eq:S3bar_surface_fit}), Fig. \ref{fig:S3_D_fit_curves} illustrates the newly identified $\bar{S}_3-\mathcal{D}$ EoS-insensitive connection for a range of fixed $\chi$ values. The model predictions exhibit excellent agreement with the underlying test data, clearly reproducing the universal behavior of the observed $\bar{S}_3-\mathcal{D}$ relation across sequences of NS configurations with fixed spin $\chi$ parameters. Therefore, the $\bar{S}_3-\mathcal{D}$ universal correlation reveals itself as a novel and robust characteristic of rapidly rotating NS sequences, demonstrating consistency across a wide range of EoSs when the spin rate $\chi$ is kept constant.

Most importantly, selecting $(\chi, \mathcal{D}, \log\bar{Q})$ as input parameters for modeling the reduced octupole $\bar{S}_3$ underscores the influence of higher-order corrections in refining the $\bar{S}_3-\bar{Q}$ universal relation. This parameter combination results in a more robust and EoS-insensitive representation of $\bar{S}_3$, enhancing its general applicability across a wide range of stellar models. Moreover, the suggested regression model is expected to maintain comparable performance when trained with alternative spin-related parameters such as $\sigma$, $\mathcal{E}$, $M \times \hat{f}$, or $R_{\mathrm{eq}} \times f/c$ instead of $\chi$. Lastly, it should be highlighted that the star's compactness $C$ can also serve as an alternative to, or be used alongside, $\mathcal{D}$ as an input feature, preserving the universality of the relation and supporting reliable $\bar{S}_3$ inference.

\begin{figure}[!htb]
    \includegraphics[width=0.46\textwidth]{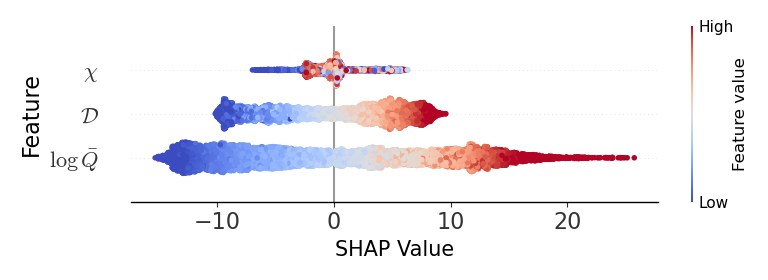}\hfill

    \includegraphics[width=0.46\textwidth]{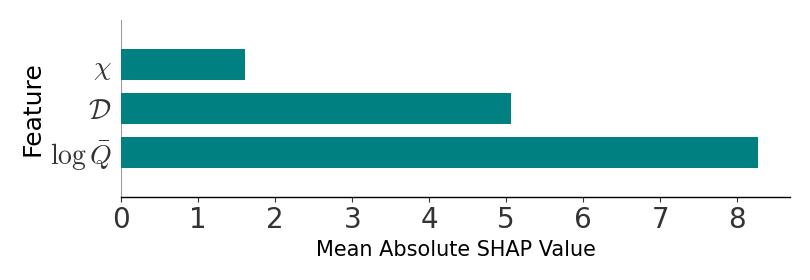}\hfill
    \caption{\label{fig:S3_shap_figs}
    Same as Fig. \ref{fig:Ibar_shap_figs} for the trained DL model (\ref{eq:S3bar_surface_fit}).
    }
\end{figure}

Consistent with previous relations, we evaluate each NS configuration in the test subset $X_1$ to examine how the selected inputs are attributed to the model’s predicted output. This evaluation provides insight into the model’s ability to estimate $\bar{S}_3$ across the defined parameter space. As illustrated in the top panel of Fig. \ref{fig:S3_shap_figs}, each parameter can contribute either positively or negatively to the model's expected value $\mathbb{E}_X(\hat{\bar{S}}_{3,\theta^\star})$. Higher (lower) values of $\mathcal{D}$ and $\log \bar{Q}$ generally correspond to positive (negative) SHAP contributions, indicating a direct correlation. In contrast, feature $\chi$ does not exhibit a clear contribution pattern, as both low and high values can influence either in a positive or negative direction. The observed SHAP-value pattern reflects known physical dependencies of $\bar{S}_3$ on NS properties. More specifically, irrespective of $\chi$, higher values of $\bar{Q}$ lead to increased octupole moments and thus positively influence the model’s expected value $\mathbb{E}_X(\hat{\bar{S}}_{3,\theta^\star})$. In addition, larger (lower) values of $\mathcal{D}$, which correlate with higher (lower) compactness and reduced deformability, are associated with smaller (larger) quadrupole moments, thereby contributing positively (negatively) to $\bar{S}_3$ expected value.
Finally, the bottom panel of Fig. \ref{fig:S3_shap_figs} displays the mean absolute SHAP values for each input feature across the test subset. The average contributions of $\chi$, $\mathcal{D}$, and $\log \bar{Q}$ are $1.60$, $5.07 $, and $8.27$, respectively.

\subsubsection{\label{subsubsec:ANN_for_M4bar}{\bf Relations for the star's reduced Hexadecapole deformation}}
It is crucial to recognize that the multipole moments have a distinct spin dependence when described in terms of stellar compactness or normalized moment of inertia. Moreover, the analysis so far has also highlighted the importance of the $\mathcal{D}$ parameter in capturing the underlying universal representation. Furthermore, as first suggested in \cite{yagi2014multipole}, this spin dependence tends to partially cancel out when one multipole moment is expressed as a function of another (see, for example, the $\bar{S}_3-\bar{Q}$ relation).

In this subsection, we investigate universal relations regarding the spacetime's Geroch-Hansen hexadecapole, which is a higher-order contribution to the sequence of non-vanishing mass multipole moments following the quadrupole deformation. In our analysis, we primarily focus on different forms that include the reduced hexadecapole $\bar{M}_4$. It is well known that $\bar{M}_4$ and $\bar{M}_4/\bar{S}_3$ are directly correlated with $\bar{Q}$, following relations that are approximately EoS-independent \cite{yagi2014effective,pappas2017accurate, yagi2017approximate}. Utilizing the three-hair property for the higher-order moments, we reexamine the universal correlation between the fraction of the higher moments $\bar{M}_4/\bar{S}_3$, and the $\bar{Q}$ through our proposed DL architecture, aiming to achieve potential enhancements.

Although we employed a high-resolution grid $\mathrm{MDIV} \times \mathrm{SDIV} = 751 \times 1501$ for stellar models with axial ratios $r_{\mathrm{pole}}/r_{\mathrm{eq}} \in [0.90, 0.98]$, the numerical extraction of the mass hexadecapole moment becomes unreliable in most cases for $r_{\mathrm{pole}}/r_{\mathrm{eq}} > 0.94$. Given the substantial computational cost associated with the employment of a higher-resolution numerical grid for constructing rapidly rotating NSs in equilibrium, we confine our analysis to stellar models with axial ratios $r_{\mathrm{pole}}/r_{\mathrm{eq}} \leq 0.94$, where the numerical extraction of the hexadecapole remains accurate. Above all, it should be highlighted that the relevant dataset consists of $1022628$ stellar models, with $75\%$ allocated for training and the remaining $25\%$ reserved for testing. Table \ref{tab:parameters_range_2} provides the parameter ranges utilized for training and testing for the models associated with the $\bar{M}_4$ and $\bar{M}_4/\bar{S}_3$ parameters.

\begin{table}[!th]
  \caption{\label{tab:parameters_range_2} Parameter ranges of the rotating NS configurations used for training and testing in the $\bar{M}_4$ and $\bar{M}_4 / \bar{S}_3$ models. }
  \begin{ruledtabular}
      \begin{tabular}{c|rr|rr} 
        \multirow{2}{*}{\textbf{Feature}} & \multicolumn{2}{c|}{\textbf{Training Set}} & \multicolumn{2}{c}{\textbf{Test Set}}\\ 
        \cline{2-5}
        & \textbf{min} & \textbf{max} & \textbf{min} & \textbf{max} \\
        \hline 
        $C\times 10^{-1}$ & 0.876 & 3.075 & 0.882 & 3.075 \\
        $\chi \times 10^{-1}$ & 2.471 & 7.818 & 2.472 &  7.803\\ 
        $\bar{Q}$ & 1.313 &14.218  & 1.314 & 14.115 \\
        $\bar{S_3}$ & 1.736 & 31.034 & 1.738 & 30.799 \\
        
        $\bar{M_4}$ &2.655  & 497.192 & 2.658 &  491.585 \\
        $\bar{M_4}/\bar{S}_3$ & 1.525 & 16.021 & 1.524 & 15.961 \\
      \end{tabular}
  \end{ruledtabular}
\end{table}

\begin{figure*}[!htb]
    \includegraphics[width=0.46\textwidth]{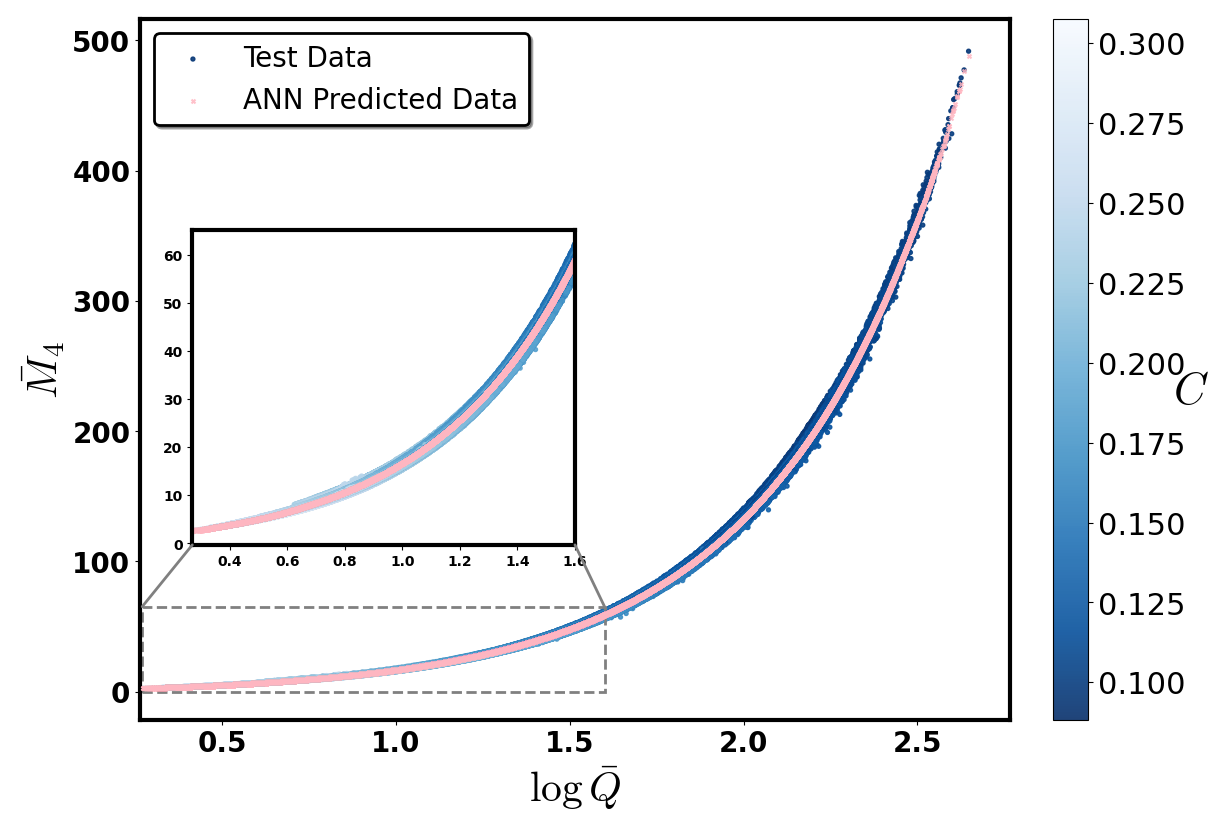}\hfill
    \includegraphics[width=0.46\textwidth]{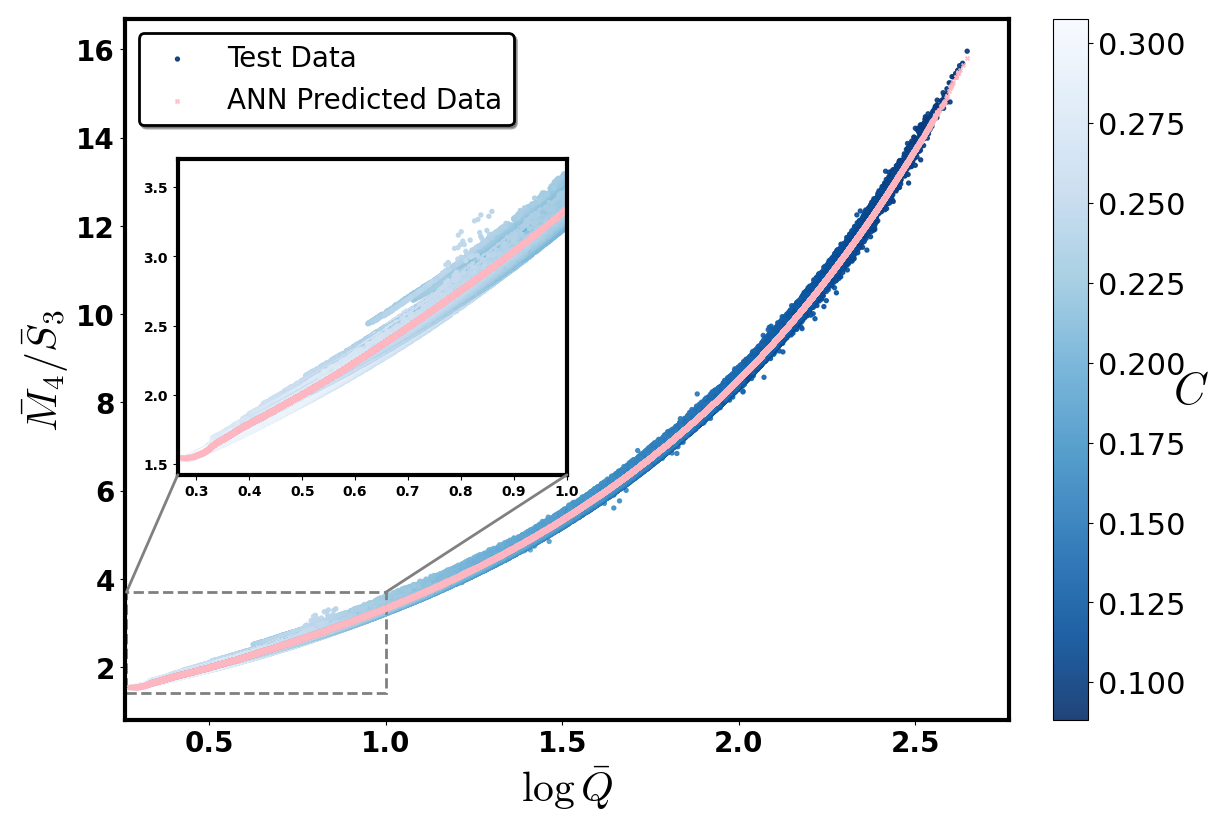}\hfill
    \includegraphics[width=0.46\textwidth]{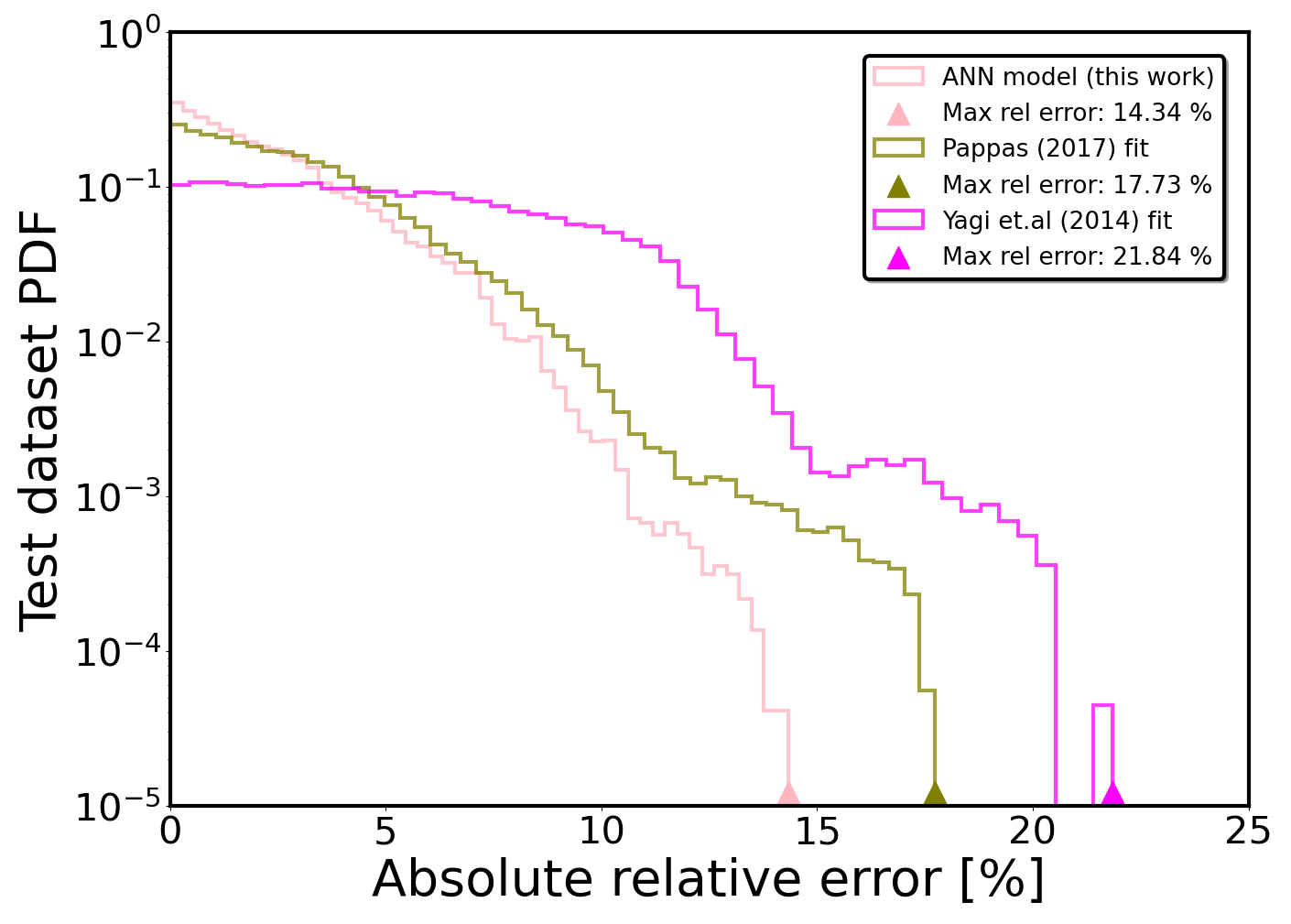}\hfill
    \includegraphics[width=0.46\textwidth]{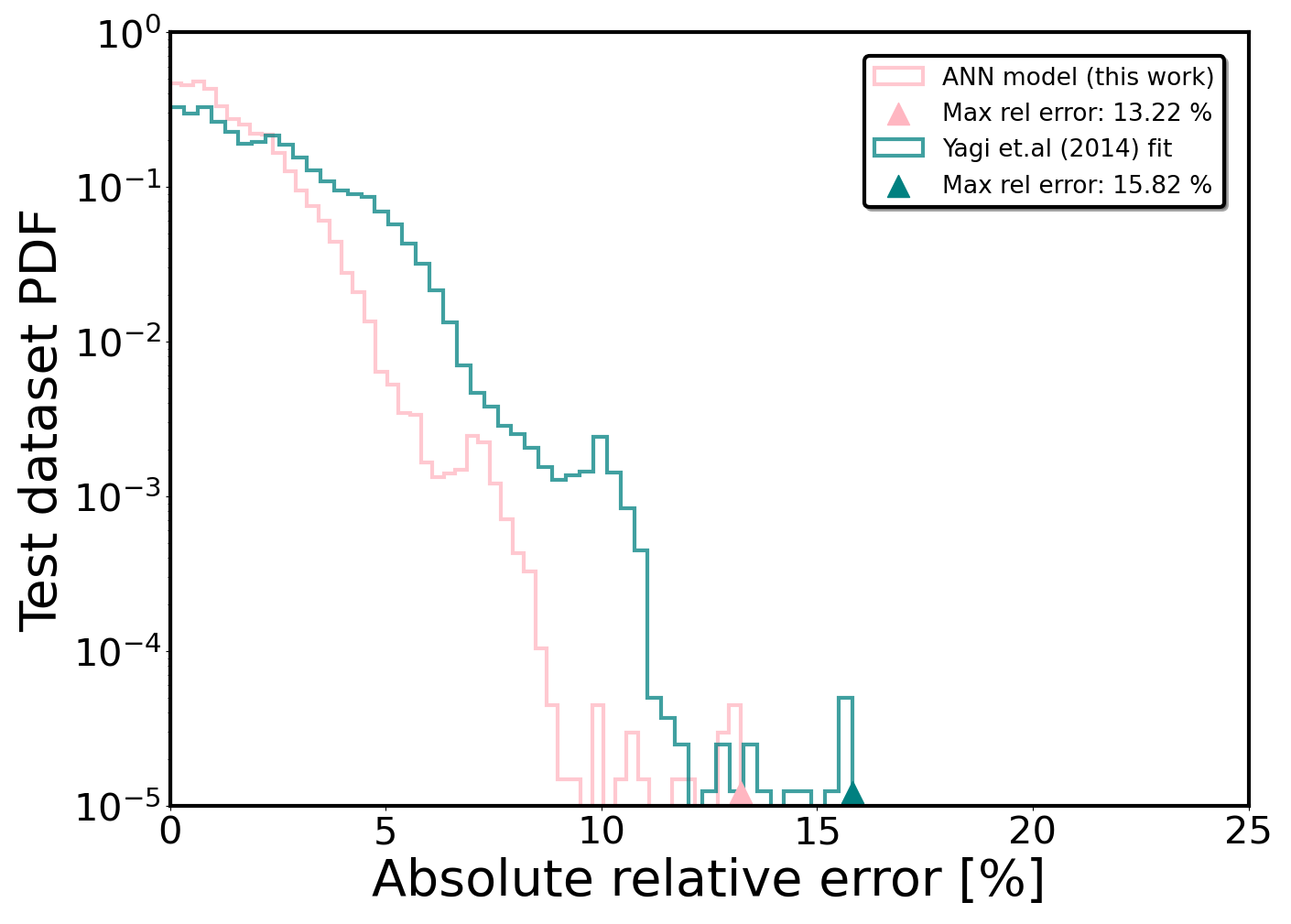}\hfill
    \caption{\label{fig:M4_M4_S3_fits_relative_errors} Top-left and top-right panels: Predicted data associated with the regression models (\ref{eq:M4_fit}) and (\ref{eq:M4_S3_fit}).  The colored variation corresponds to the star's compactness, as provided in the accompanying vertical colored bar. Bottom-left and bottom-right panels: Relative error distributions derived when evaluating the suggested DL models on the test data. Furthermore, in each case, additional relative error distributions for fitting functions proposed in the literature are illustrated with distinct colors, providing a basis for comparative analysis.}
\end{figure*}

Having determined some of the higher-order multipole moments beyond $M$ and $J$, we now focus on investigating their mutual dependencies. More specifically, we aim to refine the previously established relations involving $\bar{M}_4$, $\bar{M}_4/\bar{S}_3$, and $\bar{Q}$. Using the suggested ANN, the investigated model parametrizations that satisfactorily describe the data take the following forms,

\begin{align}
    \label{eq:M4_fit}
    \left(\bar{M}_4\right)_{\mathrm{model}} = \hat{\mathcal{M}}_{\theta^\star}(\log \bar{Q}),\\
    \label{eq:M4_S3_fit}
    \left(\frac{\bar{M}_4}{\bar{S}_3}\right)_{\mathrm{model}} = \hat{\mathbb{M}}_{\theta^\star}(\log \bar{Q}).
\end{align}
Table \ref{tab:M4bar_fits_eval_meas} summarizes the performance evaluation measures of each regression model on the corresponding test dataset.
\begin{table}[!h]
    \footnotesize
    \caption{\label{tab:M4bar_fits_eval_meas} Evaluation measures for the investigated DL models defined by Eqs. (\ref{eq:M4_fit}) and (\ref{eq:M4_S3_fit}) regarding the associated test sets.}
    \begin{ruledtabular}
        \begin{tabular}{c|ccccccc}
            \multirow{2}{*}{ANN} & MAE & MRE & MSE & $\mathrm{d}_{\text{max}}$ & MAPE & ExpVar & $R^2$ \\
            &  $\times 10^{-1}$&  $\times 10^{1}$&  $\times 10^{-2}$ &  ($\%$) &  $\times 10^{-2}$ ($\%$) &  $\times 10^{-2}$&$\times 10^{-2}$   \\    
            \hline
            Eq. (\ref{eq:M4_fit})  & 7.444 &  1.979 &  196.219 &  14.34    & 2.353  &  99.934 &  99.934 \\
            Eq. (\ref{eq:M4_S3_fit})  & 0.534 &  0.061 &  0.485 &  13.22    &   1.385&  99.916 &  99.915 \\

        \end{tabular}
    \end{ruledtabular}
\end{table}
The models' data predictions (\ref{eq:M4_fit}), (\ref{eq:M4_S3_fit}), and each model's fractional differences histogram are illustrated in Fig. \ref{fig:M4_M4_S3_fits_relative_errors}. The top and bottom left panels are associated with the $\bar{M}_4-\log \bar{Q}$ correlation, while the top and bottom right panels correspond to the $\bar{M}_4/\bar{S}_3-\log \bar{Q}$ representation. In each data representation, we use different colors to denote various values of stellar compactness within the parameter space, as indicated by the vertical colored bar.

It is evident that Eqs. (\ref{eq:M4_fit}) (\ref{eq:M4_S3_fit}) describe the test data, achieving relative errors of $\leq 14.34\%$ and $\leq 13.22\%$, respectively. These results suggest that the universality diminishes when considering multipole moment relations for higher-$\ell$ modes, as initially predicted in Ref. \cite{stein2014three}. However, as illustrated in the bottom-left and bottom-right histograms, the suggested models outperform the existing fitting functions previously proposed in the literature \cite{pappas2017accurate,yagi2014effective,yagi2014multipole}. 
Regarding the proposed approach, it should be highlighted that both ANNs yield smooth and continuous approximations of the underlying data, reflecting their effective training. Their ability to reproduce the central trends in the test data, with typical relative deviations around $\mathcal{O}(10\%)$, demonstrates good generalization as well as predictive accuracy, particularly given the quasi-universal behavior of the examined parametrizations.
In any case, considering the quasi-universal behavior of the $\bar{M}_4-\bar{Q}$ and $\bar{M}_4/\bar{S}_3-\bar{Q}$ relations, both the DL suggested models and those previously introduced in the literature approximate well the test data across all EoSs and NS models.

For the regression model given by Eq. (\ref{eq:M4_fit}), $11.01\%$ of the total NS configurations exhibit relative deviations $>5\%$. The majority of these cases correspond to NS models with $\bar{Q} \lesssim 6$, irrespective of their spin or compactness. In addition, most of the model's relative deviations that exceed $10\%$ are associated with NS configurations with quadrupole deformation within the ranges $\log\bar{Q} \in [0.62, 0.88]$ and $[1.46,1.66]$, respectively. These correspond to the EoSs {\it DS(CMF)-1 Hybr, DS(CMF)-2 Hybr, QHC21T-$D_T$}, and {\it TM1e}, where the first three belong to the hybrid EoS class and the latter to the hadronic class. Mainly, the relative errors in $\bar{M}_4-\log \bar{Q}$ representation decrease as the quadrupole deformation increases. A closer investigation of the overall error distribution across different EoS categories reveals that the hybrid class exhibits the highest relative deviation. In contrast, the hadronic and hyperonic classes display lower maximum relative errors, approximately $13.18\%$ and $7.46\%$, respectively (see e.g., Appendix \ref{app:model_characteristics} for details). The hyperonic category effectively preserves the universal behavior between the associated parameters. In any case, it is important to highlight that the $\bar{M}_4-\bar{Q}$ relation is less EoS-insensitive than the $\bar{S}_3-\bar{Q}$ correlation.

Moreover, the performance of the regression model (\ref{eq:M4_S3_fit}) is notably better, with only $0.69\%$ of predictions exhibiting relative deviations that exceed $5\%$. The majority of these cases are associated with the most compact NS configurations, characterized by $\bar{Q} \lesssim 4.32$ and compactness values in the range $C \in [0.185, 0.262]$, regardless of the star’s spin. Model predictions with relative deviations above $10\%$ correspond to a small subset of stellar configurations with $\log\bar{Q} \in [0.79, 0.86]$, associated with the hybrid EoS {\it DS(CMF)-2} Hybr. Above all, the model's predictions are consistent with the behavior observed in the model (\ref{eq:M4_fit}), where relative errors tend to increase as the quadrupole deformation $\bar{Q}$ becomes smaller. As expected, the hybrid class yields the highest relative deviations, with the hadronic and hyperonic categories demonstrating significantly lower relative errors, at around $9.29\%$ and $4.89\%$, respectively (see e.g., Appendix \ref{app:model_characteristics} for the associated violin plots). This observation suggests that hyperonic EoSs exhibit a well-defined universal behavior across the explored parameter space. Above all, we have to note that the EoS insensitive representation involving the $\bar{M_4}/\bar{S_3}-\bar{Q}$ parametrization is stronger than that of $\bar{M}_4-\bar{Q}$.

To investigate additional universal relations that refer to the mass $\bar{M}_4$ hexadecapole deformation, we also investigate parametrizations that include the star's compactness, the rotation rate, and the reduced quadrupole and octupole deformations. The relevant hypersurfaces that best describe the underlying data have functional forms,
\begin{align}
    \label{eq:M4_S3_hypersurface_fit}
    \left(\frac{\bar{M}_4}{\bar{S}_3}\right)_{\mathrm{model}} & = \hat{\mathbb{M}}^{\prime}_{\theta^\star}(\chi,C,\bar{Q}),\\
    \label{eq:M4_hypersurface_fit}
     \left(\bar{M}_4 \right)_{\mathrm{model}} & = \hat{\mathcal{M}}^{\prime}_{\theta^\star}(\chi,C,\bar{Q},\bar{S_3}).
\end{align}
Table \ref{tab:M4bar_hypersurf_eval_meas} summarizes the performance metrics of the regression models defined by Eqs. (\ref{eq:M4_S3_hypersurface_fit}) and (\ref{eq:M4_hypersurface_fit}), evaluated on their respective test datasets.
\begin{table}[!h]
    \footnotesize
    \caption{\label{tab:M4bar_hypersurf_eval_meas} Evaluation measures for the regression models defined by Eqs.  (\ref{eq:M4_S3_hypersurface_fit}) and (\ref{eq:M4_hypersurface_fit}) regarding the associated test sets.}
    \begin{ruledtabular}
        \begin{tabular}{c|ccccccc}
            \multirow{2}{*}{ANN} & MAE & MRE & MSE & $\mathrm{d}_{\text{max}}$ & MAPE & ExpVar & $R^2$ \\
            &  $\times 10^{-2}$&  $\times 10^{-1}$&  $\times 10^{-3}$ &  ($\%$) &  $\times 10^{-3}$ ($\%$) &  $\times 10^{-2}$&$\times 10^{-2}$   \\    
            \hline
            Eq. (\ref{eq:M4_S3_hypersurface_fit}) & 2.962 &  7.248  &  1.953 &  9.19 & 7.326  &  99.966 &  99.966\\
            Eq. (\ref{eq:M4_hypersurface_fit})  & 6.788 & 100.276  &  50.411 & 8.11  &    2.149 &  99.998 & 99.998  \\
        \end{tabular}
    \end{ruledtabular}
\end{table}

\begin{figure}[!htb]
    \includegraphics[width=0.46\textwidth]{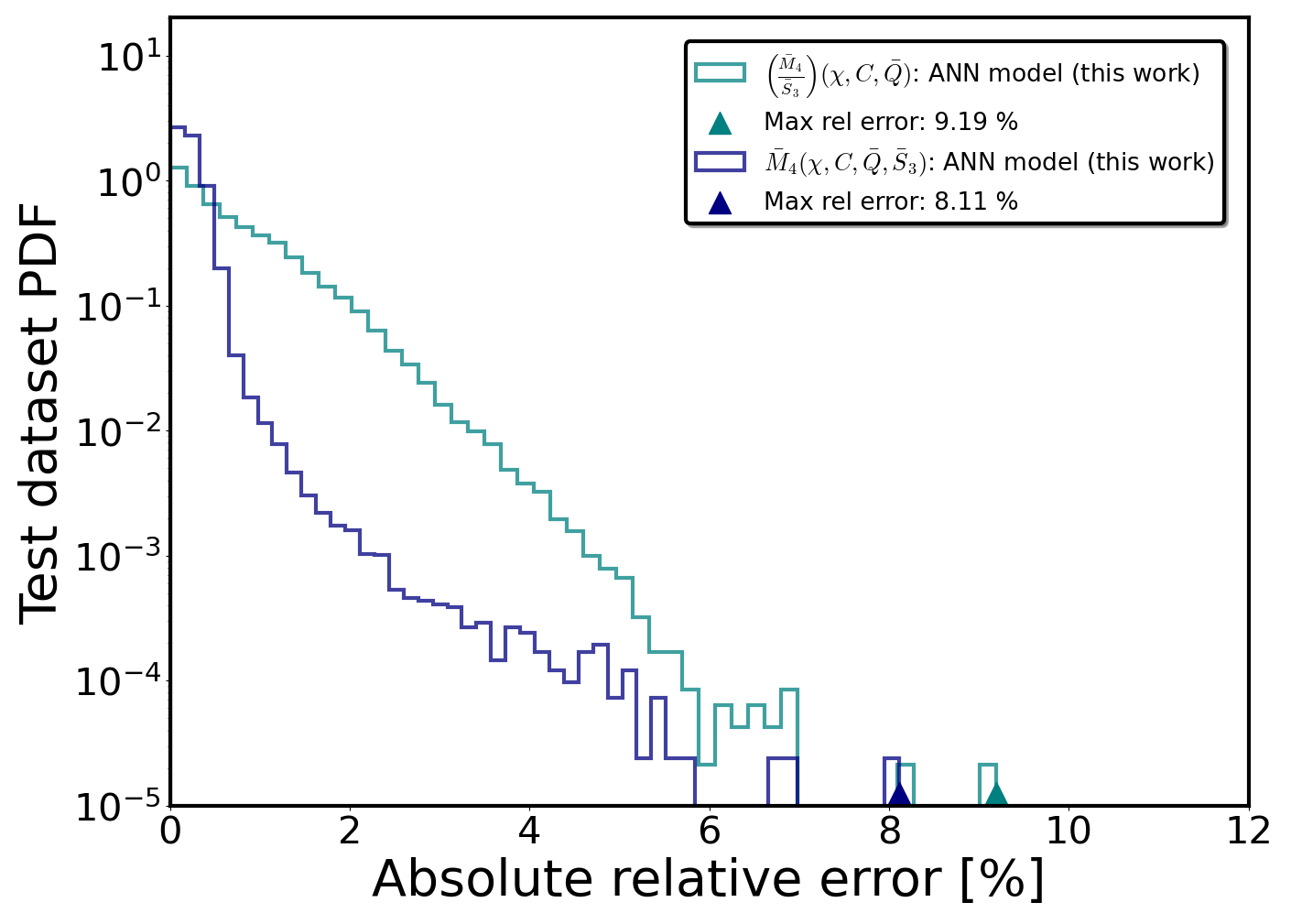}\hfill
    \caption{\label{fig:M4_hypersurfaces_relative_errors} Relative error distributions on the test dataset for the proposed ANN models described by Eqs. (\ref{eq:M4_S3_hypersurface_fit}) and (\ref{eq:M4_hypersurface_fit}), highlighting their performance.}
\end{figure}

Both ANN models (\ref{eq:M4_S3_hypersurface_fit}) and (\ref{eq:M4_hypersurface_fit}) effectively capture the underlying data patterns, achieving test set accuracies with relative errors $\leq 9.19\%$ and $\leq 8.11\%$, respectively. Depending on the inference case, incorporating global stellar parameters such as the dimensionless spin $\chi$ and compactness $C$ leads to enhanced predictive performance compared to the fitting quasi-universal representations $\bar{M}_4-\bar{Q}$ and $\bar{M}_4/\bar{S}_3-\bar{Q}$ which are based solely on quadrupole deformation. Moreover, the new suggested EoS-insensitive parametrizations exhibit robust generalization ability across the entire test set of rapidly rotating stellar configurations. Fig. \ref{fig:M4_hypersurfaces_relative_errors} presents the relative error distributions for each model in the test dataset.
It is noteworthy that only $1.06\%$ of stellar models exhibit relative deviations larger than $3\%$ for the model (\ref{eq:M4_S3_hypersurface_fit}), while the model (\ref{eq:M4_hypersurface_fit}) shows even better performance, with just $0.6\%$ of stellar models having deviations exceeding $1\%$. For the $\bar{M}_4/\bar{S}_3$ model, the majority of observed deviations are primarily found in NS configurations with $\bar{Q}\lesssim 7$, corresponding to the most compact stellar models with $C\in[0.136,0.256]$, independent of the spin. In contrast, the model for the $\bar{M}_4$ exhibits no discernible pattern in error variance throughout the ($\chi, C, \bar{Q}, \bar{S}_3$) parameter space. In both model representations, a detailed analysis of error distributions across different EoS categories reveals that the hadronic class exhibits the highest maximum relative deviations. In contrast, the hybrid and hyperonic categories show comparatively lower maximum errors—approximately $5.87\%$ and $5.25\%$ for the $\bar{M}_4/\bar{S}_3$ regression model (\ref{eq:M4_S3_hypersurface_fit}), and around $5.04\%$ and $4.40 \%$ for the $\bar{M}_4$ model (\ref{eq:M4_hypersurface_fit}), respectively (see e.g., Appendix \ref{app:model_characteristics} for the corresponding violin plots). Within each EoS class, the EoS models exhibiting the largest deviations are identical and ordered similarly—{\it TM1e, DS(CMF)-2 Hybr}, and {\it DS(CMF)-3}-for both the $\bar{M}_4/\bar{S}_3$ and $\bar{M}_4$ models.

\begin{figure*}[!thb]
    \centering
    \includegraphics[width=1\textwidth]{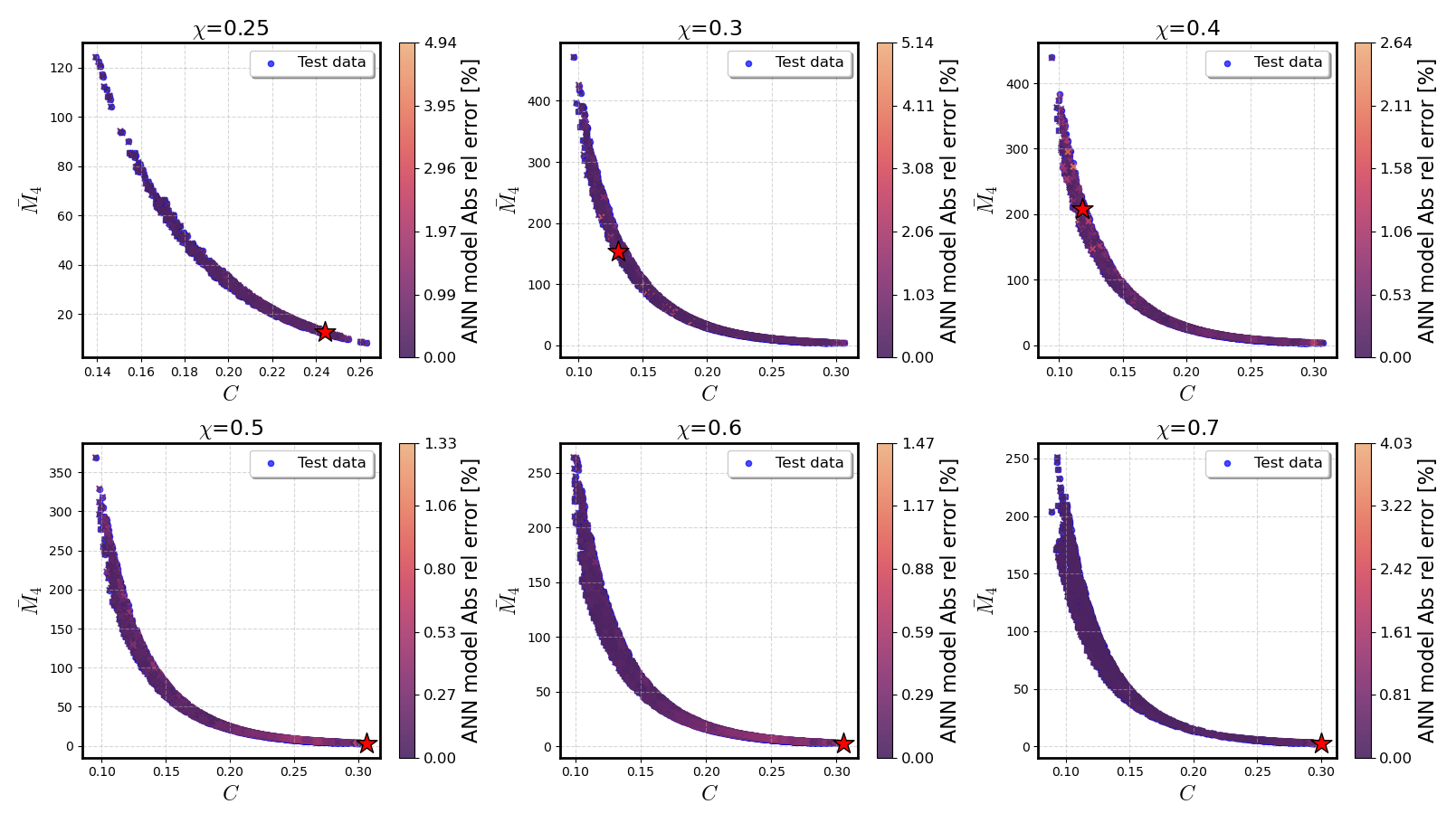}\hfill
    \caption{\label{fig:M4_C_fit_curves}
    $\bar{M}_4-C$ plane: Predicted $\bar{M}_4(C,\bar{Q},\bar{S}_3)$ data generated by the regression model (\ref{eq:M4_hypersurface_fit}), shown for a discrete set of spin values $\chi \in [0.25, 0.70]$. Data points are color-coded by the ANN model’s absolute relative error [\%], while the red star marks the configuration with the maximum relative error.
    In all cases, the predictions closely match the illustrated test data, reinforcing the reliability of the trained ANN model.} 
\end{figure*}

To further assess the reliability of the trained ANN model in Eq. (\ref{eq:M4_hypersurface_fit}), Fig. \ref{fig:M4_C_fit_curves} presents the $\bar{M}_4-C$ correlation evaluated across several fixed values of $\chi$.
The predicted $\bar{M}_4(C,\bar{Q},\bar{S}_3)$ data from the proposed regression model satisfactorily reproduce the quasi-universal patterns observed across sequences of stellar models with fixed dimensionless angular momentum $\chi$, demonstrating good agreement with the test underlying data. In any case, it is important to note that as the spin increases, the universal nature of the discussed correlation tends to weaken, particularly for configurations with lower compactness. In addition, the $\bar{M}_4/\bar{S}_3-C$ relation is similarly well captured by the regression function (\ref{eq:M4_S3_hypersurface_fit}), with its predictive representations satisfactorily reproducing the data and exhibiting the same spin-dependent weakening at lower compactness.

\begin{figure}[!htb]
    \includegraphics[width=0.46\textwidth]{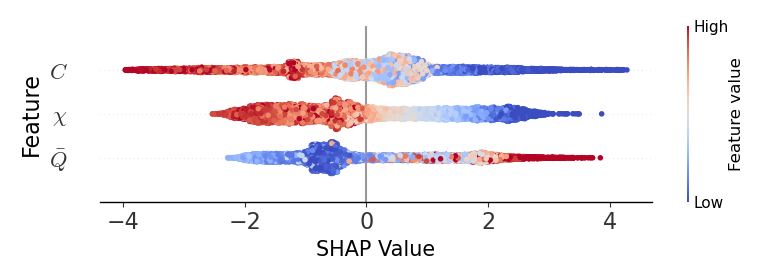}\hfill
    \includegraphics[width=0.46\textwidth]{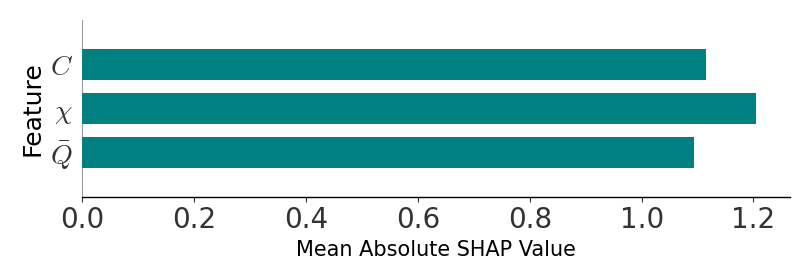}\hfill
    \caption{\label{fig:M4_S3_shap_figs}  Same as Fig. \ref{fig:Ibar_shap_figs} for the trained DL model (\ref{eq:M4_S3_hypersurface_fit}).}
\end{figure}

\begin{figure}[!htb]
    \includegraphics[width=0.46\textwidth]{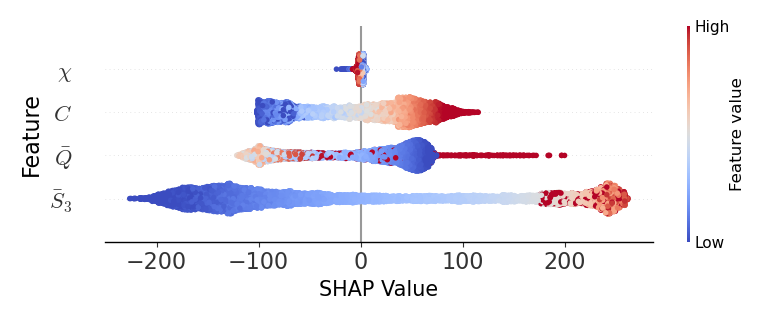}\hfill
    \includegraphics[width=0.46\textwidth]{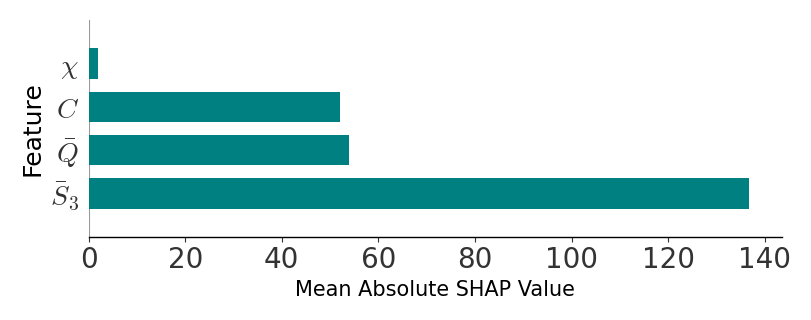}\hfill
    \caption{\label{fig:M4_hypresurf_shap_figs} Same as Fig. \ref{fig:Ibar_shap_figs} for the trained DL model (\ref{eq:M4_hypersurface_fit}).}
\end{figure}

Next, we apply the SHAP method to interpret the predictions of the suggested ``deep'' EoS-insensitive parametrizations related to the $\bar{M}_4/\bar{S}_3$ and $\bar{M}_4$ fits.  For the dataset consisting of rapidly rotating NS configurations with axial ratio $r_{\mathrm{pole}}/r_{\mathrm{eq}} \leq 0.94$, we use a subset $X^{'}_1$ which includes $10,000$ NS configurations with parameters $C\in[0.090,0.307],\chi\in[0.249,0.768], \bar{Q}\in[1.317,13.448],\bar{S}_3\in[1.742,28.910]$, to provide interpretability of each model's predictions. For both models (\ref{eq:M4_S3_hypersurface_fit}) and (\ref{eq:M4_hypersurface_fit}), we assess how the chosen parameters can impact the associated expected outputs. The corresponding SHAP values, shown in Figures \ref{fig:M4_S3_shap_figs} and \ref{fig:M4_hypresurf_shap_figs}, highlight the contribution of the star's parameters to the models' inference.

For the regression model defined in Eq. (\ref{eq:M4_S3_hypersurface_fit}), the SHAP analysis (Fig. \ref{fig:M4_S3_shap_figs}, top panel) reveals that each global parameter of the NS's ensemble can exert either a positive or negative influence on the estimator’s mean value. We see that the most (less) compact stars have a negative (positive) contribution. Notably, stars exhibiting larger quadrupole deformation tend to contribute positively relative to the model’s expected output. Conversely, the most rapidly rotating NSs are associated with negative SHAP attributions, whereas lower-spin configurations show positive ones. The bottom panel of Fig. \ref{fig:M4_S3_shap_figs} presents the mean absolute SHAP values for each input feature, quantifying their average impact across the $X^{\prime}_1$ test subset. The features $\chi$, $C$, and $\bar{Q}$ exhibit mean contributions of $1.20$, $1.11$, and $1.09$, respectively. In this case, the average SHAP contributions of the physical features are found to be comparable. As shown in previous parametrizations, this is not always the case, but it can naturally occur when correlations within the universal parametrization, together with the way the trained model’s weights distribute importance across the inputs, lead to a more balanced SHAP contribution among features. Nevertheless, the model’s input parameters enable robust and accurate inference of $\bar{M}_4/\bar{S}_3$ across a wide range of stellar configurations.

As shown in the top panel of Fig. \ref{fig:M4_hypresurf_shap_figs}, the SHAP analysis for the regression model in Eq. (\ref{eq:M4_hypersurface_fit}) demonstrates that each global parameter of the employed NS ensemble can variably influence the predicted expected value $\mathbb{E}_X(\hat{\mathcal{M}}^{\prime}_{\theta^\star})$, with contributions that may be either positive or negative depending on the specific parameter configuration. For both $\chi$ and $\bar{Q}$, low and high extreme values exhibit mixed SHAP attributions—positive and negative—relative to the model's expected value prediction. This variability suggests an absence of a clear or consistent SHAP value pattern associated with these parameters. In contrast, the influence of the $C$ and $\bar{S}_3$ parameters is more pronounced. Higher values of both $C$ and $\bar{S}_3$ tend to contribute positively, whereas lower values are predominantly associated with negative contributions. The bottom panel of Fig. \ref{fig:M4_hypresurf_shap_figs} displays the mean absolute SHAP values for each input feature, providing a quantitative measure of their average influence on the model’s predictions. The input features $\chi$, $C$, $\bar{Q}$, and $\bar{S}_3$ exhibit mean contributions of $1.99$, $51.97$, $54.03$, and $136.75$, respectively. Above all, while the mean contributions of $C$ and $\bar{Q}$ are comparable, the $\bar{S}_3$ parameter stands out with a dominant influence on the model's expected value $\mathbb{E}_X(\hat{\mathcal{M}}^{\prime}_{\theta^\star})$.

Motivated by the seemingly minor mean contribution of the feature $\chi$ observed in the SHAP analysis, we further investigate the underlying universal behavior of $\bar{M}_4$ with respect to the parameters $C$, $\bar{Q}$, and $\bar{S}_3$. In this case, the regression model that best captures the data trend is expressed as
\begin{align}
    \label{eq:M4_hypersurface_fit_v2}
     \left(\bar{M}_4 \right)_{\mathrm{model}} & = \hat{\mathcal{M}}^{\prime \prime}_{\theta^\star}(C,\bar{Q},\bar{S_3}),
\end{align}
with its predictive performance on the test set summarized in Table \ref{tab:M4bar_surface_fit_eval_meas_v2}.
\begin{table}[!h]
    \footnotesize
    \caption{\label{tab:M4bar_surface_fit_eval_meas_v2} Evaluation measures for the parametrization given by the Eq. (\ref{eq:M4_hypersurface_fit_v2}) regarding the test dataset.}
    \begin{ruledtabular}
        \begin{tabular}{ccccccc}
            MAE & MRE & MSE & $\mathrm{d}_{\text{max}}$ & MAPE & ExpVar & $R^2$ \\
            $\times 10^{-1}$& $\times 10^{-1}$ &  $\times 10^{-1}$ &  ($\%$) &  $\times 10^{-3}$ ($\%$) &  $\times 10^{-2}$&$\times 10^{-2}$   \\    
            \hline
             1.300&  99.872 &  1.070 &   7.52   &  3.134 & 99.996  &  99.996 \\

        \end{tabular}
    \end{ruledtabular}
\end{table}
In this formulation, the regression hypersurface in Eq. (\ref{eq:M4_hypersurface_fit_v2}) accurately reproduces the test data, achieving a maximum relative error of $\leq 7.52\%$. This represents an improvement with respect to the maximum fractional deviation evaluation measure compared to the model in Eq. (\ref{eq:M4_hypersurface_fit}), while maintaining a simpler parametrization. Therefore, it demonstrates consistent performance across diverse EoS models, establishing it also as a well-defined EoS-insensitive relation. In contrast to the previous model (with only $0.6\%$ of stellar configurations exceeding $1\%$ deviation), the current simplest one yields $2.38\%$ of NSs with deviations above this threshold. Moreover, as with the previous $\bar{M}_4$ DL model, the distribution of errors exhibits no apparent systematic trends, indicating a random variance across the $(C, \bar{Q}, \bar{S}_3)$ parameter space. An assessment of the model's error distribution across the employed EoS categories reveals that the hadronic class exhibits the largest maximum relative deviation, primarily attributed to the {\it TM1e} EoS model. In contrast, the hybrid and hyperonic classes exhibit notably lower relative errors, restricted to approximately $5.56\%$ and $4.32\%$, respectively. These deviations are chiefly driven by the {\it QHC19-D} and {\it DS(CMF)-3} EoS models.

\begin{figure}[!htb]
    \includegraphics[width=0.46\textwidth]{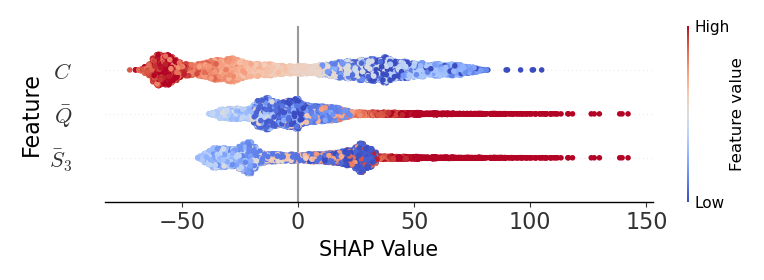}\hfill
    \includegraphics[width=0.46\textwidth]{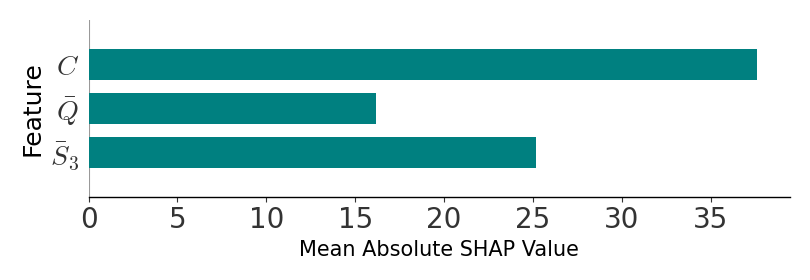}\hfill
    \caption{\label{fig:M4_hypresurf_shap_figs_v2} Same as Fig. \ref{fig:Ibar_shap_figs} for the trained DL model (\ref{eq:M4_hypersurface_fit_v2}).}
\end{figure}

It is also insightful to perform a SHAP analysis for the newly proposed $\bar{M}_4$ regression model. As illustrated in the top panel of Fig. \ref{fig:M4_hypresurf_shap_figs_v2}, the SHAP values corresponding to the ANN model (\ref{eq:M4_hypersurface_fit_v2}) reveal that the utilized input parameters of the NS ensemble influence the model's predictions in varying degrees. These contributions can be either positive or negative, contingent on the NS parameters of interest. For example, we can see that stars with higher compactness tend to exert a negative influence on the model's expected value, while less compact configurations contribute positively. Notably, stellar configurations exhibiting the largest quadrupole and octupole deformations generally have a positive impact. In contrast, configurations with lower deformation parameters display mixed contributions, which can be either positive or negative depending on the specific parameter combination.
In addition, an examination of the mean absolute SHAP values for the current $\bar{M}_4$ trained model shows that the compactness $C$, the reduced spin octupole $\bar{S}_3$, and the normalized mass quadrupole $\bar{Q}$ contribute, on average as $37.59$, $25.19$, and $16.19$, respectively (see e.g., Fig. \ref{fig:M4_hypresurf_shap_figs_v2} bottom panel). Unlike the previous $\bar{M}_4(\chi, C,\bar{Q},\bar{S}_3)$ model, compactness $C$ now plays the most significant role. On average, the model's prediction is strongly influenced by how compact the star is. The second most important feature, $\bar{S}_3$, is also closely related to the mass quadrupole moment $\bar{Q}$, reflecting known universal connections between different multipole moments in rotating NSs. In any case, the involvement of all features reflects their collective importance in capturing the underlying universal behavior across different EoSs.

\subsection{\label{subsubsec:ANN_for_Req}Relation for the star's equatorial radius}
Recent developments in relativistic astrophysics have provided analytic spacetime solutions that describe the exterior geometry of NSs in terms of their multipole moments \cite{pappas2015unified, pappas2017accurate, maselli2020new}. In parallel, several universal relations have been formulated to characterize the stellar surface and some of its properties, with the equatorial radius emerging as a key parameter that encapsulates the star’s rotational deformation \cite{silva2021surface, papi_vard_surface, morsink2007oblate, algendy2014universality}. Consequently, for a theoretical standpoint, exploring a universal relation that links the star's equatorial radius with the spacetime's mass moments ($M, M_2$) and the associated current moments ($J, S_3$) would be highly beneficial. This parametrization could serve as a higher-order refinement to the $R_{\mathrm{eq}}/M-\chi-\bar{Q}$ EoS-insensitive relation proposed in \cite{papigkiotis2023universal}.

In this investigation, we aim to establish a more refined EoS-insensitive relation that effectively connects the stellar radius to the spacetime's underlying multipole moments. This way, we delve into a parametrization that links the star's reduced equatorial radius $\bar{\mathcal{R}}$ with the parameters $\tilde{M}, \chi, \bar{Q}$, and $\bar{S_3}$.

\begin{figure}[!htb]
    \includegraphics[width=0.46\textwidth]{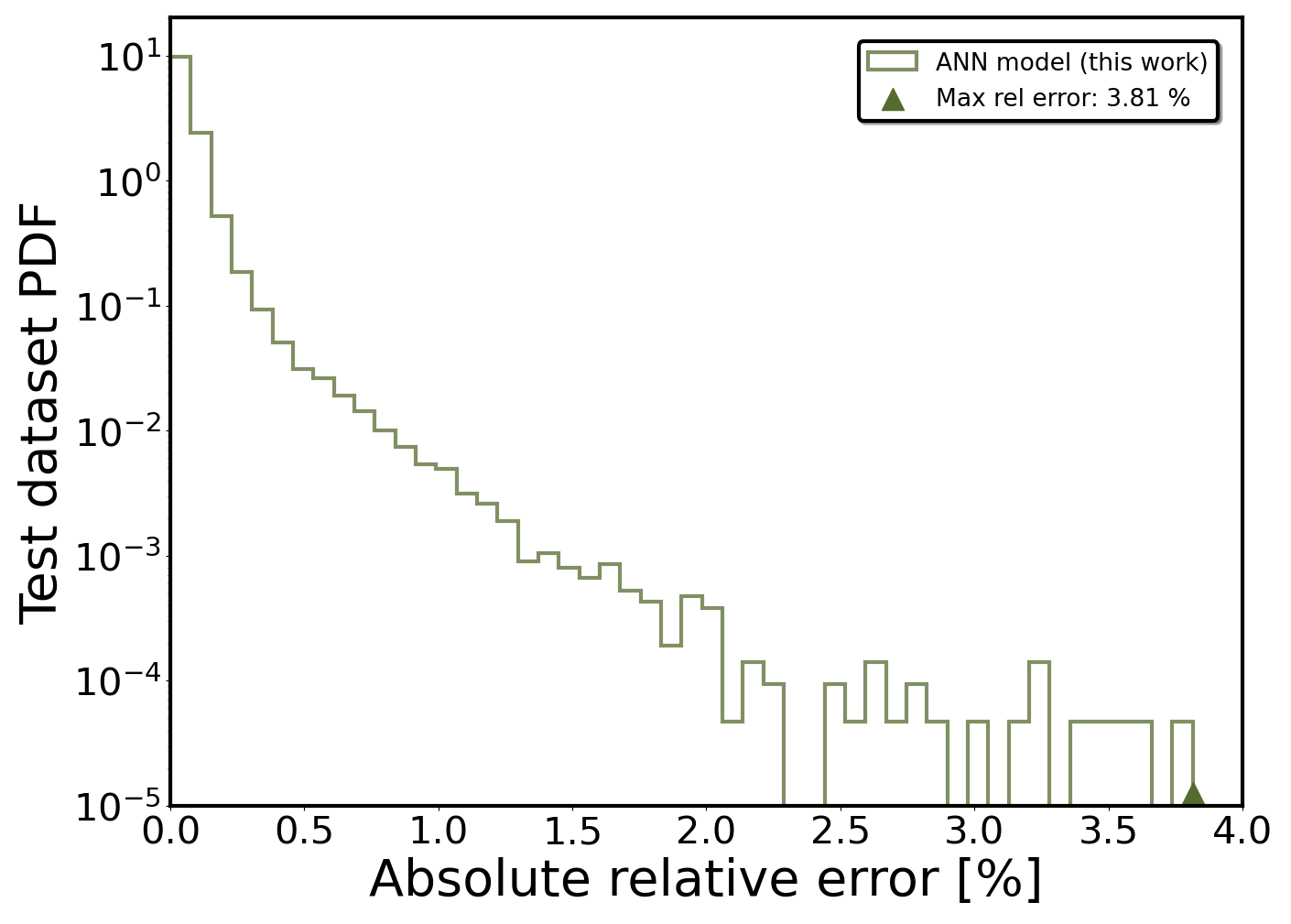}\hfill
    \caption{\label{fig:R_fit_relative_errors} Relative error distribution ($100\% \times (\bar{\mathcal{R}}_{\mathrm{model}} - \bar{\mathcal{R}})/ \bar{\mathcal{R}})$ derived using the suggested DL model (\ref{eq:R_fit}) on the test dataset.}
\end{figure}

Employing the proposed DL architecture, the trained ANN hypersurface that best verifies the data has the respective form,
\begin{equation}
    \label{eq:R_fit}
    \bar{\mathcal{R}}_{\mathrm{model}} = \hat{\bar{\mathcal{R}}}_{\theta^\star}(\bar{M},\chi,\bar{Q},\bar{S}_3)
\end{equation}
The associated statistical performance measures for the proposed optimal model, evaluated on the test set, are presented in Table (\ref{tab:R_fit_eval_meas}).
\begin{table}[!h]
    \footnotesize
    \caption{\label{tab:R_fit_eval_meas} Evaluation measures for the parametrization given by the Eq. (\ref{eq:R_fit}) regarding the test dataset.}
    \begin{ruledtabular}
        \begin{tabular}{ccccccc}
            MAE & MRE & MSE & $\mathrm{d}_{\text{max}}$ & MAPE & ExpVar & $R^2$ \\
            $\times 10^{-3}$&  $\times 10^{-1}$&  $\times 10^{-4}$ &  ($\%$) &  $\times 10^{-4}$ ($\%$) &  $\times 10^{-2}$&$\times 10^{-2}$   \\    
            \hline
            9.259  & 6.020  & 2.831  &   3.81   & 6.549  &  99.988 & 99.988  \\

        \end{tabular}
    \end{ruledtabular}
\end{table}
In addition, Fig. \ref{fig:R_fit_relative_errors} presents a histogram of the relative error distribution, offering further insight into the model’s predictive performance.
The suggested regression model (\ref{eq:R_fit}) accurately reproduces the data on the test dataset, demonstrating high precision with relative errors $\leq 3.81\%$. Therefore, its consistent performance across a broad range of EoSs highlights its robustness and reinforces its EoS-independence. In addition, it is worth emphasizing that only $0.15\%$ of the test data exhibit relative deviations $>1\%$, with the associated error variance showing no systematic pattern across the $(\bar{M}, \chi, \bar{Q}, \bar{S}_3)$ parameter space. A closer analysis of the error distribution across different EoS categories reveals that the hadronic class exhibits the largest maximum relative deviation. More specifically, the largest fractional deviation of the model is associated with the {\it TW99} EoS. In contrast, the hyperonic and hybrid categories show notably smaller maximum errors, approximately $3.45\%$ (EoS {\it OPGR(DDH$\delta$Y4)}) and $1.62\%$ (EoS {\it DS(CMF)-1 Hybr}), respectively (see e.g., Appendix \ref{app:model_characteristics} for the associated violin plots).

Above all, the suggested regression model serves as an invaluable tool for fitting functions that aim to infer the NS surface \cite{silva2021surface, morsink2007oblate, papi_vard_surface}, where $R_{\mathrm{eq}}$ serves as an input parameter, with direct applications to current observatories like NICER telescope \cite{gendreau2012neutron, arzoumanian2014neutron, gendreau2017searching} and future missions like the enhanced X-ray Timing and Polarimetry (eXTP) \cite{zhang2019enhanced}, and the Spectroscopic Time-Resolving Observatory for Broadband Energy X-rays (STROBE-X) \cite{ray2018strobe,ray2019strobe}. These forthcoming missions are expected to surpass the current capabilities of NICER in enhancing the precision of parameter estimation for NS radii.

\begin{figure}[!htb]
    \includegraphics[width=0.46\textwidth]{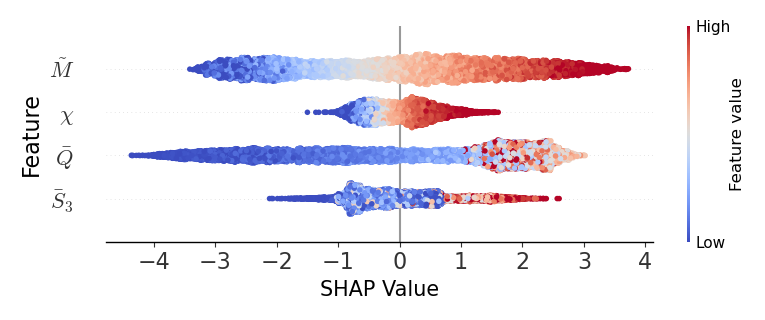}\hfill
    \includegraphics[width=0.46\textwidth]{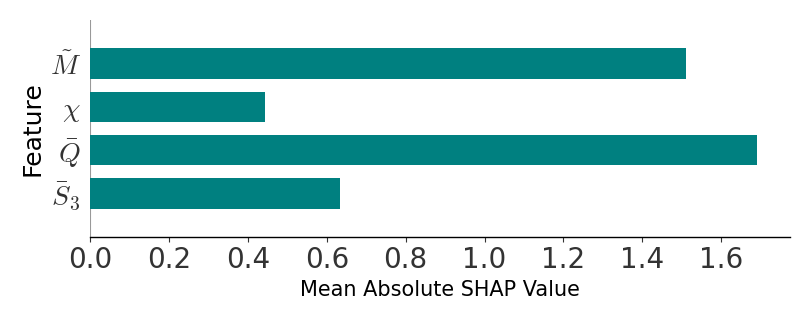}\hfill
    \caption{\label{fig:R_shap_figs} Same as Fig. \ref{fig:Ibar_shap_figs} for the trained DL model (\ref{eq:R_fit}).}
\end{figure}

Finally, for each NS configuration in the test subset $X_1$, we analyze the individual contribution of each parameter to the model's prediction. As illustrated in Fig. \ref{fig:R_shap_figs} (top panel), each feature can exert either a positive or negative influence on the model’s expected value $\mathbb{E}_X (\hat{\bar{\mathcal{R}}}_{\theta^\star})$. The highest (lowest) values of the parameters $\bar{M}, \chi, \bar{Q},\bar{S}_3$ have a positive (negative) SHAP contribution.
This behavior is consistent with the underlying NS physics. Higher values of $\chi$, $\bar{Q}$, and $\bar{S}_3$ reflect stronger rotational effects and multipole-moment deformations, which lead to an increase in the equatorial radius. Lastly, the bottom panel of Fig. \ref{fig:R_shap_figs} shows the mean absolute SHAP values for each parameter across the test subset. The average contributions of $\bar{M}$, $\chi$, $\bar{Q}$, and $ \bar{S_3}$ to the model's predictions are $1.51$, $0.44$, $1.69$, and $0.63$, respectively. All four parameters play integral roles in the inference process, each encoding key aspects of the NS's physical structure. While $\chi$ and $\bar{S}_3$ are closely tied to rotational dynamics and higher-order shape deformations, the dimensionless mass $\bar{M}$ and reduced quadrupole $\bar{Q}$ consistently emerge as dominant contributors, exhibiting the highest average impact on the model's predictions. This combination of mass and current multipole moments reflects the model’s ability to incorporate complementary higher-order multipole moments as parameters, enabling robust and accurate estimation of the equatorial radius. In any case, the employed set of parameters underscores the significance of rotationally driven multipole moments in shaping the stellar surface geometry.

\section{\label{sec:conclusions} Summary and Conclusions}

This paper systematically explores EoS-independent relations concerning some of the NS's global properties. The investigation encompasses an extensive ensemble of $1109271$ rapidly rotating NSs in $\beta$ equilibrium, utilizing $70$ realistic, tabulated cold EoSs covering a lot of theoretical microphysics composition cases.  The numerical solutions for these NSs were obtained using the RNS code \cite{rns}. The used ensemble of stellar configurations spans a broad range of compactness $C \in [0.876,3.075]\times 10^{-1}$, and rotation frequencies $f[\mathrm{kHz}] \in [0.173,1.871]$.

Throughout the investigation, a feed-forward neural network was employed for regression analysis. The fully connected deep learning architecture, detailed in Table (\ref{tab:hidden_layers_struct}), includes five hidden layers. At the input layer, min-max scaling was used as a data preprocessing step to normalize each feature to a range between $0$ and $1$. For each EoS, the data was split into a $75:25$ train/test ratio. Additionally, in most cases, a modified sigmoid function was applied at the end of the final layer. This approach markedly enhanced the network's learning efficiency and substantially decreased the network's errors in the particular case study. During training, a dynamic learning rate strategy was implemented for optimization, with the Huber loss being the objective function. Given the non-linear nature of the optimization process, the Adamax \cite{kingma2015adam} optimizer was used to determine the neural network's optimal $\theta^\star$ weights.

Furthermore, the utilization of the SHapley Additive exPlanations (SHAP) \cite{shapley1953value,maschler2020game, lundberg2017unified} method was of paramount importance in the interpretation and explainability of the predictions generated by the proposed feed-forward network. This framework elucidates the contributions of individual stars' parameters to the DL model's expected value, thereby enhancing the transparency and interpretability of its decisions. More specifically, it enabled a systematic attribution of the influence of each input feature, thereby providing valuable insights into the factors driving the model's predictions. In addition, this interpretative approach enhances our comprehension of the model's behavior, corroborates its dependability, and identifies potential avenues for enhancement.

We provide a concise summary of the main findings discussed in Sec. \ref{sec:univ_rel_with_ANNs}.
For the normalized moment of inertia $\bar{I}$, we derived an EoS-independent relation in terms of the dimensionless angular momentum $\chi$ and spacetime quadrupole deformation $\bar{Q}$. The suggested ANN model (\ref{eq:Ibar_fit}) demonstrates overall accuracy better than $2.76\%$ on the test set, effectively capturing all the rapidly rotating cases. SHAP analysis reveals that while $\chi$ aids pattern recognition, $\bar{Q}$ is more pivotal in the model's inference. For the reciprocal moment of inertia $\mathcal{D}$, we proposed two universal parametrizations: $\mathcal{D}(\chi,\log\bar{Q})$ and $\mathcal{D}(C,\chi,\log\bar{Q})$ with the associated DL models achieving relative errors of $\leq 2.70\%$ and $\leq 2.35\%$, respectively (Eqs. (\ref{eq:D_freq_fit}) and (\ref{eq:D_freq_fit_2})).  For the first parameterization, $\log\bar{Q}$ dominates the network's predictions, while for the second, $\log\bar{Q}$ and $C$ contribute comparably. It should also be emphasized that, for fixed $\chi$, model (\ref{eq:D_freq_fit_2}) reinforces the quasi-universal character of the new extracted $\mathcal{D}(C)$ relation across the considered ensemble of stellar configurations and EoSs. Additionally, we introduced a new relation $(M\times\hat{f})(\chi, C)$ (\ref{eq:Mxf_fit}) for relatively slowly rotating stars with accuracy better than $1.77\%$, where $C$ has a more pronounced effect than $\chi$ on the model's predictions.

For the spacetime multipole moments, we developed a new EoS-independent relation $\bar{Q}(C,\chi,\mathcal{D})$ (\ref{eq:Qbar_fit}) with a fractional difference better than $2.32\%$, effectively embedding the established $\bar{I}(\bar{Q})$ correlation. More specifically, the suggested model reproduces the expected quasi-universal behavior of the $\bar{Q}(C)$ relation and further supports the robustness of predictions for the newly identified $\bar{Q}(\mathcal{D})$ correlation across the test data, provided that the dimensionless spin parameter $\chi$ is held constant. Above all, SHAP analysis indicates that $\mathcal{D}$ significantly influences the model's expected value,  while $C$ and $\chi$ contribute to the model's ability to capture the underlying universal patterns. For the octupole deformation, we revisited the $\bar{S}_3(\log\bar{Q})$ relation (\ref{eq:S3bar_curve_fit}), achieving fractional difference better than $6.64\%$. In addition, we further introduced an improved new parameterization, $\bar{S}_3(\chi,\mathcal{D},\log\bar{Q})$ (\ref{eq:S3bar_surface_fit}), which demonstrates relative deviation better than $3.32\%$. In addition, the analysis uncovered a robust, previously unidentified universal correlation between $\bar{S}_3$ and $\mathcal{D}$ for NS sequences with fixed spin $\chi$ values. The associated trained model (\ref{eq:S3bar_surface_fit}) successfully captures this EoS-insensitive behavior, underscoring its effectiveness. In any case, within this EoS-insensitive framework, $\mathcal{D}$ and $\log\bar{Q}$ have a pronounced impact on the model's predictions, while $\chi$ contributes significantly to the identification of spin-related patterns. For the mass hexadecapole, standard relations $\bar{M}_4(\log\bar{Q})$ (\ref{eq:M4_fit}) and $(\bar{M}_4/\bar{S}_3)(\log\bar{Q})$ (\ref{eq:M4_S3_fit}) achieved accuracies of $14.34\%$ and $13.22\%$, respectively. The inclusion of higher-order corrections in the $\bar{M}_4/\bar{S}_3$ and $\bar{M}_4$ predictions led to improved accuracies of $9.19\%$ and $8.11\%$, as shown in Eqs. (\ref{eq:M4_S3_hypersurface_fit}) and (\ref{eq:M4_hypersurface_fit}), respectively. In the $(\bar{M}_4/\bar{S}_3)(\chi, C, \bar{Q})$ model, the input features contribute with comparable mean SHAP values, indicating a balanced influence. In contrast, for the $\bar{M}_4(\chi, C, \bar{Q}, \bar{S}_3)$ model, $\bar{S}_3$ emerges as the most influential feature, followed by moderate contributions from $C$ and $\bar{Q}$, while $\chi$ exhibits the weakest impact. In addition, the marginal contribution of $\chi$ prompted us to explore a simplified $\bar{M}_4(C,\bar{Q},\bar{S}_3)$ model. Despite its reduced parameter space, this formulation achieves a relative error $\leq7.52\%$, demonstrating satisfactory predictive accuracy. In this model, parameter $C$ emerges as the dominant feature, followed by contributions from $\bar{S}_3$ and $\bar{Q}$. While this simpler relation offers a viable and good alternative, the more comprehensive model $\bar{M}_4(\chi, C,\bar{Q},\bar{S}_3)$ remains accurate when spin information is available. Specifically, for the simpler model, $2.38\%$ of NS configurations exhibit relative deviations exceeding $1\%$, whereas this fraction decreases to $0.6\%$ for the more comprehensive formulation.

We also proposed a new EoS-independent relation $\bar{\mathcal{R}}(\bar{M}, \chi, \bar{Q}, \bar{S}_3)$ for the equatorial radius (\ref{eq:R_fit}) with relative deviations better than $3.81\%$. Among the input parameters, $\bar{M}$ and $\bar{Q}$ exert the most significant contribution on the model's expected value, while $\chi$ and $\bar{S}_3$ play comparatively smaller roles. Mostly, this accurate regression model is a crucial asset for fitting functions focused on inferring NS surfaces \cite{silva2021surface, papi_vard_surface, morsink2007oblate}. It has direct applications for current observatories such as the NICER telescope \cite{gendreau2012neutron, arzoumanian2014neutron, gendreau2017searching} and upcoming missions like the enhanced X-ray Timing and Polarimetry (eXTP) \cite{zhang2019enhanced} and the Spectroscopic Time-Resolving Observatory for Broadband Energy X-rays (STROBE-X) \cite{ray2018strobe,ray2019strobe}. These upcoming missions are expected to surpass NICER’s current capabilities, providing enhanced precision in measurements of neutron star radii.

To summarize, this study introduces several universal relations that characterize NS global properties. Some of these relations have been revisited and either confirmed or enhanced, while others are newly proposed. Most test data achieve relative errors $\leq 1\%$ for the majority of the proposed relations. A concise overview of the parametrizations utilized is presented in Table \ref{tab:relations_table}, highlighting the key formulations underlying the proposed models. Additionally, Appendix \ref{app:model_characteristics} presents violin plots depicting the distributions of relative errors for each ANN inference model across the various EoS categories considered.
\begin{table}[!ht]
    \footnotesize
    \caption{\label{tab:relations_table} Summary of the universal relations investigated in this work based on an extensive ensemble of NS configurations and $70$ tabulated EoS models of cold and ultradense nuclear matter.}
    \begin{ruledtabular}
        \begin{tabular}{c|c|c}
           
          {\bf Relation} &  {\bf Suggested ANN} & {\bf Max $\%$ Error}  \\ 
            \hline
            $\bar{I}(\chi,\bar{Q})$& Improved model Eq. (\ref{eq:Ibar_fit}) & $2.76\%$\\ 
            $\mathcal{D}(\chi,\log\bar{Q})$& Improved model Eq. (\ref{eq:D_freq_fit})&$2.70\%$ \\ 
            $\mathcal{D}(C,\chi,\log\bar{Q})$&  New model Eq. (\ref{eq:D_freq_fit_2})& $2.35\%$\\ 
            $(M\times\hat{f})(\chi, C)$& New model Eq. (\ref{eq:Mxf_fit}) & $1.77\%$\\ 
            $\bar{Q}(C,\chi,\mathcal{D})$& New model Eq. (\ref{eq:Qbar_fit})& $2.32\%$\\ 
            $\bar{S}_3(\log\bar{Q})$&  model Eq. (\ref{eq:S3bar_curve_fit}) & $6.64\%$\\ 
            $\bar{S}_3(\chi,\mathcal{D},\log\bar{Q})$& New model Eq. (\ref{eq:S3bar_surface_fit})& $3.32\%$\\ 
            $\bar{M}_4(\log\bar{Q})$& Improved model Eq. (\ref{eq:M4_fit})& $14.34\%$\\ 
            $\left(\bar{M}_4/\bar{S}_3\right)(\log\bar{Q})$& Improved model Eq. (\ref{eq:M4_S3_fit}) & $13.22\%$\\ 
            $\left(\bar{M}_4/\bar{S}_3\right)(\chi, C,\bar{Q})$& New model Eq. (\ref{eq:M4_S3_hypersurface_fit}) & $9.19\%$\\ 
            $\bar{M}_4(\chi, C,\bar{Q},\bar{S}_3$)& New model Eq. (\ref{eq:M4_hypersurface_fit})& $8.11\%$\\ 
            
            $\bar{M}_4(C,\bar{Q},\bar{S}_3$)& New model Eq. (\ref{eq:M4_hypersurface_fit_v2})& $7.52\%$\\ 

            $\bar{\mathcal{R}}(\bar{M},\chi,\bar{Q},\bar{S}_3)$& New model Eq. (\ref{eq:R_fit})& $3.81\%$\\ 
            \end{tabular}
    \end{ruledtabular}
\end{table}

By integrating deep learning into the modeling process, we enhanced existing EoS-insensitive parametrizations and proposed new, highly accurate relations that capture the complex dependencies among NS properties. While this collection serves as a proof-of-concept, these relations hold significant promise for improving the analysis of gravitational wave signals from binary neutron star mergers (e.g., the $I-Q$ relation, which plays a significant role in gravitational wave detection from the inspiral phase)
\cite{agathos2015constraining, lahaye2023efficient, samajdar2020constructing,chatziioannou2015probing,paschalidis2018implications,chatziioannou2018measuring,abbott2018gw170817,abbott2019ligo_prl,kumar2019inferring,carson2019equation,chatziioannou2020neutron,tan2020neutron,narikawa2021gravitational,dietrich2021interpreting}. 
Moreover, they may be used in the effort to constrain NS properties through electromagnetic observations, such as those obtained by the NICER telescope, as well as upcoming astrophysical missions \cite{nattila2018radiation, silva2021astrophysical, al2021combining, tan2022extreme, riley2021nicer, bogdanov2021constraining, yunes2022gravitational}.

\section*{\label{sec:acknowledgements} Acknowledgements}
G. P. wants to thank George Pappas for useful discussions. G. P. acknowledges that the research work was supported by the Hellenic Foundation for Research and Innovation (HFRI) under the 5th Call for HFRI PhD Fellowships (Fellowship
No. 20450). G. V. acknowledges partial funding support from HFRI Project No. 15940. The EoSs used in this work come from the ComPOSE database. We thank the team members of CompOSE for their development. The collection of neutron star data was conducted at the
Department of Computer Science and Engineering, University of Ioannina, Greece. Especially, we would like to thank Professor Christophoros Nikou for generously
providing access to the computing resources of the Impala cluster. The training of the proposed deep learning models was carried out using the computational resources of the Department of Physics at Aristotle University of Thessaloniki (AUTh), Greece. We gratefully acknowledge the support of the AUTh IT Center, as the results presented in this work were obtained using the university’s High-Performance Computing infrastructure. Finally, we are grateful to the anonymous referee for their insightful comments and suggestions that improved the final version of this manuscript.

\section*{\label{sec:data_availability} Data Availability}
All necessary information supporting the findings of this study is included within the article. Representative code and notebook examples, along with the associated optimal parameters $\theta^\star$ (i.e., the trained model weights), are publicly available in Ref. \cite{deep_univ_gitrepo}.


\appendix

\section{\label{app:EoS_list} Equation of state models}
Here, we present the catalog of $70$ EoS models used in this study. We utilized a realistic ensemble of tabulated cold EoSs based on the associated data and the numerical code provided by the ComPOSE database  \cite{compose, compose2022compose}. The nominal list divided by each model category is provided below,
\paragraph{{\bf Hadronic EoS models}: SLy2, SKb, SkMp, SLy9, SkI3, KDE0v, SK255, Rs, SkI5, SKa, SkOp, SLy230a, SKI2, SkI4, SkI6, KDE0v1, SK272, SLy4, D1M*, QMC-RMF2, FSU2R, GPPVA(TW), DDH$\delta$, PCSB1, PCSB2, TM1e, DS(CMF)-2, DS(CMF)-4, DS(CMF)-6, DS(CMF)-8, BSk22, BSk24, BSk25, BSk26, QMC-RMF3, QMC-RMF4, TW99, MTVTC, BL(chiral)$\_$2018.}
\paragraph{{\bf Hyperonic EoS models}: OPGR(DDH$\delta$Y4), OPGR(GM1Y5), OPGR(GM1Y6), DNS, DS(CMF)-1, DS(CMF)-3, DS(CMF)-5, DS(CMF)-7.}
\paragraph{{\bf Hybrid EoS models}: DS(CMF)-1 Hybr, DS(CMF)-2 Hybr, DS(CMF)-3 Hybr, DS(CMF)-4 Hybr, DS(CMF)-5 Hybr, DS(CMF)-6 Hybr, DS(CMF)-7 Hybr, DS(CMF)-8 Hybr, DD2-FRG (2) flavors, DD2-FRG vector interaction-(2) flavors, QHC18, QHC19-B, QHC19-C, QHC19-D, QHC21T $A_T$, QHC21T $B_T$, QHC21T $C_T$, QHC21T $D_T$, QHC21 $A_{\chi}$, QHC21 $B_{\chi}$, QHC21 $C_{\chi}$, VQCD(APR) soft, VQCD(APR) intermidiate.}

\begin{figure*}[!th]
	\includegraphics[width=1.\textwidth]{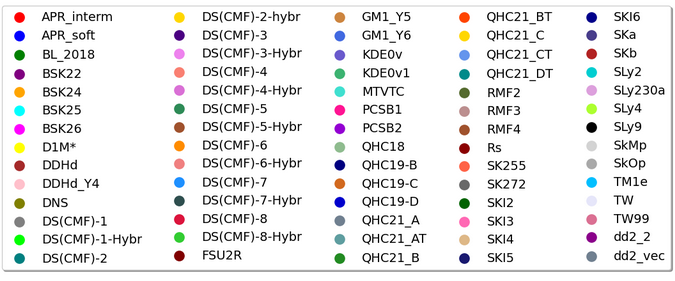}
	\caption{EoS-Color map used for the various parameter space representations given in Fig \ref{fig:params_representations} of Sec. \ref{sec:univ_rel_with_ANNs}.}%
	\label{fig:color_band}
\end{figure*}

More details about EoS samples for each category, along with their specific theoretical descriptions and the associated microphysics compositions, can be found in Appendix B of Ref. \cite{papi_vard_surface} and the CompOSE database \cite{compose, compose2022compose}. Lastly, in Fig. \ref{fig:color_band}, an EOS-color map is shown for the various parameter-space distributions demonstrated in Fig \ref{fig:params_representations} of Sec. \ref{sec:univ_rel_with_ANNs}.

\section{\label{app:train_prop_characteristics} Training properties of ANN models}

In this subsection, we present the training details of each ANN regression model that led to the final $\theta^{\star}$-best EoS-insensitive representation discussed in the main text. Table \ref{tab:ann_architectures} summarizes the training properties and outlines the learning strategy used in each case. More specifically, it details the batch size, the total number of training epochs, the learning rate schedule across fixed epoch intervals, and whether a modified sigmoid activation function was used in the final layer.

\begin{table*}[!htb]
    \caption{\label{tab:ann_architectures} Training properties for each deep EoS-insensitive training formulation, including batch size, epochs, learning rate schedule, and activation function presence.}
    \footnotesize
    \begin{tabular}{lcccccc}
        \toprule
        \textbf{Features} & \textbf{Target} & \textbf{Batch Size} & \textbf{Total Epochs} & \textbf{Epochs per $\ell r$} & \textbf{Effective $\ell r$ Schedule} & \textbf{Modified Sigmoid $\tilde{\sigma}(x)$} \\
        \midrule
        $\chi,\bar{Q}$ & $\bar{I}$ & 2048 & 1500 & $5 \times 200 \ \& \ 2\times 250$ & $\left[9 \to 7 \to 5 \to 3 \to 1 \to 0.7 \to 0.5 \right]\times 10^{-3}$ $^*$ & \xmark \\

        $\chi,\log \bar{Q}$ & $\mathcal{D}$ & 512 & 2000 & $4 \times 500$ & $\left[5 \to 3 \to 1 \to 0.5 \right]\times 10^{-4}$ & \xmark \\

        $C,\chi,\log \bar{Q}$ & $\mathcal{D}$ & 512 & 2000 & $4 \times 500$ & $\left[5 \to 3 \to 1 \to 0.5 \right]\times 10^{-4}$ & \xmark \\

        $\chi,C$ & $M \times \hat{f}$ & 16 & 2000 & $4 \times 500$ & $\left[7 \to 5 \to 3 \to 1 \right] \times 10^{-4}$ & \cmark \\

        $C,\chi,\mathcal{D}$ & $\bar{Q}$ & 256 & 2000 & $4 \times 500$ & $\left[7 \to 5 \to 3 \to 1 \right] \times 10^{-4}$ & \cmark \\

        $\log \bar{Q}$ & $\bar{S}_3$ & 1024 & 2000 & $5 \times 400$ & $\left[9\to 7 \to 5 \to 3 \to 1 \right] \times 10^{-4}$ & \cmark \\

        $\chi,\mathcal{D}, \log \bar{Q}$ & $\bar{S}_3$ & 512 & 1500 & $5 \times 300$ & $\left[9 \to 7 \to 5 \to 3 \to 1 \right] \times 10^{-3}$ & \cmark \\

        $\log \bar{Q}$ & $\bar{M}_4$ & 512 & 1000 & $4 \times 250$ & $\left[9\to 6 \to 3 \to 1 \right]\times 10^{-3}$ & \cmark \\

        $\log \bar{Q}$ & $\bar{M}_4/\bar{S}_3$ & 512 & 1000 & $4 \times 250$ & $\left[9\to 6 \to 3 \to 1 \right]\times 10^{-3}$ & \cmark \\

        $\chi,C,\bar{Q}$ & $\bar{M}_4/\bar{S}_3$ & 512 & 2200 & $4 \times 550$ & $\left[7 \to 5 \to 3 \to 1 \right] \times 10^{-4}$ & \cmark \\

        $\chi,C,\bar{Q},\bar{S}_3$ & $\bar{M}_4$ & 512 & 2200 & $4 \times 550$ & $\left[8 \to 6 \to 4 \to 2 \right] \times 10^{-4}$ & \cmark \\

         $C,\bar{Q},\bar{S}_3$ & $\bar{M}_4$ & 256 & 2000 &  $4 \times 500$ & $\left[7 \to 5 \to 3 \to 1 \right] \times 10^{-4}$ & \cmark \\

        $\bar{M},\chi,\bar{Q},\bar{S}_3$ & $\bar{\mathcal{R}}$ & 256 & 2000 & $4 \times 500$ & $\left[1 \to 0.9 \to 0.7 \to 0.5\right] \times 10^{-3}$ & \cmark \\
        \bottomrule
    \end{tabular}
    \begin{flushleft}
    $^*$ The ANN model was trained for a total of 1500 epochs, with a distinct learning rate $\eta_i$ assigned to each 200-epoch segment. This approach implements an effective, progressively decreasing learning rate schedule. Specifically, the first 200 epochs used $\eta = 9 \times 10^{-3}$, the next 200 used $\eta = 7 \times 10^{-3}$, and so on, as outlined in the table above. This strategy was consistently applied across all ANN model trainings.
    \end{flushleft}
\end{table*}

\section{\label{app:model_characteristics} Relative deviations in the test set for each regression model proposed}
\begin{figure*}[!ht]
    \centering

    \subfloat[\label{fig:violin_ibar} ANN model Eq. (\ref{eq:Ibar_fit})]{%
        \includegraphics[width=0.32\textwidth]{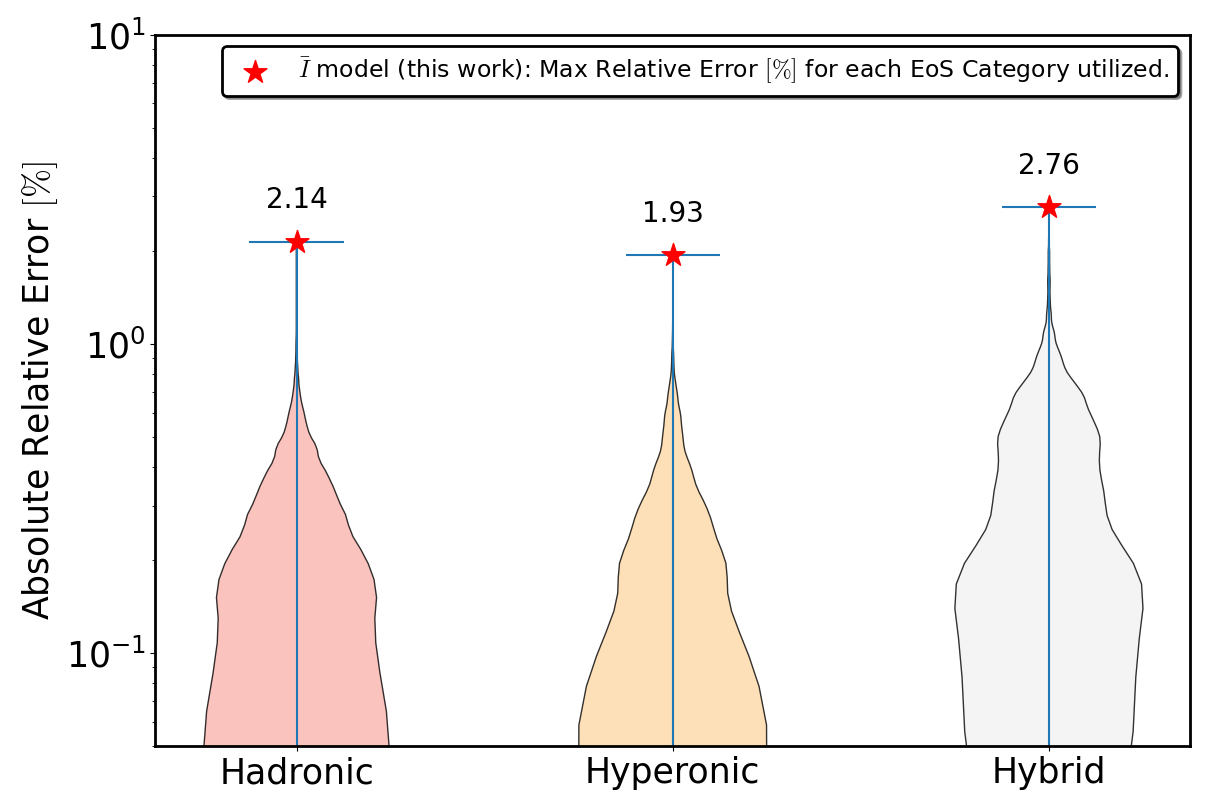}
    }
    \subfloat[ANN model Eq. (\ref{eq:D_freq_fit})]{%
        \includegraphics[width=0.32\textwidth]{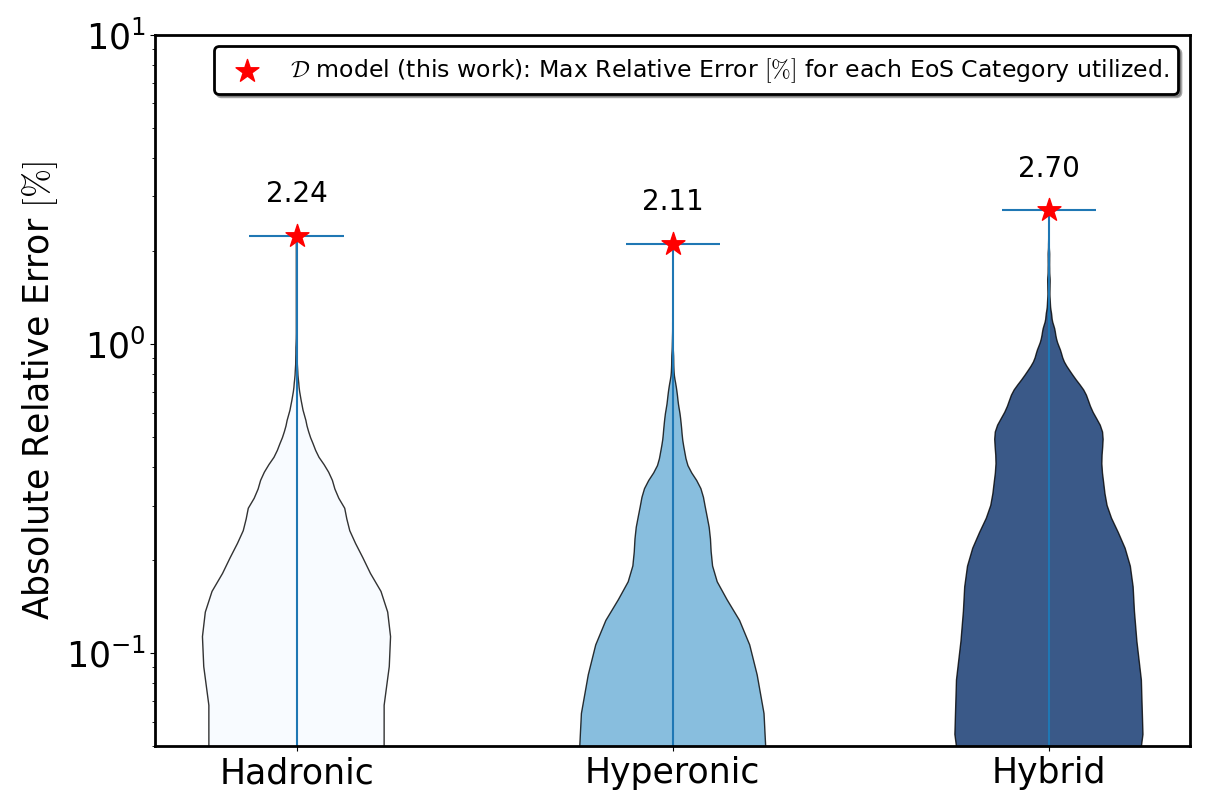}
        \label{fig:violin_d}
    }
    \subfloat[ANN model Eq. (\ref{eq:D_freq_fit_2})]{%
        \includegraphics[width=0.32\textwidth]{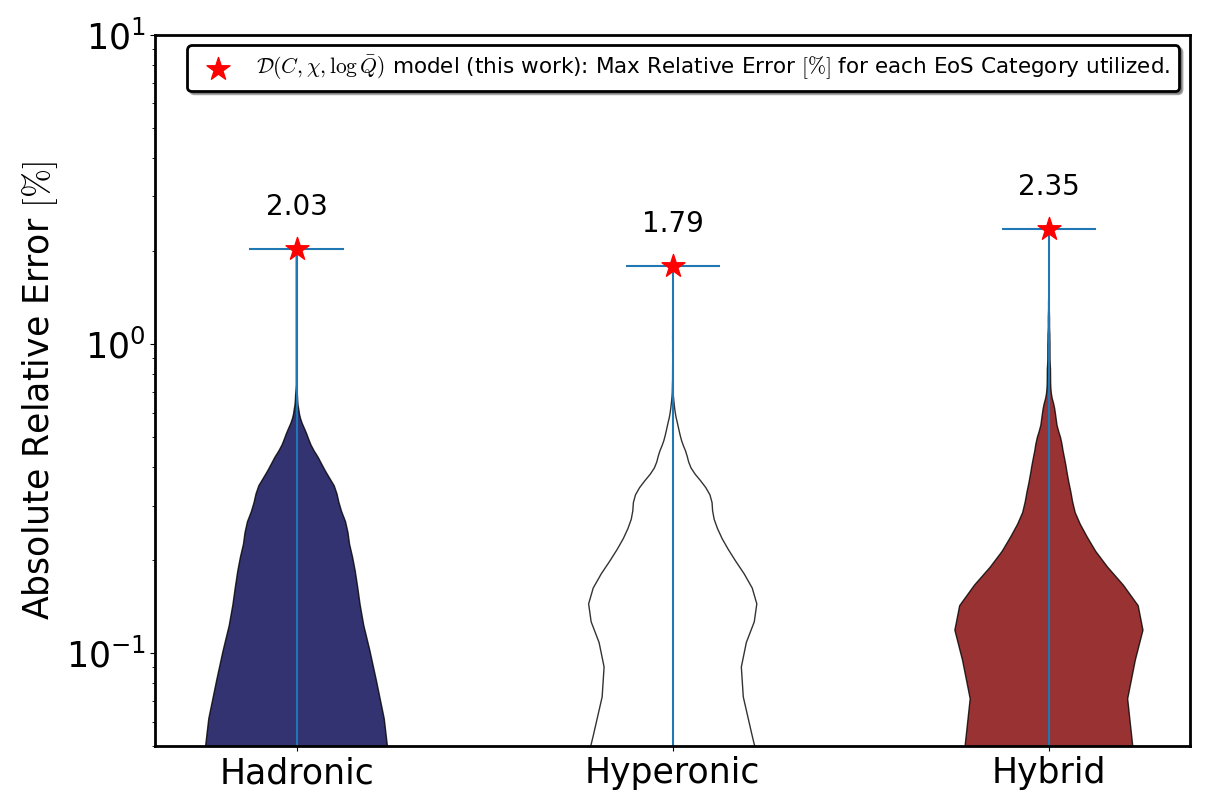}
        \label{fig:violin_d2}
    }

    \subfloat[ANN model Eq. (\ref{eq:Mxf_fit})]{%
        \includegraphics[width=0.32\textwidth]{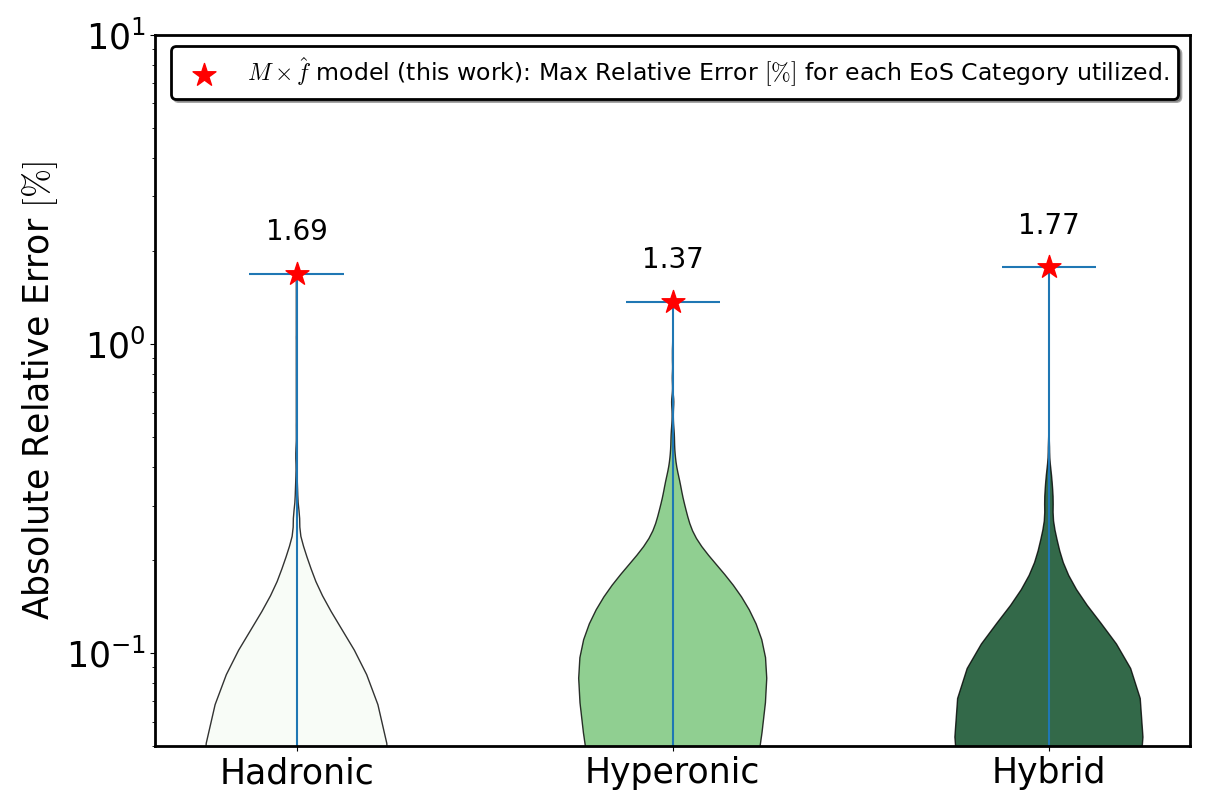}
        \label{fig:violin_mxf}
    }
    \subfloat[ANN model Eq. (\ref{eq:Qbar_fit})]{%
        \includegraphics[width=0.32\textwidth]{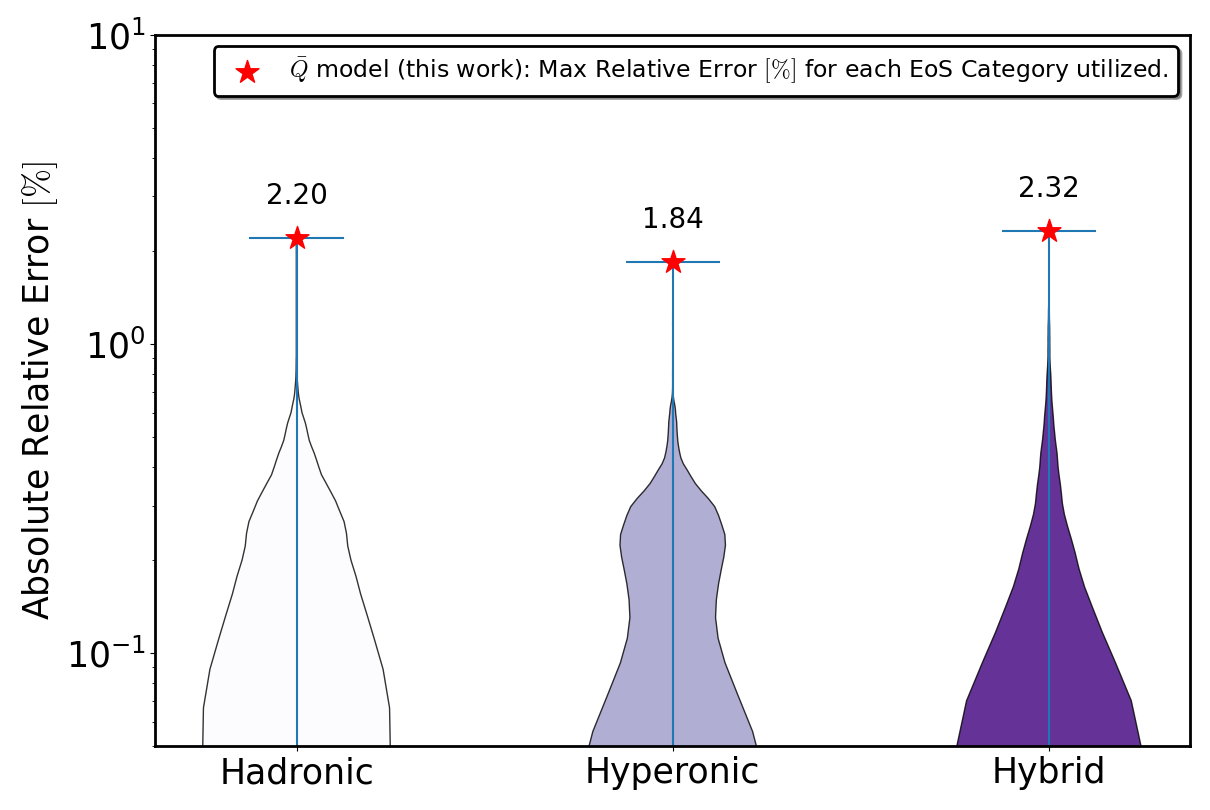}
        \label{fig:violin_qbar}
    }
    \subfloat[ANN model Eq. (\ref{eq:S3bar_curve_fit})]{%
        \includegraphics[width=0.32\textwidth]{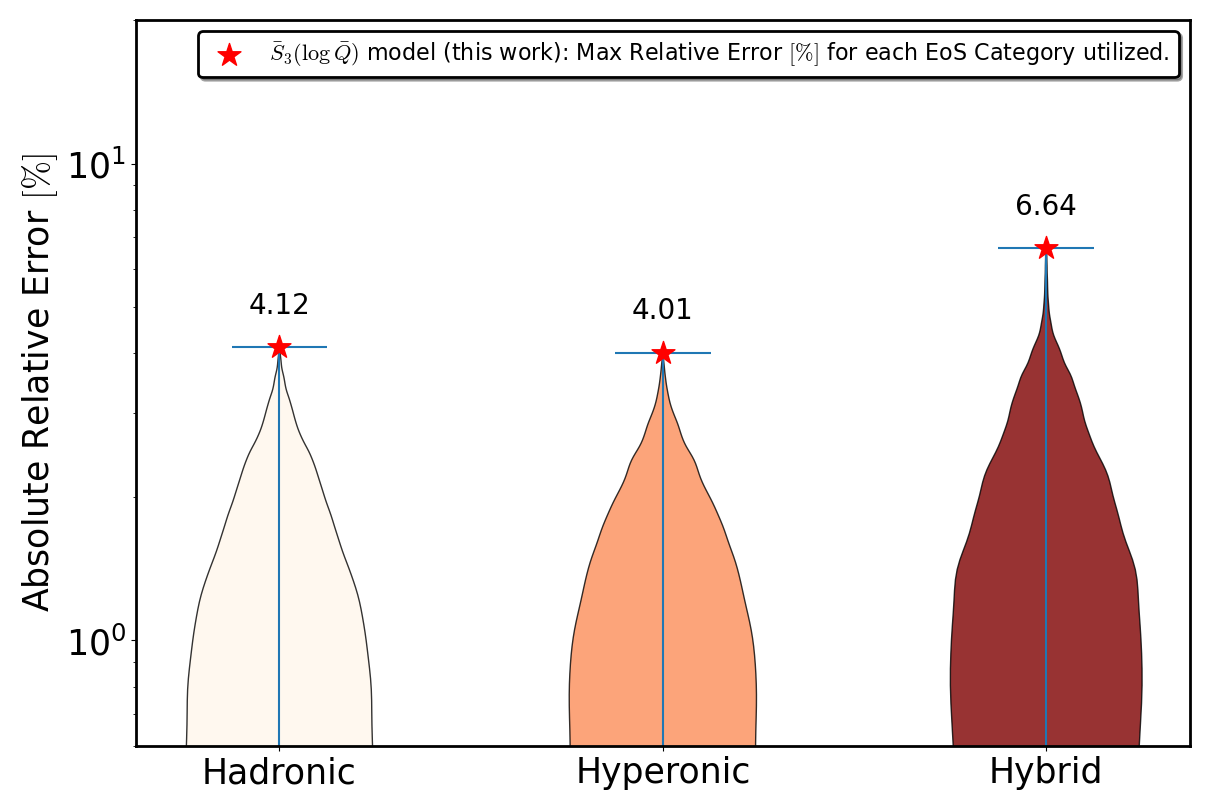}
        \label{fig:violin_s3_1}
    }

    \subfloat[ANN model Eq. (\ref{eq:S3bar_surface_fit})]{%
        \includegraphics[width=0.32\textwidth]{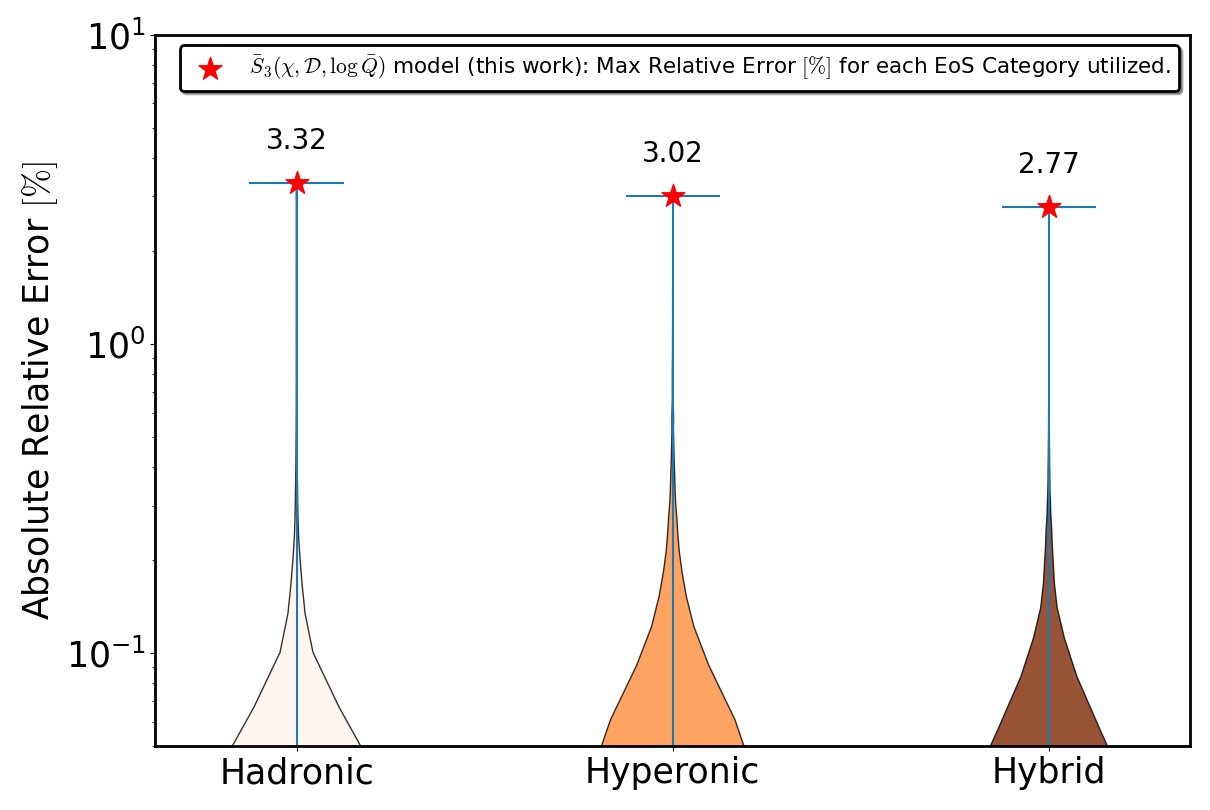}
        \label{fig:violin_s3_2}
    }
    \subfloat[ANN model Eq. (\ref{eq:M4_fit})]{%
        \includegraphics[width=0.32\textwidth]{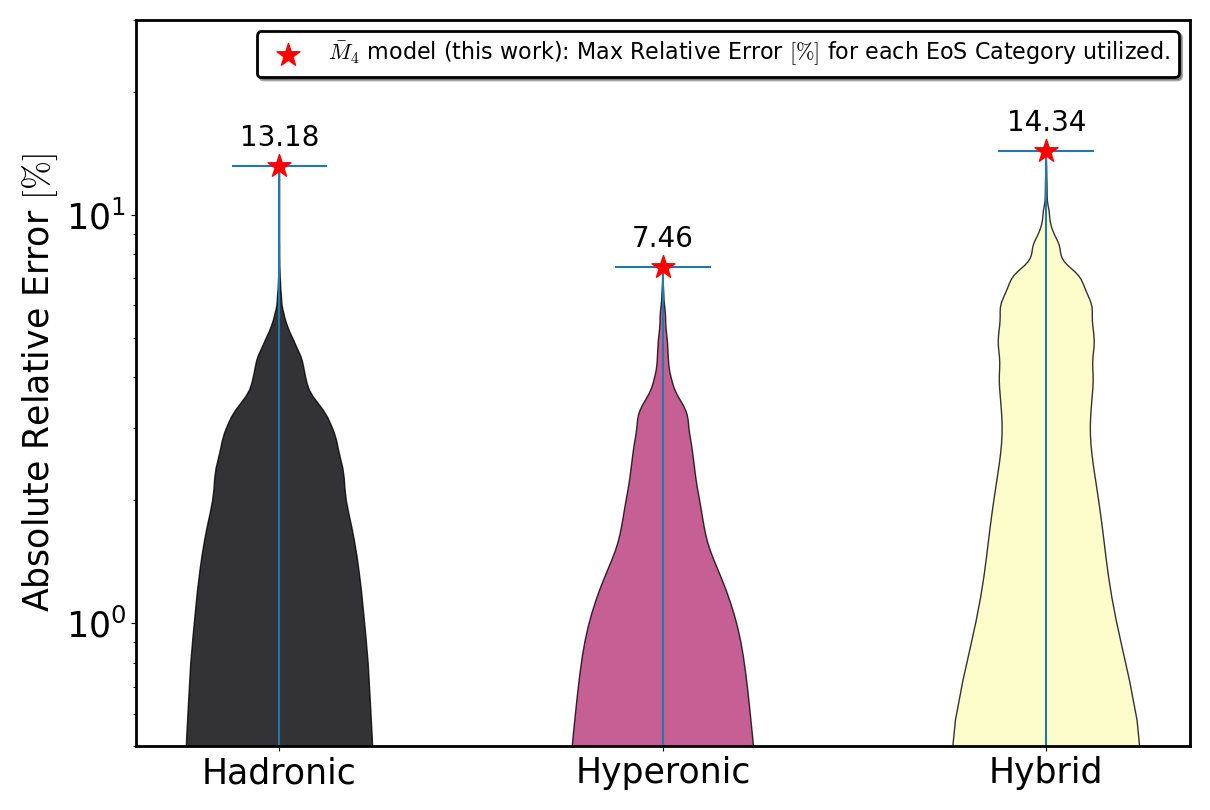}
        \label{fig:violin_m4_logq}
    }
    \subfloat[ANN model Eq. (\ref{eq:M4_S3_fit})]{%
        \includegraphics[width=0.32\textwidth]{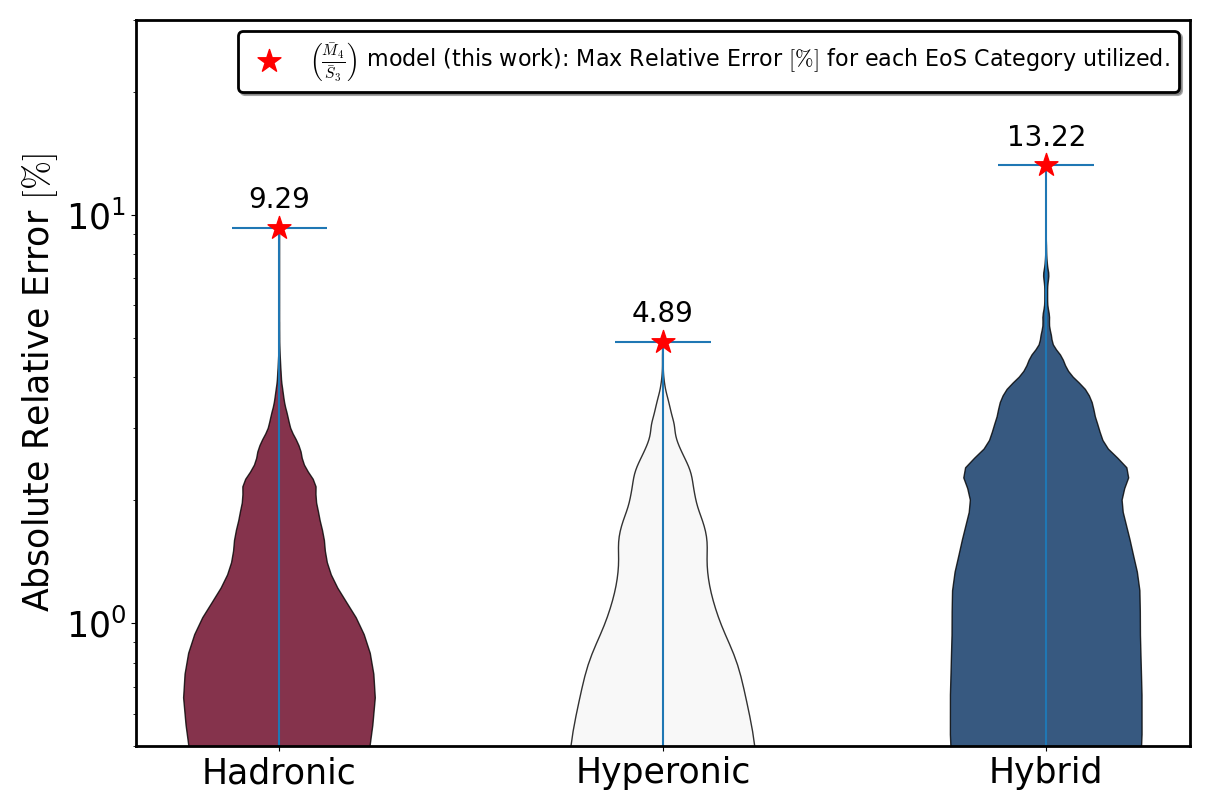}
        \label{fig:violin_m4s3_logq}
    }

    \subfloat[ANN model Eq. (\ref{eq:M4_S3_hypersurface_fit})]{%
        \includegraphics[width=0.32\textwidth]{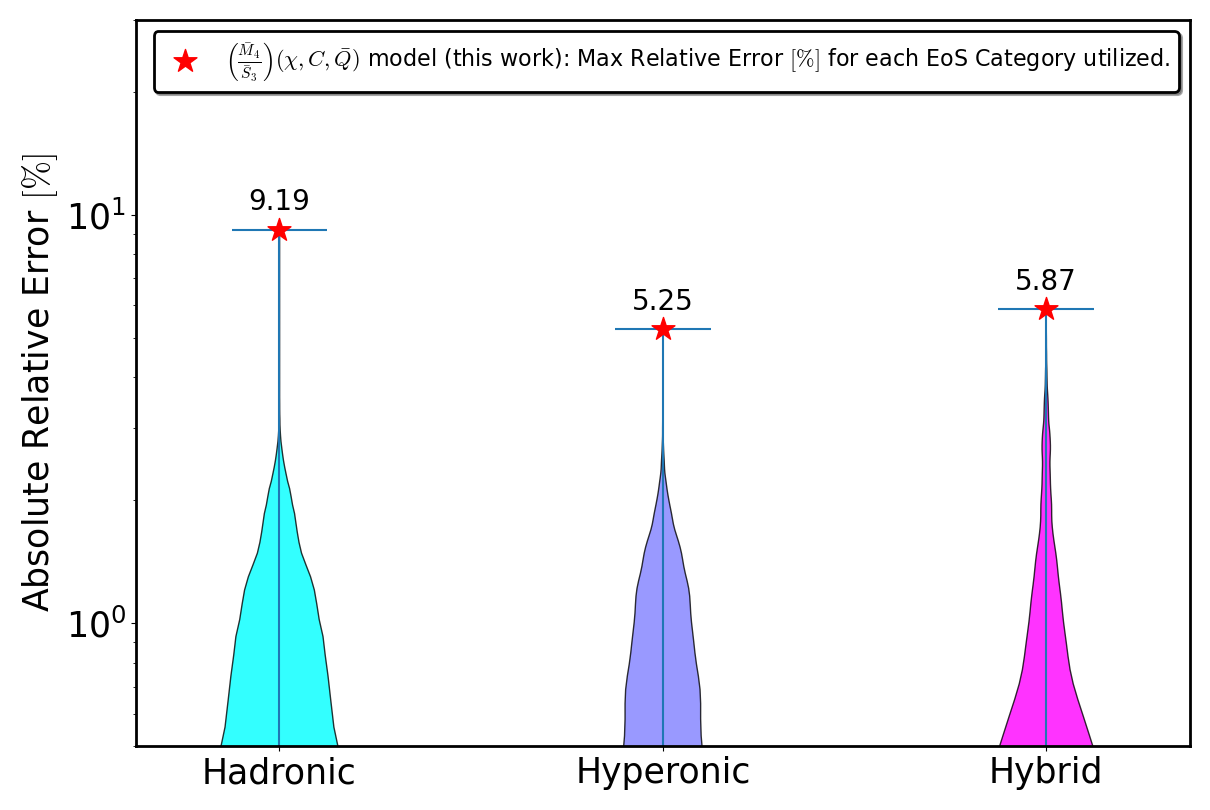}
        \label{fig:violin_m4s3}
    }
    \subfloat[ANN model Eq. (\ref{eq:M4_hypersurface_fit})]{%
        \includegraphics[width=0.32\textwidth]{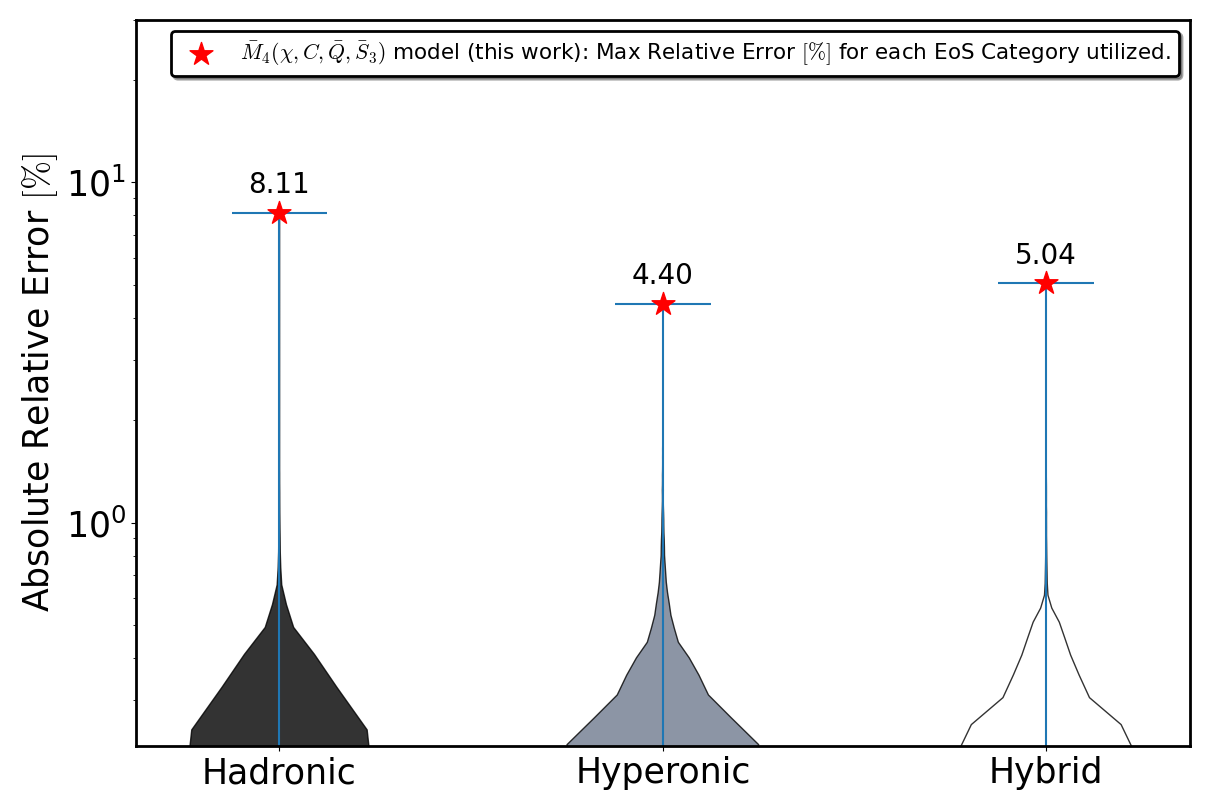}
        \label{fig:violin_m4_hypers}
    }
    \subfloat[ANN model Eq. (\ref{eq:R_fit})]{%
        \includegraphics[width=0.32\textwidth]{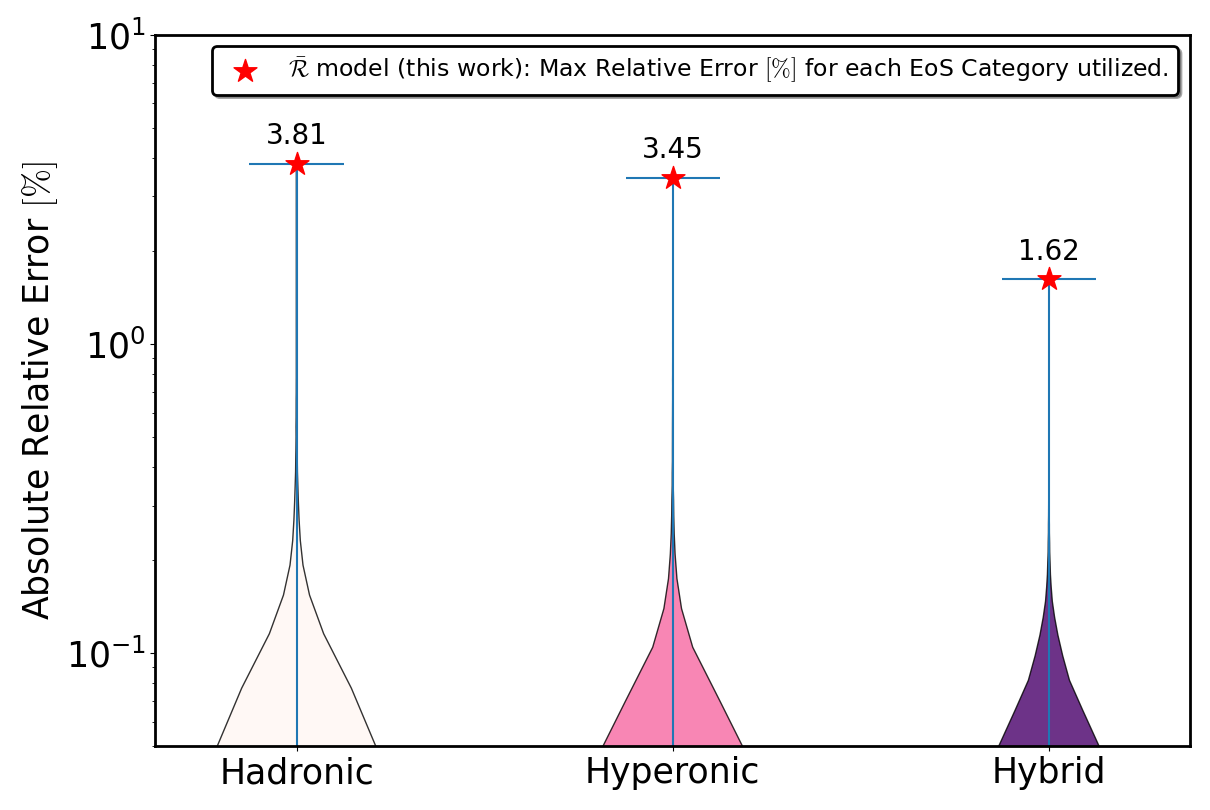}
        \label{fig:violin_req}
    }

    \caption{\label{fig:violin_plots}Violin plots illustrating the distribution of absolute $\%$ fractional difference across the associated test set for each EOS category discussed in the main text. Each panel highlights the absolute maximum relative error within the corresponding distribution.}
\end{figure*}

In this subsection, we identify the sources of relative deviations in the proposed universal relations regarding each EoS category employed. Fig. \ref{fig:violin_plots} concludes the analysis by presenting violin plots that illustrate the distribution of absolute relative errors for each ANN model examined.

\bibliography{apssamp}

\end{document}